\pdfoutput=1
\documentclass[%
superscriptaddress,
nofootinbib,
twocolumn,
amsmath,amssymb,
aps,
prd
]{revtex4-2}

\usepackage{graphicx}
\usepackage{dcolumn}
\usepackage{bm}

\def\araa{{ARA\&A}}             
\def\apj{{ApJ}}                 
\def\apjl{{ApJ}}                
\def\aap{{A\&A}}                

\def\jcap{{J. Cosmology Astropart. Phys.}}
\def\mnras{Mon.~Not.~Roy.~Astron.~Soc.}             
\def\prd{{Phys.~Rev.~D}}        
\def\prl{{Phys.~Rev.~Lett.}}    
\def\pasj{{PASJ}}               
\def\physrep{{Phys.~Rep.}}   

\newcommand{\beq}{\begin{equation}}
\newcommand{\beqa}{\begin{eqnarray}}
\newcommand{\eeq}{\end{equation}}
\newcommand{\eeqa}{\end{eqnarray}}

\newcommand{\simgt}{\lower.5ex\hbox{$\; \buildrel > \over \sim \;$}}
\newcommand{\simlt}{\lower.5ex\hbox{$\; \buildrel < \over \sim \;$}}

\newcommand{\bd}[1]{\mbox{\boldmath $#1$}}

\usepackage[usenames,dvipsnames]{xcolor}
\newcommand{\rev}[1]{{\color{black} #1}}

\begin{document}


\title{Searching for eV-mass Axion-like Particles with\\ 
Cross Correlations between Line Intensity and Weak Lensing Maps}

\author{Masato Shirasaki}
 \email{masato.shirasaki@nao.ac.jp}
\affiliation{%
National Astronomical Observatory of Japan (NAOJ), Mitaka, Tokyo 181-8588, Japan
}%
\affiliation{
The Institute of Statistical Mathematics,
Tachikawa, Tokyo 190-8562, Japan
}

\date{\today}

\begin{abstract}
We study cross correlations between line intensity and 
weak lensing maps to search for axion-like particles (ALPs).
Radiative decay of eV-mass ALPs can contribute to 
cosmic background emissions at optical and infrared wavelengths.
Line intensity mapping is a unique means of measuring the background emission at a given photon frequency.
If ALPs constitute the abundance of cosmic dark matter,
line intensity maps can correlate with large-scale structures probed by
weak gravitational lensing effects in galaxy imaging surveys.
We develop a theoretical framework to predict the cross correlation.
We then explore potentiality of probing ALPs with the cross correlation in upcoming galaxy-imaging and spectral surveys.
Assuming SPHEREx and the Vera Rubin Observatory's Legacy Survey of Space and Time (LSST),
we find that the cross correlation by the ALP decay can be greater than the astrophysical-line counterparts at wavelength of $\sim3000\, {\rm nm}$ for ALPs with a particle mass of $m_a\sim1\, \mathrm{eV}$ and a particle-to-two-photons coupling of 
$g_{a\gamma\gamma}\sim1\times10^{-11}\, \mathrm{GeV}^{-1}$.
We also predict that a null detection of the cross correlation
can place a $2\sigma$-level upper bound of $g_{a\gamma\gamma} \lower.5ex\hbox{$\; \buildrel < \over \sim \;$} 10^{-11}\, \mathrm{GeV}^{-1}$ for eV-mass ALPs,
improving the current constraint by a factor of $\sim10$.
We then make a forecast of expected constraints on ALP parameters in SPHEREx and LSST by Fisher analysis, providing a guideline of searching for the ALP decay with the large-scale structure data. 
\end{abstract}

\maketitle


\section{Introduction}

The nature of cosmic dark matter is one of the long-standing mysteries in modern astronomy and physics.
Dark matter accounts for about 80\% of the average mass density in our universe \cite{2020A&A...641A...6P}, but can not be explained by the Standard Model of particle physics.
The QCD axion is a well-motivated candidate of dark matter \cite{1983PhLB..120..127P, 1983PhLB..120..133A, 1983PhLB..120..137D},
while it is originally proprosed as a solution to the strong CP problem \cite{1977PhRvL..38.1440P, 1977PhRvD..16.1791P, 1978PhRvL..40..279W, 1978PhRvL..40..223W}.
The QCD axion is a pseudoscalar boson with an approximate shift symmetry.
More general light pseudoscalars are known as axion-like particles, or ALPs. 
ALPs are pseudo-Nambu Goldstone bosons produced through 
the breaking of global $U(1)$ symmetries, or zero modes of higher dimensional gauge fields \cite{2006JHEP...06..051S, 2010PhRvD..81l3530A, 2012JCAP...06..013A, 2012JHEP...10..146C}. 
Because ALPs are not associated with the strong CP problem,
they can exhibit a wider range of couplings and masses.
Hence, ALPs are considered to be a compelling candidate of dark matter \cite{2019PhRvD.100a5049B}.

In this paper, we are interested in the search for ALPs with astronomical datasets.
In particular, we pay special attention to radiative decay of ALPs with their particle mass 
of about 1 eV.
The eV-mass ALPs can decay into two photons with photon frequencies of $\sim10^{14}\, \mathrm{Hz}$,
corresponding to optical and infrared wavelenghs.
Galaxy surveys at optical and infrared bands are about to reach their peak.
Upcoming large surveys include the Vera Rubin Observatory's Legacy Survey of Space and Time (LSST)\footnote{\url{https://www.lsst.org/}},
The Nancy Roman Space Telescope\footnote{\url{https://roman.gsfc.nasa.gov/}},
Euclid\footnote{\url{https://sci.esa.int/web/euclid}},
and SPHEREx\footnote{\url{https://www.jpl.nasa.gov/missions/spherex/}}.

Among the upcoming surveys, SPHEREx provides a unique opportunity of studying cumulative background emissions at optical and infrared bands with line-intensity mapping techniques.
The line-intensity mapping measures photon intensity at a narrow range of frequency along a given line-of-sight direction, allowing us to explore faint astrophysical sources that are difficult to be detected individually.
If a fraction of cosmic dark matter would consist of eV-mass ALPs, 
ALP-decay photons contribute to the line intensity maps obtained by SPHEREx.
At the same time, ALPs are at work as building blocks of large-scale structures in the universe
and they distort shapes of distant galaxies by weak gravitational lensing effects (see, e.g. Ref.~\cite{Bartelmann2001} for a review).
Hence, ALPs can affect two different observables of 
line intensity maps and weak lensing effects, introducing spatial cross correlations between the two.
Weak gravitational lensing effects are now recognized as a main science driver in galaxy imaging surveys
and most upcoming imaging surveys are aimed at measuring the effects with unprecedented accuracy.
It is thus strongly motivated to study expected cross correlations between the line intensity maps and weak lensing effects in galaxy imaging surveys for the search for ALPs, enhancing the science returns from large galaxy surveys.

It would be worth noting that Ref.~\cite{2018PhRvD..98f3524C} has already proposed the ALP search by using cross correlations between line intensity maps and other large-scale structure data.
Nevertheless, they mainly focus on a biased tracer of large scale structures such as galaxies, while
we consider the weak lensing effects to extract unbiased information about underlying cosmic mass density fields. We also study possible astrophysical contributions to the cross correlation analysis, which are ignored in Ref.~\cite{2018PhRvD..98f3524C}.
Recently, Ref.~\cite{2020arXiv201200771B} proposed a statistical approach to constrain decaying dark matter with line-intensity maps alone. They studied a joint analysis of auto correlation functions and one-point probability distributions in line intensity maps. 
Our approach is thought to be complementary to that in Ref.~\cite{2020arXiv201200771B}.

The rest of the present paper is organized as follows.
In Section~\ref{sec:obs}, we describe observables of interest and relevant physical effects to the observables.
In Section~\ref{sec:stats}, we summarize a theoretical model to predict cross correlation functions between line intensity maps and weak lensing effects in galaxy shapes.
We present the results in Section~\ref{sec:results}.
Concluding remarks and discussions are given in Section~\ref{sec:con}.
Throughout, we consider a flat universe with the standard cosmological parameters
$H_0=100h\, {\rm km}\, {\rm s}^{-1}$
with $h=0.68$, the average matter density 
$\Omega_{\rm m0}=0.315$, 
the baryon density $\Omega_{\rm b0}=0.0491$,
the cosmological constant $\Omega_{\Lambda}=0.685$,
the spectral index of initial curvature fluctuations $n_s=0.97$,
and the amplitude of matter density fluctuations within 
$8\,h^{-1}\, {\rm Mpc}$, $\sigma_8=0.83$.
Note that we adopt natural units with $c=\hbar=1$,
where $c$ is the speed of light and $\hbar$ is the reduced Planck constant.

\section{Observables}\label{sec:obs}

In this section, we introduce two observables of interest, 
the line-intensity and weak-lensing maps.
The former provides the cumulative emission from sources fainter than detection limits in spectroscopic surveys, while the latter is the projected mass density distribution obtained in galaxy imaging surveys.

\subsection{Line Intensity}

The observed intensity $I$ at photon frequency $\nu$ 
along a given direction $\bd{\theta}$ is expressed as
\beqa
I(\bd{\theta}\,|\,\nu) = \frac{1}{4\pi}\int \frac{\mathrm{d}\chi}{1+z} \, 
\epsilon(\bd{r}, \nu^{\prime}, z), \label{eq:intensity}
\eeqa
where $\nu^{\prime}=\nu (1+z)$ is the photon frequency emitted at the source redshift $z$,
$\chi$ is the comoving distance to the source,
and $\epsilon(\bd{r}, \nu^{\prime}, z)$ represents the volume emissivity (i.e., the photon energy emitted per unit volume, time, and energy range)
at the three-dimensional position of $\bd{r} = (\chi(z)\, \bd{\theta}, \chi(z))$ and the frequency $\nu^{\prime}$.
Note that we define the volume emissivity in the comoving coordinate in Eq.~(\ref{eq:intensity}).
In line intensity mapping measurements, we are interested in 
any emission lines
which can be observed at the frequency $\nu$.
Because the line width is usually negligible 
compared to the spectroscopic resolution in the instrument\footnote{\rev{The virial velocity dispersion in typical galaxy-sized halos with $M=10^{12-13}\, M_\odot$ can be estimated as $\sim200\, \mathrm{km}/\mathrm{s}$. Hence, we can ignore the velocity dispersion effect in the analysis 
as long as setting the frequency resolution to $\simlt 100$.}}, 
we assume that the volume emissivity is given by
\beqa
\epsilon(\bd{r}, \nu, z) = \rho_{L}(\bd{r}, z)\, 
\delta^{(1)}_{\mathrm{D}}(\nu-\nu_0), \label{eq:emissivity}
\eeqa
where $\rho_{L}$ is the luminosity density of sources,
$\delta^{(n)}_{\mathrm{D}}$ is the $n$-dimensional Dirac delta function, 
$\nu_0$ is the rest-frame line frequency of interest.
Throughout this paper, we define the intensity in units of
$\mathrm{erg}/\mathrm{s}/\mathrm{cm}^2/\mathrm{Hz}/\mathrm{str}$.
In the following, we briefly describe observable lines at optical and infrared bands, associated with the ALP decay and some relevant astrophysical sources.

\subsubsection{ALP decay}

Axion-like particles (ALPs) are light, weakly-interacting pseudo-scalar particles. 
A common feature among ALP models is that they can decay into photons with their decay rate $\Gamma$ being
\beqa
\Gamma &=& \frac{g^2_{a\gamma\gamma}\, m^3_a}{64\pi} \\
       &=& 7.556\times 10^{-26}\, \mathrm{s}^{-1} \,
       \left(\frac{g_{a\gamma\gamma}}{10^{-10}\, \mathrm{GeV}^{-1}}\right)^{2}\,
       \left(\frac{m_a}{1\, \mathrm{eV}}\right)^{3},
\eeqa
where $m_a$ and $g_{a\gamma\gamma}$ are the mass and the particle-to-two-photon coupling constant of ALPs, respectively.
In this decay, the emitted photon frequency is given by 
$\nu_a = m_a / (4 \pi) = 1.21\times10^{14} \, \mathrm{Hz}\, (m_a / 1\, \mathrm{eV})$.

Suppose that ALPs can constitute a part of dark matter,
we can express the luminosity density due to the ALP decay as
\beqa
\rho_{L, a}(\bd{r}, z) = \Gamma\, f_a\, \rho_{\mathrm{DM}}(\bd{r}, z), \label{eq:luminosity_density_ALP}
\eeqa
where $f_{a}$ is the fraction of ALPs in the dark matter and $\rho_{\mathrm{DM}}(\bd{r}, z)$ represents the dark matter density at redshift $z$. In the following, we consider the case of $f_a = 1$ for simplicity\footnote{Our parameter forecasts can be easily generalized for $f_a<1$ if one replaces $g_{a\gamma\gamma}$ with $g_{a\gamma\gamma}\, f_{a}^{1/2}$.}. 
Using Eqs.~(\ref{eq:intensity}), (\ref{eq:emissivity}), and 
(\ref{eq:luminosity_density_ALP}), one can find
\beqa
I_a(\bd{\theta} \, |\, \nu) &=& 
\int \mathrm{d}\chi \, W_{\mathrm{DM}}(z|\nu) \, \left[1+\delta_{\mathrm{DM}}(\bd{r}, z)\right], \\
W_{\mathrm{DM}}(z|\nu) &=& \frac{\Gamma}{4\pi}\frac{\Omega_{\mathrm{DM}}
\rho_{\mathrm{crit},0}}{\nu_a\, H(z)}\, \delta^{(1)}_{\mathrm{D}}(\chi-\chi_a),
\eeqa
where $\Omega_{\mathrm{DM}}=\Omega_{\rm m0}-\Omega_{\rm b0}\sim0.26$ 
is the dimensionless dark matter density,
$\rho_{\mathrm{crit,0}}$ is the critical density at present,
$\delta_{\mathrm{DM}}$ is the dark matter density contrast,
$H(z)$ is the Hubble expansion parameter at redshift $z$,
$\chi_a$ is the comoving distance to the redshift $z_a$ which is defined by
$z_a \equiv\nu_a/\nu-1$.

\subsubsection{Astrophysical sources}

The $\mathrm{H}_{\alpha}$ line is a Balmer line 
that corresponds to a transition between energy levels 
$n=3$ to $n=2$ of neutral hydrogen.
It is commonly used to study star-forming galaxies 
and quasars at $z\simlt2$ so far.
The rest-frame wavelength of the $\mathrm{H}_{\alpha}$ line
is given by 656.28 $\mathrm{nm}$, which is suitable for studying 
faint galaxies as well as diffuse emissions from intergalactic media in ongoing line-intensity mapping surveys.

It is known that the strength of the $\mathrm{H}_{\alpha}$ line shows a tight correlation with the star-formation rate (SFR) 
of galaxies \cite{1998ARA&A..36..189K}.
The commonly used relation between the SFR and the intrinsic $\mathrm{H}_{\alpha}$ luminosity is provided as
\beqa
L_{\mathrm{H}_{\alpha}}
 =
1.26 \times 10^{41}\, (\mathrm{erg}/\mathrm{s}) \, \left(\frac{\mathrm{SFR}}{1\, M_{\odot}/\mathrm{yr}}\right),
\label{eq:Ha_K_law}
\eeqa
where $L_{\mathrm{H}_{\alpha}}$ is the intrinsic luminosity.
The observed $\mathrm{H}_{\alpha}$ luminosity 
$L^{\mathrm{obs}}_{\mathrm{H}_{\alpha}}$ 
is subject to dust extinction. We obtain the observed luminosity by accounting for the extinction with a magnitude of $A_{\mathrm{H}_{\alpha}}$,
\beqa
L^{\mathrm{obs}}_{\mathrm{H}_{\alpha}}=10^{-A_{\mathrm{H}_{\alpha}}/2.5}\, L_{\mathrm{H}_{\alpha}}. \label{eq:L_Ha_obs}
\eeqa
In this paper, we adopt $A_{\mathrm{H}_{\alpha}}=1\, \mathrm{mag}$ which is a typical value in large $\mathrm{H}_{\alpha}$ surveys \cite{2013MNRAS.428.1128S}.

To evaluate the $\mathrm{H}_{\alpha}$ luminosity density, we assume that any star formation processes occur at the centers of single dark matter halos and SFRs can be determined by a function of halo masses and redshifts.
In this halo-based model, we can express the $\mathrm{H}_{\alpha}$ luminosity density $\rho_{L,\mathrm{H}_{\alpha}}$ as
\beqa
\rho_{L,\mathrm{H}_{\alpha}}(\bd{r}, z)
&=& 
\sum_{i} \,
L^{\mathrm{obs}}_{\mathrm{H}_{\alpha}}(M_i, z) \, \delta^{(3)}_{\mathrm{D}}(\bd{r}-\bd{r}_i) \label{eq:Ha_rho1} \\
&=& 
{\cal L}_{\mathrm{H}_{\alpha}} \sum_{i}  
\dot{M}_{*}(M_i, z)
\delta^{(3)}_{\mathrm{D}}(\bd{r}-\bd{r}_i), \label{eq:Ha_rho2}
\eeqa
where 
$\dot{M}_{*}$ is the SFR in units of $M_{\odot}/\mathrm{yr}$,
$M_i$ and $\bd{r}_i$ represent a halo mass and spatial coordinate of the $i$-th halo, the index $i$ runs over all the halos at redshift $z$ in a unit volume,
and we introduce 
${\cal L}_{\mathrm{H}_{\alpha}} = 1.26 \times 10^{41}\, (\mathrm{erg}/\mathrm{s}) \times 10^{-A_{\mathrm{H}_{\alpha}}/2.5}$.

We start from computing the average luminosity density based on Eq.~(\ref{eq:Ha_rho1}). It is given by
\beqa
\bar{\rho}_{L,\mathrm{H}_{\alpha}}(z) 
= 10^{-A_{\mathrm{H}_{\alpha}}/2.5} \, \int \mathrm{d}L_{\mathrm{H}_{\alpha}}
\, \frac{\mathrm{d}n}{\mathrm{d}L_{\mathrm{H}_{\alpha}}}\, L_{\mathrm{H}_{\alpha}}, \label{eq:mean_Ha_rho}
\eeqa
where $\mathrm{d}n/\mathrm{d}L_{\mathrm{H}_{\alpha}}$
is the $\mathrm{H}_{\alpha}$ luminosity function (LF)\footnote{Note that one can express $\bar{\rho}_{L,\mathrm{H}_{\alpha}}(z)$
by integrating the halo mass function and 
the SFR-to-halo-mass relation as well.
Nevertheless, we think Eq.~(\ref{eq:mean_Ha_rho}) would 
give a more reasonable estimate, because 
the SFR-to-halo-mass relation is poorly constrained at present.}.
The mean intensity of the $\mathrm{H}_{\alpha}$ emission is then computed as
\beqa
\bar{I}_{\mathrm{H}_{\alpha}}(z) = \frac{1}{4\pi}\frac{1}{\nu_{\mathrm{H}_{\alpha}} H(z)} 
\bar{\rho}_{L,\mathrm{H}_{\alpha}}(z), \label{eq:mean_Ha_I}
\eeqa
where $\nu_{\mathrm{H}_{\alpha}}$ is the rest-frame frequency of the $\mathrm{H}_\alpha$ line.
Ref.~\cite{2017arXiv171109902S} provides a phenomenological model of Eq.~(\ref{eq:mean_Ha_I}) using Eq.~(\ref{eq:mean_Ha_rho}) and the current constraint of LFs at different redshifts. Their model is expressed as 
\beqa
&&\bar{I}_{\mathrm{H}_{\alpha}}(z) = 
3.326\times10^{-9} \, \frac{0.027+0.28z}{1+(z/4.8)^{5.3}} \,
10^{-A_{\mathrm{H}_{\alpha}}/2.5} \nonumber \\
&&
\qquad
\times \left(\frac{\nu^{-1}_{\mathrm{H}_{\alpha}}\, H^{-1}(z)}{h^{-1}\, \mathrm{Mpc}\, \mathrm{Hz}^{-1}}\right)\, \mathrm{erg}/\mathrm{s}/\mathrm{cm}^2/\mathrm{Hz}/\mathrm{str}.
\eeqa
Finally, we express the observed $\mathrm{H}_{\alpha}$ intensity as
\beqa
I_{\mathrm{H}_{\alpha}}(\bd{\theta} \, |\, \nu) &=& 
\int \mathrm{d}\chi \, W_{\mathrm{\mathrm{H}_{\alpha}}}(z|\nu) \, \left[1+\delta_{\mathrm{\mathrm{H}_{\alpha}}}(\bd{r}, z)\right], \\
W_{\mathrm{\mathrm{H}_{\alpha}}}(z|\nu) &=& \bar{I}_{\mathrm{H}_{\alpha}}(z)\,  \delta^{(1)}_{\mathrm{D}}(\chi-\chi_{\mathrm{H}_{\alpha}}),
\eeqa
where $\chi_{\mathrm{H}_{\alpha}}$ is the comoving distance to the redshift $z_{\mathrm{H}_{\alpha}}$ which is defined by
$z_{\mathrm{H}_{\alpha}} \equiv \nu_{\mathrm{H}_{\alpha}}/\nu-1$,
and $\delta_{\mathrm{\mathrm{H}_{\alpha}}}$ is the density contrast in the $\mathrm{H}_{\alpha}$ luminosity density.

In actual observations, several lines can contaminate the observed $\mathrm{H}_{\alpha}$ intensity maps because the measurement relies on spectroscopy at single wavelengths. The contamination lines include
ionized oxygen $[\mathrm{O}_\mathrm{I\hspace{-.1em}I}]$ 372.7 nm and 
$[\mathrm{O}_\mathrm{I\hspace{-.1em}I\hspace{-.1em}I}]$ 500.7 nm lines, 
the hydrogen $\mathrm{H}_{\beta}$ 486.1 nm,
and Lyman $\alpha$ (Ly$\alpha$) 121.6 nm lines\footnote{
In principle, 
$[\mathrm{N}_\mathrm{I\hspace{-.1em}I}]$ 658.3 nm/654.8 nm and $[\mathrm{S}_\mathrm{I\hspace{-.1em}I}]$ 671.7 nm/673.1 nm doublet lines
also contribute to the $\mathrm{H}_{\alpha}$ intensity maps. Nevertheless,
the SPHEREx is not able to distinguish the $\mathrm{H}_{\alpha}$ line between these doublet lines in frequency spaces. In this paper, we include the contribution from these doublet lines as follows in Ref.~\cite{2017arXiv171109902S}.
We assume that the $\mathrm{N}_\mathrm{I\hspace{-.1em}I}$ line contributes 22\% of the sum the $\mathrm{H}_{\alpha}$ and $\mathrm{N}_\mathrm{I\hspace{-.1em}I}$ line intensities \cite{2003A&A...401.1063A}, while 
the $\mathrm{S}_\mathrm{I\hspace{-.1em}I}$ doublet line intensity is set to 12\% of the $\mathrm{H}_{\alpha}+\mathrm{N}_\mathrm{I\hspace{-.1em}I}+\mathrm{S}_\mathrm{I\hspace{-.1em}I}$ line intensity \cite{2016MNRAS.460.3587M}. 
}.
To take into account these interlopers, we assume the published relations between line luminosity and the SFR. For a given line denoted as $Q$,
we express the luminosity-SFR relation as
\beqa
L^{\mathrm{obs}}_{Q} = {\cal L}_{Q} \left(\frac{\mathrm{SFR}}{1\, M_{\odot}/\mathrm{yr}}\right), \label{eq:L_Q_obs}
\eeqa
where $L^{\mathrm{obs}}_Q$ is the observed luminosity for the line $Q$,
and ${\cal L}_{Q}$ is the scale luminosity in units of $\mathrm{erg}/\mathrm{s}$. Note that the term ${\cal L}_{Q}$ 
includes the dust attenuation effect as $10^{-A_{Q}/2.5}$ 
where $A_{Q}$ is the extinction for the line $Q$.
Using Eqs.~(\ref{eq:Ha_K_law}), (\ref{eq:L_Ha_obs}), (\ref{eq:mean_Ha_rho}) and (\ref{eq:L_Q_obs}),
one can find
\beqa
\bar{\rho}_{L,Q} = \frac{{\cal L}_Q}{{\cal L}_{\mathrm{H}_{\alpha}}}\bar{\rho}_{L, \mathrm{H}_{\alpha}},
\eeqa
where $\bar{\rho}_{L,Q}$ is the luminosity density for the line $Q$.
Table~\ref{tab:luminosy_to_SFR} summarizes the model of ${\cal L}_Q$ in this paper.

\begin{table}[!t]
\begin{center}
\begin{tabular}{|c|c|c|c|}
\tableline
 Line &
 $\lambda_Q$ (nm) &
 ${\cal L}_Q\, (\mathrm{erg/s})$ &
 $A_Q\, (\mathrm{mag})$
 \\ \hline
$\mathrm{H}_\alpha$
& 656.28 & $1.26\times10^{41}\times 10^{-A_{\mathrm{H}_\alpha}/2.5}$ $^($\footnotemark[1]$^)$ & 1 $^($\footnotemark[2]$^)$ \\
$\mathrm{O}_\mathrm{I\hspace{-.1em}I\hspace{-.1em}I}$
& 500.7 & $1.32\times10^{41}\times 10^{-A_{\mathrm{O}_\mathrm{I\hspace{-.1em}I\hspace{-.1em}I}}/2.5}$ $^($\footnotemark[3]$^)$ & 1.35 $^($\footnotemark[4]$^)$ \\
$\mathrm{H}_\beta$
& 486.1 & $4.43\times10^{40}\times 10^{-A_{\mathrm{H}_\beta}/2.5}$ $^($\footnotemark[5]$^)$ & 1.35 $^($\footnotemark[6]$^)$\\
$\mathrm{O}_\mathrm{I\hspace{-.1em}I}$
& 372.7 & $7.18\times10^{40}\times10^{-A_{\mathrm{O}_\mathrm{I\hspace{-.1em}I}}/2.5}$ $^($\footnotemark[7]$^)$ & 0.62 $^($\footnotemark[8]$^)$ \\
Ly$\alpha$
& 121.6 & $1.10\times10^{42}\times10^{-A_{\mathrm{Ly}\alpha}/2.5}$ $^($\footnotemark[9]$^)$ & 0
\\ \tableline
\end{tabular}
\caption{
\label{tab:luminosy_to_SFR}
Characteristics of astrophysical lines in this paper.
For a given line $Q$, $\lambda_{Q}$ represents the rest-frame wavelength,
${\cal L}_{Q}$ is the scale luminosity in the luminosity-SFR relation in Eq.~(\ref{eq:L_Q_obs}), and $A_{Q}$ is the extinction.
}
\footnotetext[1]{Ref.~\cite{1998ARA&A..36..189K}.}
\footnotetext[2]{Ref.~\cite{2013MNRAS.428.1128S}.}
\footnotetext[3]{Ref.~\cite{2007ApJ...657..738L}.}
\footnotetext[4]{Ref.~\cite{2015MNRAS.452.3948K}.}
\footnotetext[5]{Using Eq.~(\ref{eq:Ha_K_law}) and the recombination emission line ratios.}
\footnotetext[6]{Refs.~\cite{2000ApJ...533..682C, 2015ApJ...800..108S}.}
\footnotetext[7]{Using a ratio of $0.57$ between the $\mathrm{O}_\mathrm{I\hspace{-.1em}I}$ and $\mathrm{H}_{\alpha}$ fluxes observed in local galaxies \cite{1998ARA&A..36..189K}.}
\footnotetext[8]{Ref.~\cite{2013MNRAS.430.1042H}.}
\footnotetext[9]{Ref.~\cite{1998ARA&A..36..189K}.}
\end{center}
\end{table}

Hence, the total intensity from astrophysical sources is written as
\beqa
I_{\mathrm{astro}}(\bd{\theta}\,|\, \nu)
&=& \sum_{Q} \int \mathrm{d}\chi \, W_{Q}(z|\nu) \, \left[1+\delta_{\mathrm{astro}}(\bd{r}, z)\right], \label{eq:I_astro}\\
W_{Q}(z|\nu) &=& 
\frac{1}{4\pi}\frac{\bar{\rho}_{Q}}{\nu_{Q}\,H(z)}\,  \delta^{(1)}_{\mathrm{D}}(\chi-\chi_{Q}) \nonumber \\
&=&
\frac{{\cal L}_Q}{{\cal L}_{\mathrm{H}_{\alpha}}}
\frac{\nu_{\mathrm{H}_{\alpha}}}{\nu_Q}
\bar{I}_{\mathrm{H}_{\alpha}}(z)\, 
\delta^{(1)}_{\mathrm{D}}(\chi-\chi_{Q})
\eeqa
where 
$Q = \{\mathrm{H}_{\alpha}, 
[\mathrm{O}_\mathrm{I\hspace{-.1em}I}],
[\mathrm{O}_\mathrm{I\hspace{-.1em}I\hspace{-.1em}I}],
\mathrm{H}_{\beta},
\mathrm{Ly}\alpha\}$,
$\nu_{Q}$ is the rest-frame frequency for the line $Q$,
$\chi_{Q}$ is the comoving distance to the redshift $z_{Q}$ 
which is defined by $z_{Q} \equiv \nu_{Q}/\nu-1$,
$\delta_{\mathrm{astro}}$ is the density contrast in the luminosity density of astrophysical sources. Because the luminosity density is always expressed as in Eq.~(\ref{eq:Ha_rho2}) within our framework, statistics of $\delta_{\mathrm{astro}}$ are determined by the SFR alone, not depending on kinds of lines.
To be specific, the field $\delta_{\mathrm{astro}}$ is given by
\beqa
1+\delta_{\mathrm{astro}}(\bd{r}, z) &=& 
{\cal N}^{-1}_{\mathrm{astro}}(z)
\sum_{i}  
\dot{M}_{*}(M_i,z)
\nonumber \\
&&
\qquad \qquad \qquad
\times\delta^{(3)}_{\mathrm{D}}(\bd{r}-\bd{r}_i) \label{eq:delta_astro} \\
{\cal N}_{\mathrm{astro}}(z) &=&
\int \mathrm{d}M\, \frac{\mathrm{d}n}{\mathrm{d}M}\, 
\dot{M}_{*}(M_i,z),
\eeqa
where
$\mathrm{d}n/\mathrm{d}M$ is the halo mass function.

\subsection{Weak Lensing}

\subsubsection{Cosmic shear}

Weak gravitational lensing effect is usually characterized by
the distortion of image of a source object by the 
following 2D matrix:
\beq
A_{ij} = \frac{\partial \beta^{i}}{\partial \theta^{j}}
           \equiv \left(
\begin{array}{cc}
1-\kappa -\gamma_{1} & -\gamma_{2}  \\
-\gamma_{2} & 1-\kappa+\gamma_{1} \\
\end{array}
\right), \label{distortion_tensor}
\eeq
where $\bd{\theta}$ represents the observed position of a source object,
$\bd{\beta}$ is the true position, 
$\kappa$ is the convergence, and $\gamma$ is the shear.
In the weak lensing regime (i.e., $\kappa, \gamma \ll 1$), 
each component of $A_{ij}$ can be related to
the second derivative of the gravitational potential $\Phi$
\cite{Bartelmann2001}.
Using the Poisson equation and the Born approximation, 
one can express the weak lensing convergence field as the weighted 
integral of matter overdensity field $\delta_{\rm m}(\bd{x})$:
\beq
\kappa(\bd{\theta})
= \int_{0}^{\chi_{H}} {\rm d}\chi \ W_{\kappa}(\chi)\, \delta_{\rm m}(\bd{r},z), \label{eq:kappa_delta}
\eeq
where 
$\delta_{\rm m}$ is the matter overdensity field, 
$\chi_{H}$ is the comoving distance up to $z\rightarrow \infty$
and $W_{\kappa}(\chi)$ is called lensing kernel.
For a given redshift distribution of source galaxies,
the lensing kernel is expressed as
\beq
W_{\kappa}(\chi) = \frac{3}{2} \,\Omega_{\rm m0}H^2_{0}\, (1+z(\chi)) \int_{\chi}^{\chi_{H}} {\rm d}\chi^{\prime} p(\chi^{\prime})\frac{\chi(\chi^{\prime}-\chi)}{\chi^{\prime}}, \label{eq:lens_kernel}
\eeq
where 
$p(\chi)$ represents the redshift distribution of source galaxies
normalized to $\int_{0}^{\chi_H} {\rm d}\chi \, p(\chi) =1$.
In this paper, we assume that $p(\chi)$ is expressed as
\beqa
p(\chi) \propto \left(\frac{\mathrm{d}\chi}{\mathrm{d}z}\right)^{-1}\, z^{\alpha}\, \exp\left[-\left(\frac{z}{z_0}\right)^{\beta}\right], \label{eq:dndz_source}
\eeqa
where we adopt $\alpha=1.27$, $\beta=1.02$, and $z_0=0.50$. 
These parameters are expected in a realistic situation in the LSST \cite{2013MNRAS.434.2121C}.
Note that the median source redshift in this model is given by 0.83.

\subsubsection{Intrinsic alignments of galaxies}

Galaxies are thought to be formed at high density regions 
in the universe, or gravitationally-bound dark matter halos.
Such dark matter halos are grown by anisotropic mass accretion through cosmic filaments.
Because filamentary structures can introduce tidal interaction of dark matter halos and surrounding 
density fields, orientation of halo shapes can not be randomly distributed, but aligned with the filaments \cite{1991ApJ...369..287W, 1997MNRAS.290..411T}.

Possible large-scale alignments of galaxies' shape even in the absence of gravitational lensing, referred to as intrinsic alignments (IAs), can introduce a systematic uncertainty in any cross correlation analyses based on weak lensing \cite{2015PhR...558....1T}.
To account for the IA effect on the cross correlation between line intensity and weak lensing maps, we work with the linear alignment model \cite{2001MNRAS.320L...7C, 2004PhRvD..70f3526H}.
In this paper, we set a relevant IA field to lensing convergence so that it can reproduce the E-mode power spectrum of shapes in the linear alignment model \cite{2019PhRvD.100f3514F}. 
The IA field $\kappa_{\rm IA}$ is then given by
\beqa
\kappa_{\rm IA}(\bd{\theta}) &=& \int \mathrm{d}\chi\, W_{\rm IA}(\chi)\, \delta_{\rm m}(\bd{r}, z), \\
W_{\rm IA}(\chi) &=& -{\cal A}_{\rm IA} C_1 \frac{\Omega_{\rm m0}\rho_{{\rm crit}, 0}}{D(z)}\left(\frac{1+z}{1+z_0}\right)^{\eta_{\rm IA}} p(\chi),
\eeqa
where ${\cal A}_{\rm IA}$ controls the IA amplitude, 
$\eta_{\rm IA}$ is a free parameter to characterize redshift evolution, $C_1= 5 \times 10^{-14}\, h^{-2}\, M_{\odot}\, {\rm Mpc}^3$ is a normalization constant, 
$z_0=0.62$ is the pivot redshift,
and $D(z)$ is the normalized linear growth factor, 
so that $D(0) = 1$.
We adopt an estimate of $\eta_{\rm IA}=3$ based on observations of SuperCOSMOS Sky Survey \cite{2002MNRAS.333..501B} and Subaru Hyper Suprime Cam \cite{2019PASJ...71...43H}, while 
we set ${\cal A}_{\rm IA}=1$ for our fiducial model.
Note that some recent weak-lensing measurements reported a marginal non-zero ${\cal A}_{\rm IA}$ \cite{2018PhRvD..98d3528T, 2019PASJ...71...43H, 2020PASJ...72...16H}, but the exact value should depend on the selection of source galaxies used in weak lensing analyses.

\subsection{Relevant redshift ranges}

\begin{figure}[!t]
\includegraphics[clip, width=0.95\columnwidth]{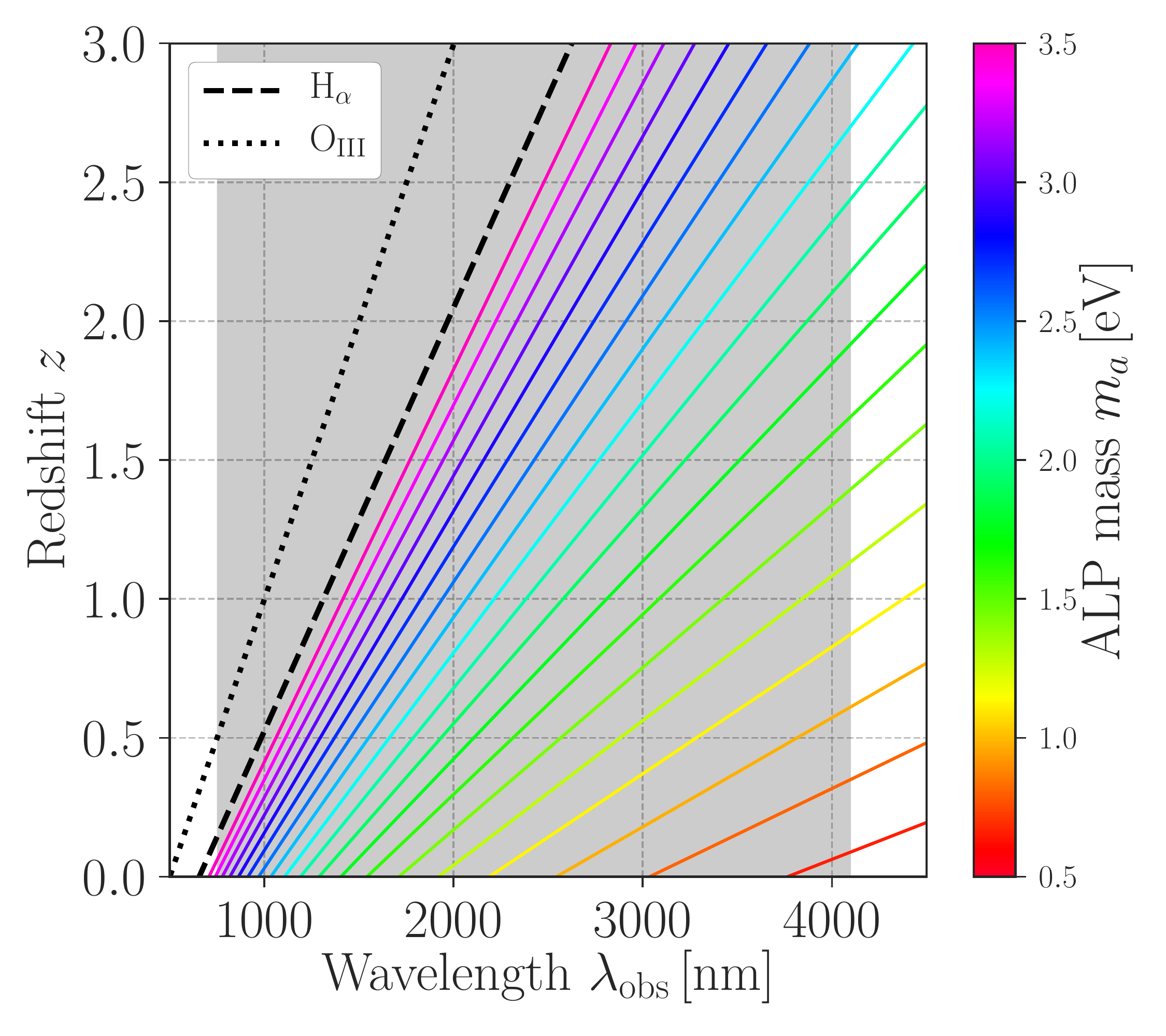}
\caption{\label{fig:fig1} 
The relation between observable wavelengths and line redshifts in the SPHEREx.
The gray filled region shows the range of wavelengths in the SPHEREx, while
the dashed and dotted lines represent the wavelength-redshift relation for 
$\mathrm{H}_{\alpha}$ and $\mathrm{O}_{\rm I\hspace{-.1em}I\hspace{-.1em}I}$ lines, respectively.
The colored solid lines show the relation for ALP decay lines with various ALP masses $m_a$.
ALPs with smaller $m_a$ emit photons with a longer wavelentgh as shown in redder lines in this figure.
}
\end{figure}

We here describe effective redshift ranges to the cross correlation analysis between line intensity 
and weak lensing maps.
For the line intensity maps, we consider the SPHEREx survey observing an all sky in optical and near-infrared wavelengths.
The SPHEREx will provide line intensity maps 
at wavelengths of $750-4100$ nm with a frequency resolution being $R=41.5$ \cite{2014arXiv1412.4872D}.

Figure~\ref{fig:fig1} shows the relation between observed wavelengths $\lambda_{\mathrm{obs}}$ and line redshifts in the SPHEREx. Different colored solid lines represent line redshifts by ALP decays with a variety of ALP masses $m_a$.
For the wavelength range in the SPHEREx, we expect that ALP decays with $0.5 \simlt m_a \, [\mathrm{eV}] \simlt 3.5$ would be relevant to the cross correlation.
For $m_a \simgt 3.5 \, \mathrm{eV}$, 
the wavelength by ALP-decay line becomes shorter than
those of astrophysical lines such as $\mathrm{H}_{\alpha}$ and $\mathrm{O}_{\rm I\hspace{-.1em}I\hspace{-.1em}I}$.
Because the weak lensing can probe large-scale structures at $z\sim 0.5$ in modern galaxy surveys, the ALP decay with $m_a \simgt 3.5 \, \mathrm{eV}$ is not expected to induce a significant correlation between the lensing and the SPHEREx intensity maps.
On the other hand, the wavelength coverage in the SPHEREx does not allow us to detect the line emission from the ALP decay with $m_a \simlt 0.5 \, \mathrm{eV}$.

\begin{figure}[!t]
\includegraphics[clip, width=1.1\columnwidth]{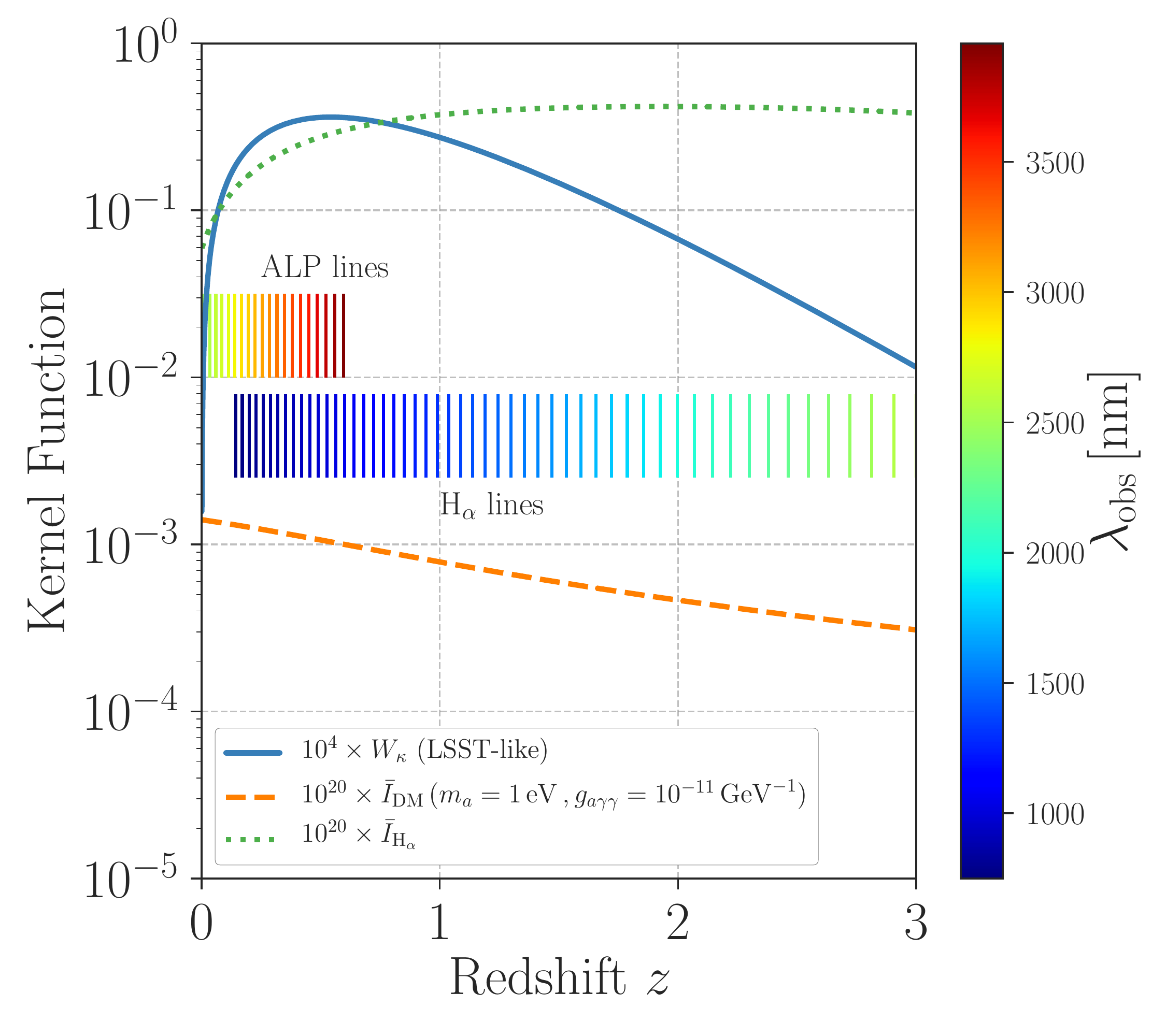}
\caption{\label{fig:fig2} 
Relevant redshift ranges in the line intensity and weak lensing maps.
In this figure, we assume an LSST-like imaging survey for the lensing map.
The blue solid line shows the lensing kernel function $W_{\kappa}$ for the LSST-like survey,
while the green dotted line represents our fiducial model of the mean intensity for 
$\mathrm{H}_{\alpha}$ line.
The orange dashed line is the mean intensity as a function of redshift $z$ for ALPs with 
a mass of $m_a = 1\, \mathrm{eV}$ and 
a particle-to-photons coupling of $g_{a\gamma\gamma} = 1\times10^{-11}\, \mathrm{GeV}^{-1}$.
Note that the mean intensity is defined in units of $\mathrm{erg}/\mathrm{s}/\mathrm{cm}^2/\mathrm{Hz}/\mathrm{str}$.
Different line segments at the middle in this figure represent relevant redshifts to 
line-intensity measurements for given observed wavelentghs $\lambda_{\mathrm{obs}}$.
The right color bar covers the range of $\lambda_{\mathrm{obs}}$ in SPHEREx.
For instance, when using a line intensity map at $\lambda_{\mathrm{obs}}\sim 2500\, \mathrm{nm}$,
one will probe large-scale structures at $z\sim 0.1$ by the ALP decay, while 
the $\mathrm{H}_{\alpha}$ line provides astrophysical information at $z\sim3$.
Because the lensing is efficient to extract the information about structures at $z\sim0.3$,
the cross correlation with the line intensity and weak lensing maps allows us to separate 
ALP-decay signals from the total line line intensity when $\lambda_{\mathrm{obs}}$ is limited to be greater than $\sim2500\, \mathrm{nm}$.
}
\end{figure}

Figure~\ref{fig:fig2} summarizes the relevant redshift range to the cross correlation with line-intensity and weak-lensing maps.
In this figure, the blue solid line represents the lensing kernel defined in Eq.~(\ref{eq:lens_kernel}) for the expected source-redshift distribution in the LSST. The LSST can be the most efficient to probe the large-scale cosmic mass density at $z\simeq0.3-0.6$. For comparison, 
the orange dashed line shows the function of 
\beqa
\bar{I}_{\rm DM}(z) = \frac{\Gamma}{4\pi}\frac{\Omega_{\mathrm{DM}}
\rho_{\mathrm{crit},0}}{\nu_a\, H(z)},
\eeqa
where this provides the redshift dependence on the mean line intensity by the ALP decay. In Figure~\ref{fig:fig2}, we consider a representative example as ALPs with $m_{a}=1\, \mathrm{eV}$ and $g_{a\gamma\gamma}=1\times 10^{-11}\, \mathrm{GeV}^{-1}$.
In addition, the green dotted line shows the mean ${\rm H}_{\alpha}$ line intensity defined in Eq.~(\ref{eq:mean_Ha_I}).
For the line intensity mapping, we measure the cumulative emission from any photon sources at a given wavelegth $\lambda_{\rm obs}$. The line segments at the middle of Figure~\ref{fig:fig2} highlight effective redshifts in 
the line-intensity map as a function of $\lambda_{\rm obs}$.
According to the relation between line redshifts and $\lambda_{\rm obs}$ in Figure~\ref{fig:fig1}, the ALP-decay line can {\it always} provide lower-redshift information than the ${\rm H}_{\alpha}$ line 
as long as we limit the particle mass to be $m_a\simlt 3.5\, \mathrm{eV}$.
Because the lensing measurement mainly carries the information at $z\simlt 1$ in the LSST, the cross correlation between the SPHEREx line intensity and the LSST weak lensing allows us to extract the ALP-decay signal alone in an efficient way when one measures the correlation as a function of $\lambda_{\rm obs}$.

\section{Statistics}\label{sec:stats}

We here define the statistic of interest, i.e. the cross correlation between the line intensity and weak lensing maps.

\subsection{Cross Correlation Functions}

For a given set of two different random fields $X(\bd{\theta})$ and $Y(\bd{\theta})$, the cross correlation function is defined as
\beqa
\xi_{XY}(\theta) = \langle X(\bd{\phi})Y(\bd{\phi}+\bd{\theta}) \rangle, \label{eq:xi}
\eeqa
where $\langle \cdots \rangle$ represents an ensemble average. The Fourier transform of Eq.~(\ref{eq:xi}) is known as the cross power spectrum and it is given by
\beqa
\langle \tilde{X}(\bd{\ell})\tilde{Y}(\bd{\ell}^{\prime})\rangle
= (2\pi)^2\, C_{XY}(\ell)\,\delta_\mathrm{D}^{(2)}(\bd{\ell}-\bd{\ell}^{\prime}), \label{eq:cross_pk}
\eeqa
where 
$C_{XY}$ is the cross power spectrum, $\tilde{X}$ represents the Fourier transform of the field $X$, and so on.
In this paper, we use the cross power spectrum to study the cross correlation between the line intensity and weak lensing maps.

As shown in Section~\ref{sec:obs}, the observed maps are decomposed into
\beqa
I_\mathrm{obs}(\bd{\theta}\,|\, \nu) &=& I_{a}(\bd{\theta}\,|\, \nu) + I_\mathrm{astro}(\bd{\theta}\,|\, \nu) + I_\mathrm{N}(\bd{\theta}\,|\, \nu), \label{eq:I_obs} \\
\kappa_\mathrm{obs}(\bd{\theta}) &=& \kappa(\bd{\theta}) 
+ \kappa_\mathrm{IA}(\bd{\theta}) + \kappa_\mathrm{N}(\bd{\theta}), \label{eq:kappa_obs}
\eeqa
where we denote $I_{\rm obs}$ and $\kappa_{\rm obs}$ as the observed line intensity and weak lensing maps, respectively.
Apart from the components as described in Section~\ref{sec:obs}, we include observational noise terms in Eqs.~(\ref{eq:I_obs}) and (\ref{eq:kappa_obs}).
The noise in the line intensity map $I_{\rm N}$ will be dominated by photo noise from zodiacal light in the SPHEREx \cite{2014arXiv1412.4872D}, which is diffuse and nearly uniform across the field of view.
On the other hand, the noise in the weak lensing map $\kappa_\mathrm{N}$
mainly arises from the intrinsic scatter in galaxy shapes.

Hence, the cross power spectrum between $I_{\rm obs}$ and $\kappa_{\rm obs}$ has four different contributions as
\beqa
C^{(\mathrm{obs})}_{\mathrm{LIM}-\kappa}(\ell|\nu)
&=& C_{\mathrm{ALP}-\kappa}(\ell|\nu)
+ C_{\mathrm{ALP}-\mathrm{IA}}(\ell|\nu) \nonumber \\
&&
\, \,
+ C_{\mathrm{astro}-\kappa}(\ell|\nu)
+ C_{\mathrm{astro}-\mathrm{IA}}(\ell|\nu), \label{eq:cross_pk_obs}
\eeqa
where 
$C_{\mathrm{ALP}-\kappa}$ is the cross correlation between the ALP-decay line and cosmic shear,
$C_{\mathrm{ALP}-\mathrm{IA}}$ is the cross correlation between the ALP-decay line and IA terms,
$C_{\mathrm{astro}-\kappa}$ is the cross correlation between the astrophysical lines and cosmic shear,
and 
$C_{\mathrm{astro}-\mathrm{IA}}$ is the cross correlation between the astrophysical lines and IA terms.
Note that the noise terms $I_\mathrm{N}$ and $\kappa_\mathrm{N}$ are assumed to be uncorrelated with other fields.
In the following, we evaluate each term in the right hand side of Eq.~(\ref{eq:cross_pk_obs}).

\subsection{Model}

\subsubsection{ALP decay line and cosmic shear}

The term $C_{\mathrm{ALP}-\kappa}$ represents the cross correlation between the ALP decay line and cosmic shear.
Note that the ALP decay line exactly traces the dark matter distribution in the universe (see Eq.~[\ref{eq:luminosity_density_ALP}]), while cosmic shear, or lensing convergence $\kappa$ provides an unbiased estimate of projected total mass density with some weight function along a line of sight (see Eq.~[\ref{eq:kappa_delta}]).
Using the Limber approximation \cite{Limber:1954zz}, we can express the cross power spectrum as
\beqa
C_{\mathrm{ALP}-\kappa}(\ell|\nu) &=& 
\int\mathrm{d}\chi \frac{W_{\mathrm{DM}}(z|\nu)\, W_\kappa(z)}{\chi^2} P_{\mathrm{DM-m}}\left(\frac{\ell}{\chi}, z(\chi)\right) \nonumber \\
&=& \frac{\bar{I}_{\mathrm{DM}}(\chi_a)\, W_{\kappa}(\chi_a)}{\chi^2_a}\, P_{\mathrm{DM-m}}\left(\frac{\ell}{\chi_a}, z_a\right),
\eeqa
where $P_{\mathrm{DM}-m}(k,z)$ is the three-dimensional cross power spectrum 
between the dark matter and total mass density fields.
Throughout this paper, we assume that the dark-matter overdensity field is expressed as $(\Omega_{\mathrm{DM}}/\Omega_{\mathrm{m0}})\, \delta_{\rm m}$.
Then, the cross power spectrum is given by
\beqa
P_{\mathrm{DM-m}}(k,z) = \frac{\Omega_{\mathrm{DM}}}{\Omega_{\mathrm{m0}}}\, P_{\mathrm{m}}(k,z),
\eeqa
where $P_{\mathrm{m}}(k,z)$ is the three-dimensional non-linear matter power spectrum.
In the following, we adopt a simulation-calibrated fitting formula \cite{2012ApJ...761..152T} to compute $P_{\mathrm{m}}(k,z)$.
It would be worth noting that we ignore possible baryonic effects on $P_{\mathrm{DM}-m}(k,z)$, while it may induce a $\simlt 30\%$ level systematic error in our forecasts on the constraint of $\Gamma$ (see Ref.~\cite{2019OJAp....2E...4C} for a review of baryonic effects on large-scale structures).
Since the decay rate $\Gamma$ is proportional to $g_{a\gamma\gamma}^2$, 
we expect that the constraint of $g_{a\gamma\gamma}$ can be affected by the baryonic effects with a level of $\simlt 15\%$ at most.
One needs to account for this systematic error in real measurements, but 
it does not play a central role in our forecasts.

\subsubsection{ALP decay line and intrinsic alignments}

Similarly, the cross correlation between the ALP decay line and IA terms is given by
\beqa
C_{\mathrm{ALP}-\mathrm{IA}}(\ell|\nu) &=& 
\frac{\bar{I}_{\mathrm{DM}}(\chi_a)\, W_\mathrm{IA}(\chi_a)}{\chi^2_a} \nonumber \\
&& \qquad \times
P_{\mathrm{DM-m}}\left(\frac{\ell}{\chi_a}, z_a\right),
\eeqa
where this term is strongly degenerate with $C_{\mathrm{ALP}-\kappa}(\ell|\nu)$ at a given $\nu$, because ether is proportional to $P_{\mathrm{DM-m}}$ at the same redshift of $z_a$.
Nevertheless, one can constrain an unknown IA parameter ${\cal A}_{\mathrm{IA}}$ in the function of $W_\mathrm{IA}$ and break the degeneracy when using the frequency-dependence on $C_{\mathrm{ALP}-\mathrm{IA}}(\ell|\nu)$.
The frequency dependence on $C_{\mathrm{ALP}-\mathrm{IA}}(\ell|\nu)$ can not be same as that of 
$C_{\mathrm{ALP}-\kappa}(\ell|\nu)$, because the kernel functions $W_{\kappa}$ and $W_\mathrm{IA}$ exhibit the different redshift-dependence.

\subsubsection{Astrophysical lines and cosmic shear} \label{subsubsec:astroline_vs_kappa}

Under the Limber approximation with Eqs.~(\ref{eq:I_astro}) and (\ref{eq:kappa_delta}),
we can write the term $C_{\mathrm{astro}-\kappa}$ as
\beqa
C_{\mathrm{astro}-\kappa}(\ell|\nu) &=& \sum_{Q}\, \frac{{\cal L}_Q}{{\cal L}_{\mathrm{H}_{\alpha}}}\frac{\nu_{\mathrm{H}_{\alpha}}}{\nu_Q}\, \frac{\bar{I}_{\mathrm{H}_{\alpha}}(z_Q)\, W_{\kappa}(z_Q)}{\chi^2_Q} \nonumber \\
&& 
\qquad
\times P_{\mathrm{astro-m}}\left(\frac{\ell}{\chi_Q}, z_Q\right),
\eeqa
where $P_{\mathrm{astro-m}}(k,z)$ is the three-dimensional power spectrum between two fields 
of $\delta_\mathrm{astro}$ and $\delta_\mathrm{m}$ (see Eq.~[\ref{eq:delta_astro}] for our definition of $\delta_\mathrm{astro}$).

In this paper, we compute $P_{\mathrm{astro-m}}(k,z)$ based on a halo-model approach \cite{2002PhR...372....1C}. Within the halo model, the power spectrum can be decomposed into two terms. One is the one-halo term arising from two-point correlations in single dark matter halos,
and another is the two-halo term given by two-point correlations between neighboring halos.
Each term can be expressed as
\beqa
P_{\mathrm{astro-m}}(k,z) = P^\mathrm{1h}_{\mathrm{astro-m}}(k,z) + P^\mathrm{2h}_{\mathrm{astro-m}}(k,z), \label{eq:P_astro_m}
\eeqa
where $P^\mathrm{1h}_{\mathrm{astro-m}}$ and $P^\mathrm{2h}_{\mathrm{astro-m}}$
are the one-halo and the two-halo terms, respectively, and
\beqa
P^\mathrm{1h}_{\mathrm{astro-m}}(k,z) &=&
{\scriptstyle {\cal N}^{-1}_\mathrm{astro}(z)}
\int\mathrm{d}M \frac{\mathrm{d}n}{\mathrm{d}M} \frac{\rho_{h}(k|M,z)}{\bar{\rho}_\mathrm{m}} 
\nonumber \\
&&
\qquad \qquad \qquad \qquad
\times
\dot{M}_{*}(M,z), \\
P^\mathrm{2h}_{\mathrm{astro-m}}(k,z) &=&
\left(
\int\mathrm{d}M \frac{\mathrm{d}n}{\mathrm{d}M} b_h(M,z) 
\frac{\rho_{h}(k|M,z)}{\bar{\rho}_\mathrm{m}}
\right)
\nonumber \\
&\times&
\left(
{\scriptstyle {\cal N}^{-1}_\mathrm{astro}(z)}
\int\mathrm{d}M \frac{\mathrm{d}n}{\mathrm{d}M} b_h(M,z) 
\dot{M}_{*}(M,z)
\right)
\nonumber \\
&\times&
P_L(k,z),
\eeqa
where 
$\rho_{h}(k|M,z)$ is the Fourier transform of a spherical halo density profile,
$\bar{\rho}_\mathrm{m}=\rho_\mathrm{crit,0}\Omega_\mathrm{m0}$,
$b_{h}$ is the linear halo bias and $P_{L}$ is the linear matter power spectrum.
In the halo-model computations, we define the halo mass as 
the mass of a spherical overdensity with 200-times the mean density of the universe.
For the halo density profile, we adopt the the analytical Navarro-Frenk-White (NFW) profile \cite{1996ApJ...462..563N}, where we use the concentration-mass-redshift relation in Ref.~\cite{2015ApJ...799..108D}.
We also use the fitting formulas of the halo mass function \cite{2008ApJ...688..709T} and the linear halo bias \cite{2010ApJ...724..878T}.
The key ingredient in the computation of Eq.~(\ref{eq:P_astro_m}) is the SFR as a function of halo masses and redshifts, $\dot{M}_{*}(M,z)$.
Although this quantity is still uncertain, we adopt a simulation-based fitting formula 
as in Ref.~\cite{2017arXiv171109902S} for our baseline model.
Their model of $\dot{M}_{*}(M,z)$ has a form of 
\beqa
\dot{M}_{*}(M,z)
= 10^{a(z)} \left(\frac{M}{M_1}\right)^{b(z)} 
\left(1+\frac{M}{M_2(z)}\right)^{c(z)},
\eeqa
where $M_1 = 10^{8}\, M_{\odot}$ and there exist four parameters 
of $a(z), b(z), c(z)$ and $M_2(z)$ in the model.
Ref.~\cite{2017arXiv171109902S} have provided 
the parameters at $z=0, 0.8, 1, 2.2$ and $4.8$ by analyzing semi-analytic galaxy catalogs in Ref.~\cite{2013MNRAS.428.1351G}.
We infer $a(z), b(z), c(z)$ and $M_2(z)$ at a given $z$
by employing a linear interpolation of the fitting results at $z=0, 0.8, 1, 2.2$ and $4.8$.
Note that $b\sim2.6$, $c\sim-3$ and $M_2$ ranges from $10^{11-12}\, M_{\odot}$ over redshifts.
Hence, a typical halo mass in the SFR activity (i.e. the mass of the halo which most efficiently produces stars) is set to be $M\sim10^{12-13}\, M_{\odot}$ in this model.
Although our baseline model would provide an order-of-magnitude estimate for $P_{\mathrm{astro-m}}$, 
it is based on the semi-analytic model of galaxy formation
and poorly constrained by observations.
To account for theoretical uncertainties in our model, we introduce two dimensionless parameters 
${\cal A}_\mathrm{astro}$ and $\eta_\mathrm{astro}$, which control the overall amplitude and redshift evolution in $P_{\mathrm{astro-m}}$.
To be specific, we set
\beqa
P_{\mathrm{astro-m}}(k,z) \rightarrow {\cal A}_\mathrm{astro} (1+z)^{\eta_\mathrm{astro}}\, P_{\mathrm{astro-m}}(k,z),
\eeqa
when computing $C_{\mathrm{astro}-\kappa}(\ell|\nu)$.
For our fiducial model, we adopt ${\cal A}_\mathrm{astro} = 1$ and $\eta_\mathrm{astro} = 0$.

\subsubsection{Astrophysical lines and intrinsic alignments}

The cross correlation between the astrophysical lines and the IA field is then expressed as
\beqa
C_{\mathrm{astro}-\mathrm{IA}}(\ell|\nu) &=& \sum_{Q}\, \frac{{\cal L}_Q}{{\cal L}_{\mathrm{H}_{\alpha}}}\frac{\nu_{\mathrm{H}_{\alpha}}}{\nu_Q}\, \frac{\bar{I}_{\mathrm{H}_{\alpha}}(z_Q)\, W_\mathrm{IA}(z_Q)}{\chi^2_Q} \nonumber \\
&& 
\quad
\times {\cal A}_{\rm astro}\, (1+z_Q)^{\eta_\mathrm{astro}} \nonumber \\
&&
\quad
\times
P_{\mathrm{astro-m}}\left(\frac{\ell}{\chi_Q}, z_Q\right).
\eeqa

\subsection{Survey specifications and statistical errors}\label{subsec:survey}

\begin{figure*}
\includegraphics[clip, width=2\columnwidth]{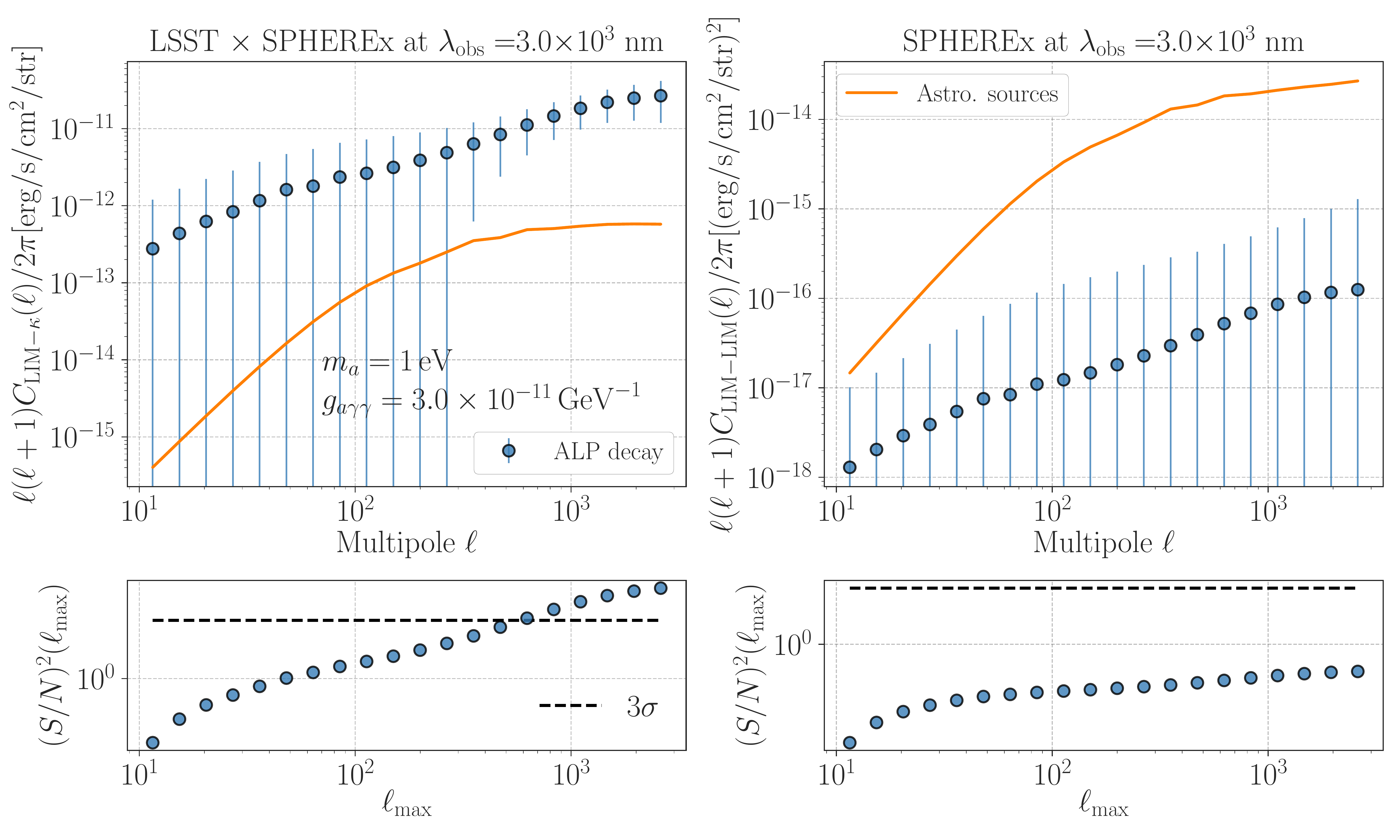}
\caption{\label{fig:fig3}
Expected signals of the cross-correlation function 
with the line intensity and weak lensing maps
as well as the auto correlation of the line intensity map.
In this figure, we assume an LSST-like imaging survey for lensing and an SPHEREx-like spectroscopic survey 
at wavelength $\lambda_{\mathrm{obs}}=3000\, \mathrm{nm}$.
Details of the survey specification are found in Section~\ref{subsec:survey}.
The left panels show the cross power spectra and its cumulative signal-to-noise ratio.
In the left top, blue points with error bars show the ALP-decay signal for an ALP mass of 
$m_a = 1\, \mathrm{eV}$ and a particle-to-photons coupling of 
$g_{a\gamma\gamma}=3\times10^{-11}\, \mathrm{GeV}^{-1}$.
For comparison, the orange line shows our fiducial model of astrophysical contributions to 
the cross power spectrum. The left bottom panel represents 
the cumulative signal-to-noise ratio of the ALP-decay signal as a function of the maximum multipole $\ell_{\mathrm{max}}$.
The black dashed line in the bottom indicates a $3\sigma$-level detection.
The right panels are similar to the left, but those are for the auto power spectrum of the intensity map.
This figure highlights that the cross power spectrum with the line intensity and weak lensing maps can be dominated by the ALP-decay signal, while the auto power spectrum of the intensity maps will be largely determined by the astrophysical sources.
}
\end{figure*}

We here describe observational parameters of line intensity and weak lensing surveys in our analysis and summarize how to set expected statistical uncertainties of Eq.~(\ref{eq:cross_pk_obs}).

For line intensity maps, we assume a hypothetical SPHEREx-like survey covering a full sky.
Among several observational modes in SPHEREx, 
we consider the all-sky survey mode with an angular resolution of 0.1 arcmin and a $5\sigma$-level point-source sensitivity of 18.5 AB magnitude per pixel \cite{2014arXiv1412.4872D}. 
Note that there exists the ``deep" survey mode in SPHEREx 
and it will has a much greater point-source sensitivity.
Hence, the deep survey mode would allow us to perform robust line-intensity mapping measurements, while the sky coverage is planed to be $200\, \mathrm{deg}^2$ only.
Because the statistical uncertainty in our cross correlation analysis scales with 
the inverse of the survey area, the deep survey mode is not suitable 
to search for large-scale cross correlation signals.
We also assume the frequency resolution of $R=41.5$ at wavelengths of $750-4100$ nm.

For weak lensing maps, we assume a LSST-like imaging survey covering a sky of 18000 $\mathrm{deg}^2$. In weak lensing measurements, we assume the source number density of $26\, \mathrm{arcmin}^{-2}$, the intrinsic scatter of galaxy ellipticity per components being 0.26, 
and the source redshift distribution as in Eq.~(\ref{eq:dndz_source}).
These survey parameters have been investigated in Ref.~\cite{2013MNRAS.434.2121C}.

Given survey configurations, we can evaluate the statistical uncertainty of Eq.~(\ref{eq:cross_pk_obs}) by using covariance matrices. Assuming that the observable fields follow Gaussian, we can write the covariance matrix as
\beqa
\bd{C}(\ell, \nu | \ell^{\prime}, \nu^{\prime}) 
&\equiv& \mathrm{Cov}\left[C^{(\mathrm{obs})}_{\mathrm{LIM}-\kappa}(\ell|\nu), C^{(\mathrm{obs})}_{\mathrm{LIM}-\kappa}(\ell^{\prime}|\nu^{\prime})\right] \nonumber \\
&=&
\frac{\delta^{K}_{\ell \ell^{\prime}}}{\rev{2 \ell} \, \Delta \ell\, f_\mathrm{sky}}
\Bigl[C^{(\mathrm{obs})}_{\mathrm{LIM-LIM}}(\ell|\nu, \nu^{\prime}) C^{(\mathrm{obs})}_{\kappa-\kappa}(\ell) \nonumber \\
&&
+C^{(\mathrm{obs})}_{\mathrm{LIM}-\kappa}(\ell|\nu)
C^{(\mathrm{obs})}_{\mathrm{LIM}-\kappa}(\ell^{\prime}|\nu^{\prime})
\Bigr], \label{eq:cov}
\eeqa
where 
$f_\mathrm{sky}$ is the fraction of sky coverage used in the cross correlation analysis,
$\delta^{K}_{ij}$ is the Kronecker delta symbol,  
$C^{(\mathrm{obs})}_{\mathrm{LIM-LIM}}(\ell|\nu, \nu^{\prime})$ represents the cross power spectrum between two intensity maps with different frequency bands, and
$C^{(\mathrm{obs})}_{\kappa-\kappa}(\ell)$ is the auto power spectrum of the observed convergence field (Eq.~[\ref{eq:kappa_obs}]).
\rev{The derivation of Eq.~(\ref{eq:cov}) is found in Appendix~\ref{apdx:cov}.}
Note that $f_\mathrm{sky}$ is set to be $18,000/41,252.96 =0.436$ for our case. 
In Eq.~(\ref{eq:cov}), we employ a binning in the multipole $\ell$ for the measurements of cross power spectra with the bin width being $\Delta \ell$. In this setup, the covariance matrix becomes non-zero only when two different $\ell$ bins have an identical value.
We summarize our model of $C^{(\mathrm{obs})}_{\mathrm{LIM-LIM}}(\ell|\nu, \nu^{\prime})$
and $C^{(\mathrm{obs})}_{\kappa-\kappa}(\ell)$ in Appendix~\ref{apdx:cov}.

\section{Results}\label{sec:results}

In this section, we show our main results in this paper.
When computing power spectra, we employ a logarithmic binning in 
the range of $10 < \ell < 3000$ with the number of bins being 20.
Also, we compute all power spectra with respect to $\nu \, I(\bd{\theta}|\nu)$ 
in units of $\mathrm{erg}/\mathrm{s}/\mathrm{cm}^2/\mathrm{str}$.

\subsection{Expected Signal}

Let us first consider expected signals in the cross correlation analysis.
As a representative example, we assume ALPs with a mass of $m_a = 1\, \mathrm{eV}$ and a particle-to-photons coupling of $g_{a\gamma\gamma}=3\times10^{-11}\, \mathrm{GeV}^{-1}$.
The top left panel in Fig.~\ref{fig:fig3} shows our baseline model for the cross power spectrum between line intensity maps and weak lensing convergence.
In the figure, we suppose the line intensity map at $\lambda_\mathrm{obs}=3000\, \mathrm{nm}$.
The blue points with error bars represent the ALP-decay signal $C_{\mathrm{ALP}-\kappa}$, while the orange line stands for the astrophysical contribution $C_{\mathrm{astro}-\kappa}$.
In the left bottom, we characterize the detectability of the signal $C_{\mathrm{ALP}-\kappa}$ by introducing the cumulative signal-to-noise ratio.
The signal-to-noise ratio at a given photon frequency is defined as
\beqa
\left(S/N\right)^2(\ell_\mathrm{max}) 
&=& 
\sum_{\ell_i, \ell_j \le \ell_\mathrm{max}} 
C_{\mathrm{ALP}-\kappa}(\ell_i | \nu)\, \bd{C}^{-1}_{\rm null}(\ell_i, \nu | \ell_j, \nu)
\nonumber \\
&&
\qquad \qquad
\times
C_{\mathrm{ALP}-\kappa}(\ell_j | \nu), \label{eq:s2n_cross_null}
\eeqa
where $\bd{C}_{\rm null}$ is the covariance matrix when we set the decay rate to be $\Gamma=0$.
Hence, Eq.~(\ref{eq:s2n_cross_null}) can be used to compute the significance of the ALP-decay signal over a null detection hypothesis.
The blue points in the left bottom panel in Fig.~\ref{fig:fig3} shows the signal-to-noise ratio as a function of maximum multipoles $\ell_\mathrm{max}$ in the analysis, 
while the dashed line represents a $3\sigma$-level significance.

On the other hand, the right panels in Fig.~\ref{fig:fig3} summarize the results for the auto power spectrum in the intensity map.
We compute the expected auto correlation signal as in Appendix~\ref{apdx:cov}.
In the top right panel, the orange line shows the auto power spectrum arising from clustering of astrophysical lines (referred to as $C_{\mathrm{astro-astro}}$ in Appendix~\ref{apdx:cov}).
The blue points with error bars are the counterpart of the ALP decay line, 
$C_{\mathrm{ALP-ALP}}$ as defined in Eq.~(\ref{eq:pk_ALP_ALP}). The error bars in the top right panel are computed as
\beqa
&&
\mathrm{Cov}\left[C^{(\mathrm{obs})}_{\mathrm{LIM-LIM}}(\ell|\nu), C^{(\mathrm{obs})}_{\mathrm{LIM-LIM}}(\ell^{\prime}|\nu)\right] \nonumber \\
&&
\qquad
=
\frac{\delta^{K}_{\ell \ell^{\prime}}}{\ell \Delta \ell\, f_\mathrm{sky}}
C^{(\mathrm{obs})}_{\mathrm{LIM-LIM}}(\ell|\nu) C^{(\mathrm{obs})}_{\kappa-\kappa}(\ell), \label{eq:cov_auto}
\eeqa
where
details of $C^{(\mathrm{obs})}_{\mathrm{LIM-LIM}}$ and $C^{(\mathrm{obs})}_{\kappa-\kappa}$ are given in Appendix~\ref{apdx:cov}.
The signal-to-noise ratio of $C_{\mathrm{ALP-ALP}}$ can be computed in a similar way to Eq.~(\ref{eq:s2n_cross_null}) and it is shown in the bottom right panel in Fig.~\ref{fig:fig3}.

The comparisons between left and right panels in Fig.~\ref{fig:fig3} clarify that
the ALP decay can induce a significant correlation signal compared to the counterpart of astrophysical lines in the cross correlation analysis, but not in the auto correlation of line intensity maps.
This can be understood as the cross correlation between line intensity and weak lensing maps
can probe lower-redshift large-scale structures and the ALP decay line mainly comes from those lower-redshift structures.
As shown in Fig.~\ref{fig:fig2}, the auto correlation function will be dominated by astrophysical lines and the astrophysical lines effectively probe higher redshifts at $z\simgt2$ as long as the observed wavelength is set to be $\lambda_{\mathrm{obs}}\simgt2000\, \mathrm{nm}$.

\begin{figure}[!t]
\includegraphics[clip, width=0.95\columnwidth]{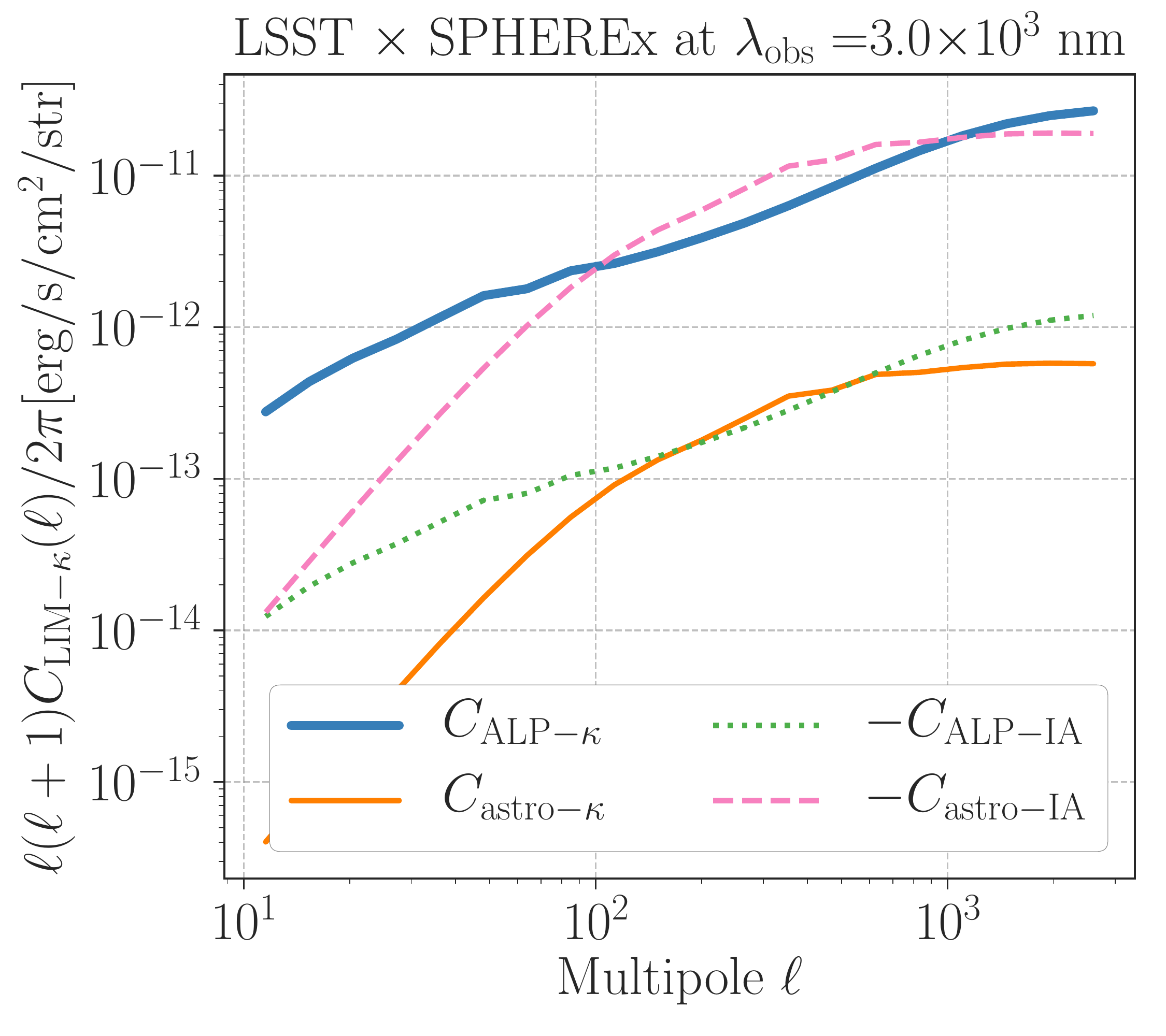}
\caption{\label{fig:fig4} 
Separated contributions to the observed cross power spectrum between line intensity 
and weak lensing maps. We here suppose the LSST-like source galaxy distribution for lensing
and the line intensity map at $\lambda_\mathrm{obs}=3000\, \mathrm{nm}$.
For the ALP decay, we consider the ALPs with a mass of $m_a=1\, \mathrm{eV}$
and a particle-to-photons coupling of $g_{a\gamma\gamma}=3\times10^{-11}\, \mathrm{GeV}^{-1}$.
The blue thick line shows the cross correlation between the ALP decay line and weak lensing
convergence, while the orange thin line stands for the counterpart of astrophysical lines.
The green dotted and pink dashed lines represent the IA-induced signals arising from
the ALP decay and astrophysical lines, respectively.
Note that the IA-induced signal becomes negative over the angular scales of interest.
}
\end{figure}

Apart from gravitational lensing effects, we will have the cross correlation arising from the IA effects of galaxy shapes. Fig.~\ref{fig:fig4} summarizes our baseline model of the cross power spectrum between line intensity and weak lensing maps.
Different colored lines in the figure show four separated terms in the observed power spectrum as in Eq.~(\ref{eq:cross_pk_obs}).
Note that the IA-induced terms ($C_\mathrm{astro-IA}$ and $C_\mathrm{ALP-IA}$)
will show a negative value because major axes in galaxy shapes are expected to 
align in the radial direction toward high density regions \cite{2001MNRAS.320L...7C, 2004PhRvD..70f3526H}.
Our baseline model shows that the cross correlation between astrophysical lines 
and intrinsic galaxy shapes can dominate the ALP-decay-induced lensing signal at $\ell\simlt1000$,
while the term $C_{\mathrm{ALP}-\kappa}$ still has a significant contribution to the observed signal at small angular scales of $\ell\simgt 1000$.
We explore the detectability of $C_{\mathrm{ALP}-\kappa}$ in the presence of the IA effects
in Section~\ref{subsec:fisher}.

\subsection{Signal-to-noise ratios}\label{subsec:s2n}

The lensing kernel function is relatively broad in redshifts and the effective redshift for the ALP decay can vary as we change frequency bands in the intensity map (see Fig.~\ref{fig:fig2}).
Hence, combining the cross power spectra among multi-frequency bands can further improve the detectability of the ALP decay line.
The total signal-to-noise ratio of the ALP-decay signal 
in the range of observed frequency 
$\nu_\mathrm{min} \le \nu \le \nu_\mathrm{max}$ can be defined as
\beqa
&&(S/N)^2_\mathrm{tot}(\ell_\mathrm{max}|\nu_\mathrm{min}, \nu_\mathrm{max})
= 
\sum_{\nu_\alpha, \nu_\beta} 
\sum_{\ell_i, \ell_j \le \ell_\mathrm{max}} 
C_{\mathrm{ALP}-\kappa}(\ell_i | \nu_\alpha) \nonumber \\
&&
\qquad \qquad
\times
\bd{C}^{-1}_{\rm null}(\ell_i, \nu_\alpha | \ell_j, \nu_\beta)
\, C_{\mathrm{ALP}-\kappa}(\ell_j | \nu_\beta), \label{eq:tot_s2n_cross_null}
\eeqa
where we impose $\nu_\mathrm{min} \le \nu_{\alpha}, \nu_{\beta} \le \nu_\mathrm{max}$ to take the sum. 
In our case, we assume the SPHEREx-like situation 
with the wavelength coverage of $750-4100$ nm, corresponding to $\nu_\mathrm{min} = 73.1\, \mathrm{THz}$ and $\nu_\mathrm{max} = 399.7\, \mathrm{THz}$.
When setting the frequency resolution being $R=41.5$, we find that 
70 intensity maps are available in total.

\begin{figure}[!h]
\includegraphics[clip, width=0.95\columnwidth]{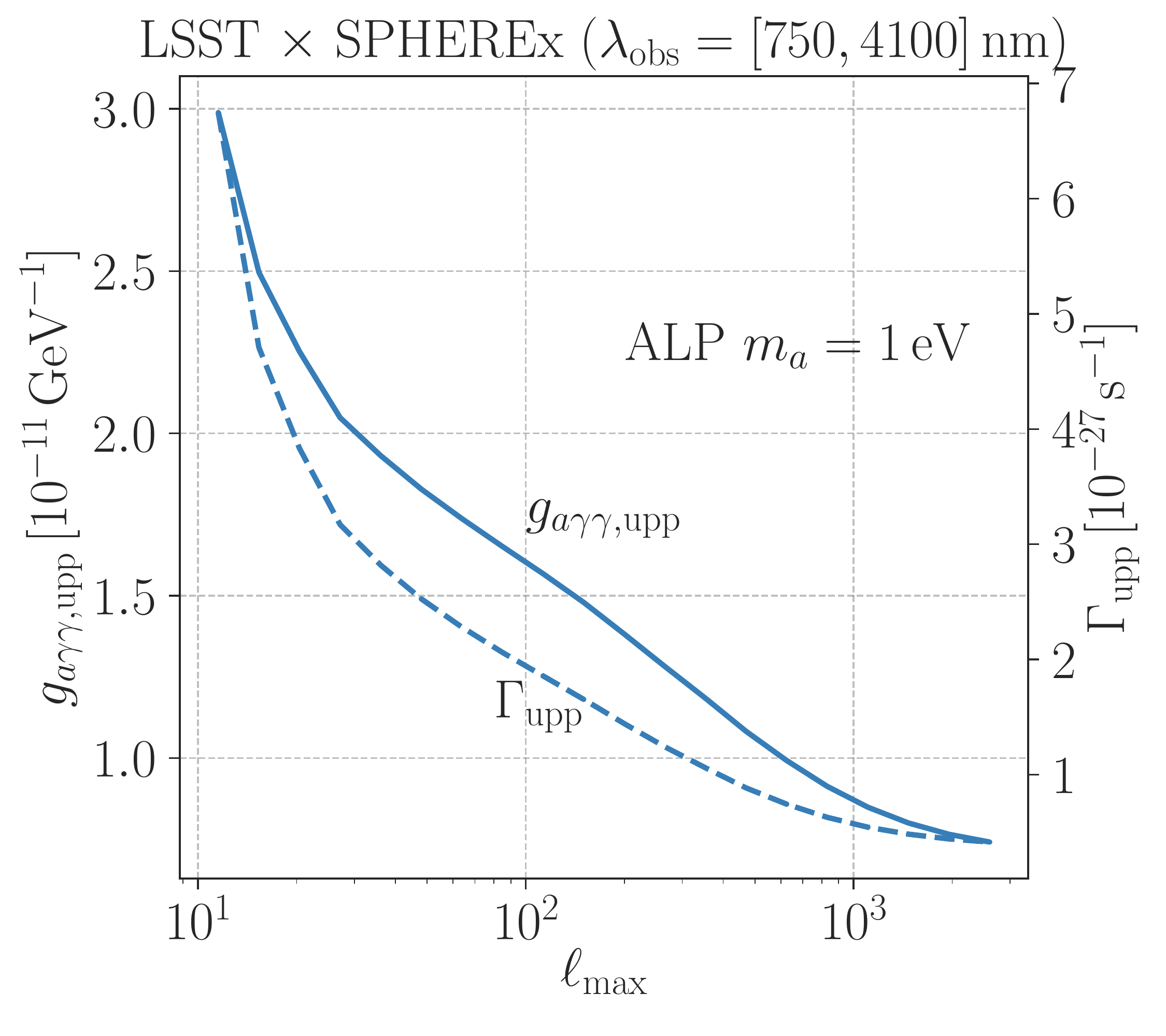}
\caption{\label{fig:fig5} 
Total signal-to-noise ratio of the cross power spectrum 
between the line intensity and weak lensing maps
for ALPs with a particle mass of $m_a=1\, \mathrm{eV}$.
For a fixed mass $m_a$, the total signal-to-noise ratio is given by $(S/N)_\mathrm{tot} = (g_{a\gamma\gamma}/g_{a\gamma\gamma, \mathrm{upp}})^2 = \Gamma/\Gamma_{\rm upp}$, 
where $g_{a\gamma\gamma}$ is the particle-to-photons coupling constant 
for ALPs and $\Gamma$ is the ALP decay rate.
The solid line and dashed lines show $g_{a\gamma\gamma, \mathrm{upp}}$ and $\Gamma_{\rm upp}$ as a function of the maximum multipole $\ell_{\rm max}$ in the cross-correlation analysis, respectively.
In this figure, we combine the cross power spectra defined 
with multiple-wavelength intensity maps in the range of $\lambda_{\rm obs}=750-4100\, \mathrm{nm}$.
Also, we assume that a sky of 18,000 ${\rm deg}^{2}$ is 
available for the analysis.
}
\end{figure}

For a given $m_a$, we can express the total signal-to-noise ratio as
\beqa
(S/N)_\mathrm{tot}(\ell_\mathrm{max})
&=& \left[\frac{\Gamma}{\Gamma_\mathrm{upp}(\ell_\mathrm{max})}\right] \\
&=& \left[\frac{g_{a\gamma\gamma}}{g_{a\gamma\gamma, \mathrm{upp}}(\ell_\mathrm{max})}\right]^2.
\eeqa
Fig.~\ref{fig:fig5} shows $\Gamma_\mathrm{upp}$ and $g_{a\gamma\gamma, \mathrm{upp}}$ for ALPs with a particle mass of $m_a = 1\, \mathrm{eV}$.
We find that combining the cross power spectra at multi-frequency bands allows us to 
probe the particle-to-photon coupling down to $g_{a\gamma\gamma} \sim 1\times 10^{-11}\, \mathrm{GeV}^{-1}$ if we can have a precise model of the cross power spectra at 
$\ell\simlt 800$.

\begin{figure}[!h]
\includegraphics[clip, width=0.95\columnwidth]{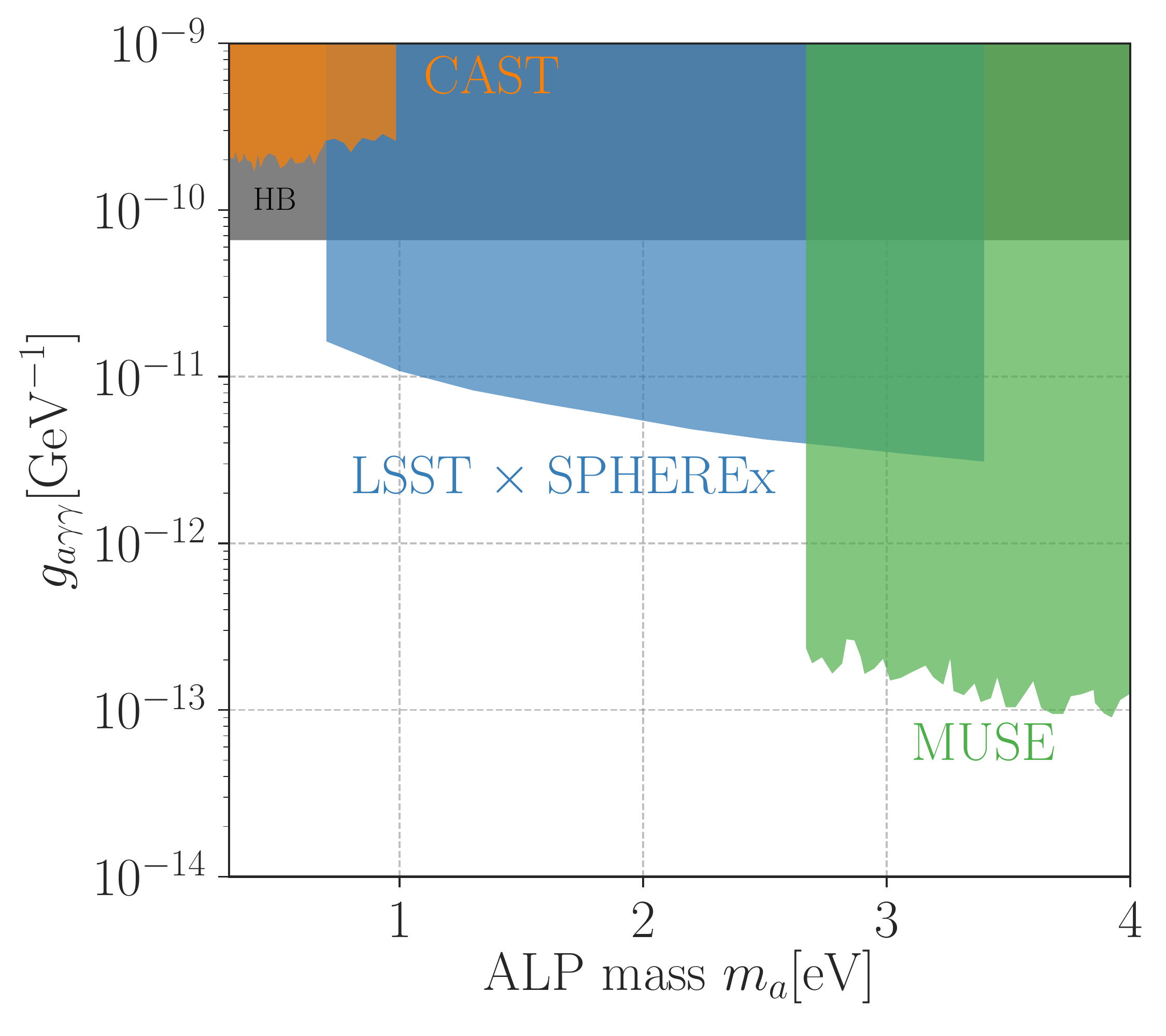}
\caption{\label{fig:fig6}
$2\sigma$-level upper limits of the ALP-to-photons coupling $g_{a\gamma\gamma}$
as a function of ALP mass $m_a$. The blue region can be excluded by the cross correlation analysis 
with SPHEREx line intensity maps and LSST lensing. For comparison, we show existing constraints in the literature; the gray, orange, and green regions have been excluded by 
the study of stellar evolution of horizontal branch (HB) to red giants in globular clusters \cite{2014PhRvL.113s1302A},
the CERN Axion Solar Telescope (CAST) \cite{2014PhRvL.112i1302A},
spectroscopic observations of the dwarf spheroidal galaxy Leo T by 
the Multi Unit Spectroscopic Explorer (MUSE) \cite{2020arXiv200901310R}, respectively.
}
\end{figure}

Fig.~\ref{fig:fig6} summarizes expected $2\sigma$ upper limits of $g_{a\gamma\gamma}$
for different particle masses $m_a$ if the cross correlation analysis in LSST and SPHEREx
is consistent with a null detection.
Note that we set $\ell_\mathrm{max}=2000$ in Fig.~\ref{fig:fig6}.
In the figure, we also show several observational constraints in the literature for comparison.
Those include the study of stellar evolution of horizontal branch (HB) to red giants in globular clusters \cite{2014PhRvL.113s1302A},
``axion helioscope'' observations by the CERN Axion Solar Telescope (CAST) \cite{2014PhRvL.112i1302A},
spectroscopic observations of the dwarf spheroidal galaxy Leo T by 
the Multi Unit Spectroscopic Explorer (MUSE) \cite{2020arXiv200901310R}.
The blue filled region in Fig.~\ref{fig:fig6} represents the expected constraint by the LSST-SPHEREx cross correlation analysis, enabling us to improve the existing constraint at 
$m_a=1-2.7\, \mathrm{eV}$ by a factor of $6-10$.
\rev{For the interested readers, we also provide a forecast in terms of 
the dark matter lifetime $\Gamma$ in Appendix~\ref{apdx:lifetime}.}

\subsection{Forecast of ALP parameter constraints}\label{subsec:fisher}

\begin{figure}[!t]
\includegraphics[clip, width=1.2\columnwidth, bb=0 0 698 598]{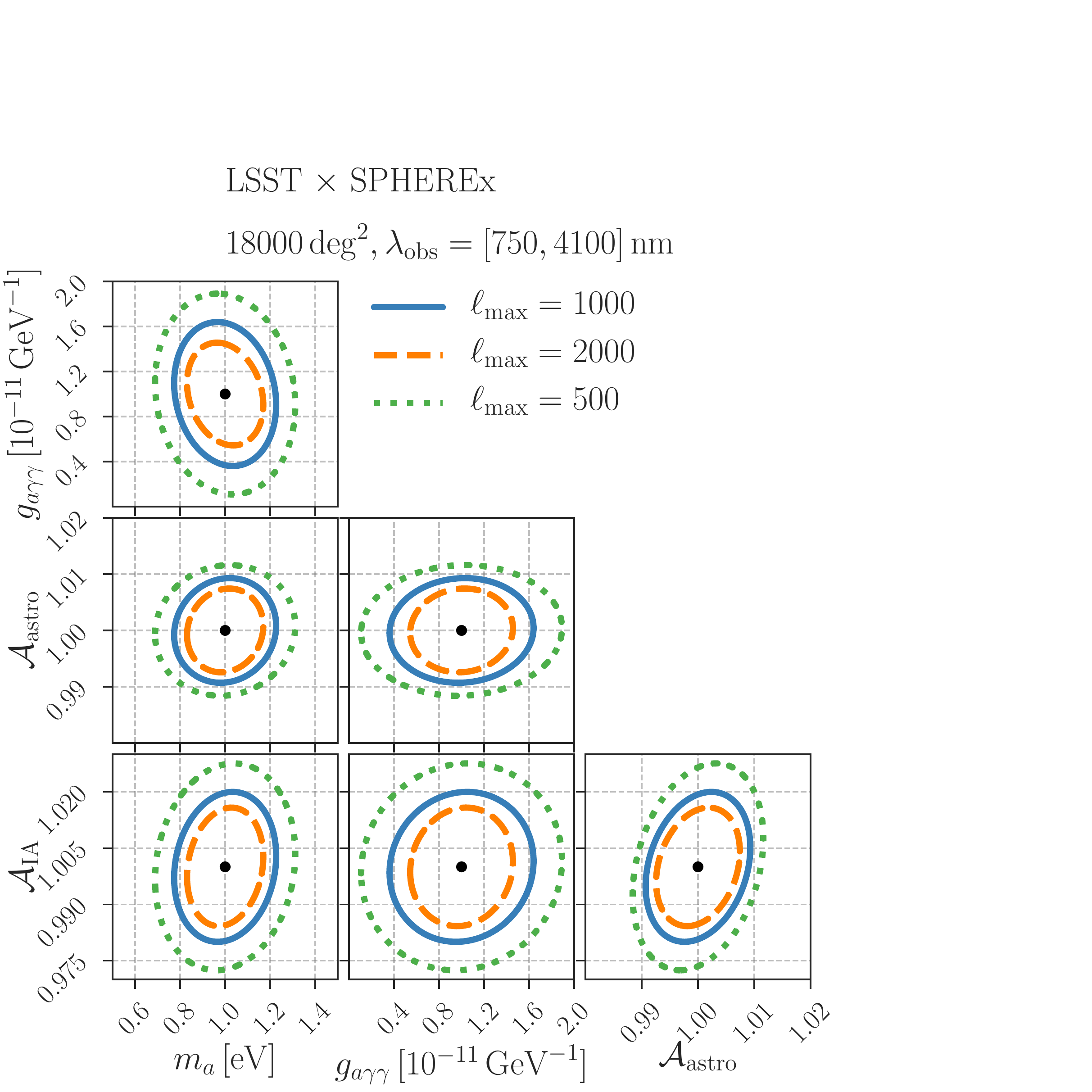}
\caption{\label{fig:fig7} 
Forecast of parameter constraints on ALP.
In this figure, we assume the cross correlation analysis 
in hypothetical LSST-like and SPHEREx-like surveys 
covering 18000 ${\rm deg}^2$.
In each panel, the solid error circle represents a 68\% confidence level when one uses the cross power spectra at $\ell \le \ell_{\rm max}=1000$.
The orange dashed and green dotted lines show the confidence level 
for $\ell_{\rm max}=2000$ and 500, respectively.
}
\end{figure}

In actual observations, we have several expected components in the cross correlation between
line intensity and weak lensing maps as in Eq.~(\ref{eq:cross_pk_obs}).
As shown in Fig.~\ref{fig:fig4}, the cross correlation caused by the astrophysical lines and the IA effect of galaxy shapes $C_{\mathrm{astro-IA}}$ is expected to prevent us from inferring 
the ALP parameters from the observed power spectra.
To study impacts of $C_{\mathrm{astro-IA}}$ on constraints of the ALP parameters,
we perform a Fisher analysis for cross correlation analyses in LSST and SPHEREx.
The Fisher matrix is commonly used to assess expected constraints of parameters of interest with given observables. In our case, it is computed as
\beqa
F_{\alpha\beta} &=& 
\sum_{i,j,p,q}\frac{\partial C^{(\rm obs)}_{\mathrm{LIM}-\kappa}(\ell_i | \nu_p)}{\partial s_\alpha}\, \bd{C}^{-1}(\ell_i, \nu_p | \ell_j, \nu_q)\, \nonumber \\
&&
\qquad \qquad \qquad \qquad
\times\frac{\partial C^{(\rm obs)}_{\mathrm{LIM}-\kappa}(\ell_j | \nu_q)}{\partial s_\beta},
\eeqa
where $C^{(\rm obs)}_{\mathrm{LIM}-\kappa}(\ell_i | \nu_p)$ represents the observed power spectrum at the $i$-th $\ell$ bin for the intensity map with the observed frequency being $\nu_p$, $\bd{C}$ is the covariance matrix of $C^{(\rm obs)}_{\mathrm{LIM}-\kappa}$ as in Eq.~(\ref{eq:cov}), and $s_\alpha$ is the $\alpha$-th parameter.
The inverse of the Fisher matrix provides an estimate of the error covariance for two parameters as
\beqa
F^{-1}_{\alpha\beta} = \langle \Delta s_\alpha \Delta s_\beta \rangle,
\eeqa
where $\Delta s_\alpha$ is the statistical uncertainty of parameter $s_\alpha$.

\begin{table*}[!t]
\begin{center}
\scalebox{0.90}[0.90]{
\begin{tabular}{|l|c|c|c|c|}
\tableline
&
Optimistic ($\ell_\mathrm{max}=1000$)
&
Optimistic ($\ell_\mathrm{max}=2000$)
&
Pessimistic ($\ell_\mathrm{max}=1000$)
&
Pessimistic ($\ell_\mathrm{max}=2000$)
\\ \hline
$\Delta m_a\, (\mathrm{eV})$ 
& 0.150 (0.146) & 0.112 (0.107) & 0.183 (0.146) & 0.146 (0.107)
\\ \hline
$\Delta g_{a\gamma\gamma}\, (10^{-11}\, \mathrm{GeV}^{-1})$ 
& 0.423 (0.416) & 0.302 (0.292) & 0.672 (0.416) & 0.552 (0.292)
\\ \hline
$\Delta {\cal A}_\mathrm{astro} \times 10^{3}$ 
& 6.15 (5.93) & 4.92 (4.74) 
& 10.4 (5.93) & 8.27 (4.74)
\\ \hline
$\Delta \eta_\mathrm{astro}$ 
& - & - & 0.0301 (0.0171) & 0.0191 (0.00904)  
\\ \hline
$\Delta {\cal A}_\mathrm{IA} \times 10^{2}$ 
& 1.32 (1.26) & 1.04 (0.993) 
& 8.32 (1.26) & 6.99 (0.993)
\\ \hline
$\Delta \eta_\mathrm{IA}$ 
& - & - & 0.110 (0.0169) & 0.0924 (0.0131) 
\\ \tableline
\end{tabular}
}
\caption{
\label{tab:ALP_const} 
Summary of the Fisher forecast of the ALP and astrophysics by the cross power spectra between line intensity and weak lensing maps.
We assume the effective survey area to be $18,000\, \mathrm{deg}^{2}$.
We consider the LSST-like source galaxy distribution (Eq.~[\ref{eq:dndz_source}])
and the intensity maps at $\lambda_{\mathrm{obs}}=750-4100\, \mathrm{nm}$ 
with the frequency resolution being 41.5. 
We refer the ``optimistic" scenario when the redshift evolution in the astrophysical line intensity and the IA effects are precisely known in advance 
(i.e. $\eta_\mathrm{astro}$ and $\eta_\mathrm{IA}$ are fixed).
On the other hand, the pessimistic scenario assumes that $\eta_\mathrm{astro}$ and $\eta_\mathrm{IA}$ are unknown.
In each table cell, the number without brackets show 
the $1\sigma$ constraint of single parameter when we marginalize other parameters, 
while the one in brackets is the un-marginalized counterpart.
Note that the fiducial parameters are set to $(m_a, g_{a\gamma\gamma}, {\cal A}_\mathrm{astro}, \eta_\mathrm{astro}, {\cal A}_\mathrm{IA}, \eta_\mathrm{IA}) = (1\, \mathrm{eV}, 1\times10^{-11}\, \mathrm{GeV}^{-1}, 1, 0, 1, 3)$.
}
\end{center}
\end{table*}

In our Fisher analysis, we consider the following parameters to vary: 
$\bd{s} = \{m_a, g_{a\gamma\gamma}, {\cal A}_\mathrm{IA}, \eta_\mathrm{IA}, {\cal A}_\mathrm{astro}, \eta_\mathrm{astro}\}$.
The fiducial values of our parameters is set to $\bd{s}_\mathrm{fid}=\{1\, \mathrm{eV}, 1\times10^{-11}\, \mathrm{GeV}^{-1}, 1, 3, 1, 0\}$.
When computing the Fisher matrix, we limit the range of multipoles as $10<\ell\le \ell_\mathrm{max}$ and examine three cases of $\ell_\mathrm{max}=500, 1000$, and 2000.
Changing $\ell_\mathrm{max}$ in our Fisher matrix tells us how smaller-scale information will help to improve the ALP constraints\footnote{
It would be worth noting that our theoretical model of cross power spectra is still phenomenological and model uncertainties at small scales are not fully understood yet.
Hence, the analysis with larger $\ell_\mathrm{max}$ will provide more stringent constraints, while it may induce some biased estimations of the ALP parameters in actual observations.}.
For reference, our cosmological model predicts
\beqa
k_\mathrm{max} &=& \frac{\ell_\mathrm{max}}{\chi(z)} \nonumber \\
&\simeq& \left(\frac{\ell_\mathrm{max}}{1000}\right)\, \left(\frac{z}{0.3}\right)^{-1} \left(1+\frac{z}{2}\right) \, h\mathrm{Mpc}^{-1}, \label{eq:kmax}
\eeqa
where $k^{-1}_\mathrm{max}$ gives an estimate of the smallest physical scale at a fixed $\ell_\mathrm{max}$
and redshift $z$. Eq.~(\ref{eq:kmax}) has a 10\%-level accuracy in the range of $0.05\le z\le 2$. 
Hence, the power spectrum with $\ell_\mathrm{max}=1000$ 
can extract the information of structures at $z=0.3-0.4$ with their size of $\simgt\mathrm{Mpc}$.

In our analysis, we use multi-frequency information of line intensity maps, allowing 
us to study the redshift evolution in our cross power spectra.
Because the effective redshift in the ALP-decay signal is different from the astrophysical counterparts, the multi-frequency information plays an essential role 
in breaking degeneracies among our cross power spectra.
Nevertheless, we expect that the efficiency of multi-frequency cross correlation depends on our prior knowledge of the redshift evolution 
in astrophysics-related parts.
In our case, the redshift evolution in the astrophysical line intensity and the IA effects of galaxy shapes controls how the multi-frequency analysis works.
To see impacts of prior knowledge in the redshift evolution,
we consider two scenarios.
One is the optimistic scenario 
when we assume that the parameters of $\eta_\mathrm{astro}$ and $\eta_\mathrm{IA}$ 
are perfectly known, 
and another is the pessimistic scenario making these two parameters unknown.
In realistic situations, 
we will have some informative prior of these two parameters from other analyses.
For instance, tomographic cosmic shear analyses can place reasonable constraints of $\eta_\mathrm{IA}$, while analyses of line intensity maps with the SPHEREx deep survey mode provide cleaner information of astrophysical sources and allow us to put constraints of $\eta_\mathrm{astro}$.
Hence, our Fisher analysis sets two extreme cases in the cross correlation analysis 
and gives some sense of how tight constraints of the ALP parameters will 
be obtained in more realistic situations.

Fig.~\ref{fig:fig7} summarizes an expected constraint in two-dimensional planes for the optimistic scenario with different $\ell_\mathrm{max}$.
We find that the multi-frequency information can successfully break degeneracies 
among the ALP-decay and astrophysical signals 
if we precisely know the redshift evolution in the astrophysical signals.
Note that cross correlation analyses at single frequencies can not separate the terms 
of $C_{\mathrm{ALP}-\kappa}$ and $C_{\mathrm{ALP-IA}}$ in practice.
The multi-frequency information can give tight constraints of ${\cal A}_\mathrm{IA}$
as well as ${\cal A}_\mathrm{astro}$, improving the ALP constraints.
Note that the astrophysical-line terms will probe higher-redshift structures 
than the ALP-decay terms do.
This can be seen from the difference in $\ell$-dependence of each term.
Using larger $\ell_\mathrm{max}$ will help to constrain the $\ell$-dependence of the observed power spectra, making the parameter constraints tighter.

For the pessimistic scenario, the parameter degeneracy between ${\cal A}_\mathrm{IA}$
and $\eta_\mathrm{IA}$ is significant as shown in Appendix~\ref{apdx:full_fisher} (see Fig.~\ref{fig:fig9}).
This strong degeneracy can degrade the constraints of the ALP parameters 
by a factor of $\sim1.5$.
We expect that the degeneracy between ${\cal A}_\mathrm{IA}$
and $\eta_\mathrm{IA}$ can be partly broken if we further apply a tomographic analysis
in lensing observables \cite{1999ApJ...522L..21H}, while we leave it for future studies.
Table~\ref{tab:ALP_const} provides a summary of our Fisher analysis.


\section{Conclusion and Discussions}\label{sec:con}

In this paper, we have studied cross correlation functions between line intensity and weak lensing maps to search for decaying axion-like particles (ALPs).
The ALP decay introduces an emission line with its frequency given by $\nu_a = 1.21\times 10^{14}\, (m_a/1\, \mathrm{eV})\, \mathrm{Hz}$, where $m_a$ is the particle mass of ALPs.
If a fraction of cosmic dark matter is made up of the ALPs, 
higher mass-density regions in the universe can emit larger line intensity by the ALP decay, as well as induce a more prominent distortion of background galaxy shapes by gravitational lensing effects.
We explored correlation signals caused by the ALP decay and weak gravitational lensing effects in large scale structures.
For this purpose, we developed a theoretical framework to predict the cross correlation functions by taking into account several astrophysical lines and intrinsic alignments (IAs) 
of galaxy shapes.
Suppose that the ALP explains the whole abundance of dark matter, 
our findings are summarized as follows.

\begin{itemize}
\item
Our baseline model shows that the cross correlation with an eV-mass ALP decay and weak lensing effects can dominate the astrophysical-line counterparts when observed wavelengths are set to $\lambda_\mathrm{obs} \sim 2000-3000\, \mathrm{nm}$. 
This is because weak lensing effects in modern galaxy surveys can probe the large-scale structures at redshifts of $z=0.3-0.6$, but 
astrophysical line sources populate higher-$z$ structures and can not introduce significant correlations.

\item
Assuming upcoming line intensity measurements by SPHEREx
and weak lensing measurements by the Large Survey of Space and Time (LSST), we found that a null detection of the cross correlation can place stringent upper limits of ALP-to-two-photons coupling $g_{a\gamma\gamma}\simlt10^{-11}\, \mathrm{GeV}^{-1}$ at $m_a=1-3.5\, \mathrm{eV}$.
The expected upper limits improve existing constraints by the study of stellar evolution of horizontal branch to red giants 
\cite{2014PhRvL.113s1302A} by a factor of $\sim6-10$. 

\item
Main systematic effects in our ALP search arise from the correlation between astrophysical lines and the IA effect. This makes the observed correlation signals negative depending on the IA amplitude. We performed a Fisher analysis to study detectability of the ALP-induced signal in the presence of the correlations between IA effects and astrophysical lines. Our Fisher analysis showed that the ALP decay with $m_a=1\, \mathrm{eV}$ and $g_{a\gamma\gamma}=10^{-11}\, \mathrm{GeV}^{-1}$ can be constrained with a 68\% confidence level of $\Delta m_a \sim 0.1-0.2\, \mathrm{eV}$ and $\Delta g_{a\gamma\gamma} \sim 0.3-0.7\times10^{-11}\, \mathrm{GeV}^{-1}$ by SPHEREx and LSST. 
The constraint depends on maximum multipoles in the analysis and prior information about the redshift evolution in astrophysical effects.

\end{itemize}

The cross correlation between line intensity and weak lensing maps can 
test various hints of ALP signals reported so far.
For example, Ref.~\cite{2014PhRvL.113s1302A} analyzed 39 galactic globular clusters and reported that
the observed number ratio of stars in horizontal over red giant branch of old stellar clusters differs from an astronomical prediction with a $2\sigma$ level. This discrepancy can be interpreted as the existence of the ALP decay with $g_{a\gamma\gamma} = 0.45^{+0.12}_{-0.16}\times10^{-10}\, \mathrm{GeV}^{-1}$, where the error represents a 68\% confidence level.
Recently, Ref.~\cite{2020arXiv201209179C} studied the ALP decay as a potential origin
of the anisotropy of the near-infrared background intensity and found that the ALP decay with 
$m_a=2.3-3\, \mathrm{eV}$ and $g_{a\gamma\gamma} = 1.1-1.6\times10^{-10}\, \mathrm{GeV}^{-1}$
can explain the excess in the observed power spectrum of background intensity at sub-arcmin scales.
These hints can be robustly examined by the cross correlation analysis with SPHEREx and LSST.

We expect that our results will provide a guideline of searching for the ALP decay with large-scale structure data.
Weak lensing analyses with several source redshift bins can improve constraints of the IA amplitude and offer further improvements of the ALP constraints.
Apart from weak lensing measurements in galaxy imaging surveys, 
there exist several tracers of large scale structures in the universe.
Those include number density of galaxies and secondary anisotropies in the cosmic microwave background (CMB) \cite{2008RPPh...71f6902A}.
Spectroscopic observations of galaxies enable us to map the large-scale structure in three-dimensional space, while such maps
are known to be a biased tracer of underlying cosmic dark matter \cite{2018PhR...733....1D}
and are subject to redshift space distortion effects \cite{1980lssu.book.....P, 1998ASSL..231..185H, 2004PhRvD..70h3007S}.
Among the secondary effects in CMB, gravitational lensing effects in CMB, referred to as CMB lensing, can provide cosmological information about higher-redshift matter density distributions at $z=2-4$ \cite{2006PhR...429....1L}.
The cross correlation with line intensity maps and CMB lensing allows us to place meaningful constraints of astrophysics but may be less sensitive to the ALP decay.
Cross correlations among line intensity maps and large-scale structures 
would be interesting not only for the ALP search but also constraints of cosmological parameters.
We leave cosmological applications for our future study.


\begin{acknowledgments}
We thank Naoki Yoshida and Kana Moriwaki for useful comments.
This work is supported by MEXT KAKENHI Grant Numbers of 19K14767 and 20H05861.
Numerical computations were carried out on Cray XC50 
at the Center for Computational Astrophysics in NAOJ.
\end{acknowledgments}

\appendix

\section{Auto power spectra in line intensity and weak lensing maps}
\label{apdx:cov}

\subsection{Gaussian covariance of binned power spectra}

\rev{Let $A$, $B$, $X$, and $Y$ be two-dimensional Gaussian random fields.
The estimator of the cross power spectrum between two fields $A$ and $B$ is given by
\beqa
\hat{C}_{AB}(\ell_i) = \frac{1}{N_\mathrm{mode}(\ell_i)}\sum_{\bd{\ell};\ell \in \ell_i}\, 
\tilde{A}(\bd{\ell}) \tilde{B}(\bd{\ell}), \label{eq:power_est}
\eeqa
where we employ a linear binning in multipoles $\ell$ with the bin width being $\Delta \ell$,
$\ell_{i}$ represents the $i$-th bin of $\ell$,
and $N_\mathrm{mode}(\ell_i)$ is the number of Fourier 
modes used for the power spectrum estimation at the $i$-th bin.
The number of Fourier modes can be computed as 
\beqa
N_\mathrm{mode}(\ell_i) 
&=& \sum_{\bd{\ell};\ell \in \ell_i} 1 \nonumber \\
&\simeq& \frac{2\pi \ell_i\, \Delta \ell}{(2\pi)^2/\Omega_S} = 2\ell_i \, \Delta\ell \, f_\mathrm{sky}, \label{eq:nmode}
\eeqa
where $\Omega_S$ is the survey area and $f_\mathrm{sky} = \Omega_S / (4\pi)$.
The ensemble average of Eq.~(\ref{eq:power_est}) is equal to its expected value $C_{AB}(\ell_i)$.}

\rev{We then consider the covariance matrix between two estimators of $\hat{C}_{AB}(\ell_i)$
and $\hat{C}_{XY}(\ell_j)$. The covariance is defined by
\beqa
\mathrm{Cov}\left[\hat{C}_{AB}(\ell_i), \hat{C}_{XY}(\ell_j)\right]
&=& 
\langle \hat{C}_{AB}(\ell_i) \hat{C}_{XY}(\ell_j) \rangle \nonumber \\
&&
- C_{AB}(\ell_i) C_{XY}(\ell_j). \label{eq:def_cov}
\eeqa
For the Gaussian fields, it holds that
\beqa
&&
\langle \tilde{A}(\bd{\ell}) \tilde{B}(\bd{\ell}) 
\tilde{X}(\bd{\ell}^{\prime}) \tilde{Y}(\bd{\ell}^{\prime}) \rangle
= 
\langle \tilde{A}(\bd{\ell}) \tilde{B}(\bd{\ell}) \rangle
\langle \tilde{X}(\bd{\ell}^{\prime}) \tilde{Y}(\bd{\ell}^{\prime}) \rangle 
\nonumber \\
&&
\qquad \qquad \qquad \qquad
+
\langle \tilde{A}(\bd{\ell}) \tilde{X}(\bd{\ell}^\prime) \rangle
\langle \tilde{B}(\bd{\ell}) \tilde{Y}(\bd{\ell}^{\prime}) \rangle 
\nonumber \\
&&
\qquad \qquad \qquad \qquad
+
\langle \tilde{B}(\bd{\ell}) \tilde{X}(\bd{\ell}^\prime) \rangle
\langle \tilde{A}(\bd{\ell}) \tilde{Y}(\bd{\ell}^{\prime}) \rangle.
\eeqa
Note that 
\beqa
&&\sum_{\bd{\ell};\ell \in \ell_i}\sum_{\bd{\ell}';\ell' \in \ell_j}
\langle \tilde{A}(\bd{\ell}) \tilde{X}(\bd{\ell}^\prime) \rangle
\langle \tilde{B}(\bd{\ell}) \tilde{Y}(\bd{\ell}^{\prime}) \rangle \nonumber \\
&=& \delta^{K}_{ij} \sum_{\bd{\ell};\ell \in \ell_i} 
\langle \tilde{A}(\bd{\ell}) \tilde{X}(\bd{\ell}) \rangle
\langle \tilde{B}(\bd{\ell}) \tilde{Y}(\bd{\ell}) \rangle \nonumber \\
&=& \delta^{K}_{ij} \sum_{\bd{\ell};\ell \in \ell_i} 
C_{AX}(\ell_i)\, C_{BY}(\ell_i) \nonumber \\
&=& \delta^{K}_{ij} \, N_\mathrm{mode}(\ell_i) \, C_{AX}(\ell_i)\, C_{BY}(\ell_i),
\eeqa
where $\delta^{K}_{ij}$ is the Kronecker delta symbol.
Hence, we arrive at
\beqa
&&\langle \hat{C}_{AB}(\ell_i) \hat{C}_{XY}(\ell_j) \rangle -
C_{AB}(\ell_i) C_{XY}(\ell_j) \nonumber \\
&=&
\frac{\delta^{K}_{ij}}{N_\mathrm{mode}(\ell_i)}
\Bigg{[} C_{AX}(\ell_i) C_{BY}(\ell_i) \nonumber \\
&&
\qquad \qquad \qquad \qquad
+C_{BX}(\ell_i) C_{AY}(\ell_i) \Biggr{]}. \label{eq:gauss_cov_general}
\eeqa
If one sets $A(\bd{\theta})=I(\bd{\theta}|\nu)$, 
$X(\bd{\theta})=I(\bd{\theta}|\nu')$, $B(\bd{\theta})=Y(\bd{\theta})=\kappa(\bd{\theta})$ 
where $I(\bd{\theta}|\nu)$ is the observed intensity at the frequency $\nu$ and $\kappa(\bd{\theta})$ is the lensing convergence, 
Eq.~(\ref{eq:gauss_cov_general}) leads to Eq.~(\ref{eq:cov}).}

\subsection{Each component in Eq.~(\ref{eq:cov})}

\rev{As in Eq.~(\ref{eq:cov}), the covariance matrix for the cross power spectrum $C^{(\mathrm{obs})}_{\mathrm{LIM}-\kappa}(\ell|\nu)$ includes two components of the cross power spectrum between two different intensity maps $C^{(\mathrm{obs})}_{\mathrm{LIM-LIM}}(\ell|\nu, \nu^{\prime}) $ and the auto power spectrum of the observed lensing field $C^{(\mathrm{obs})}_{\kappa-\kappa}(\ell)$. 
We here describe our model of these two power spectra.}

The cross power spectrum between two different intensity maps can be computed within the halo model as introduced in Section~\ref{subsubsec:astroline_vs_kappa}. 
Using the Limber approximation, we formally write 
\beqa
&&
C^{(\mathrm{obs})}_{\mathrm{LIM-LIM}}(\ell|\nu, \nu^{\prime}) 
=
C_{\mathrm{astro-astro}}(\ell|\nu, \nu^{\prime}) \nonumber \\
&&
\qquad \qquad
+
C_{\mathrm{astro-ALP}}(\ell|\nu, \nu^{\prime}) \nonumber \\
&&
\qquad \qquad
+
C_{\mathrm{ALP-ALP}}(\ell|\nu, \nu^{\prime}) +
C_{\mathrm{N}}(\ell|\nu, \nu^{\prime}), \label{eq:pk_LIM_LIM} 
\eeqa
and each term is given by
\beqa
&&
C_{\mathrm{astro-astro}}(\ell|\nu, \nu^{\prime})
=
\sum_{Q, Q^{\prime}} \int \mathrm{d}\chi\, \frac{W_Q(z|\nu)\, W_{Q^{\prime}}(z|\nu^\prime)}{\chi^2} \nonumber \\
&&
\qquad \qquad 
\times
{\cal A}^2_\mathrm{asto} \, (1+z)^{2\eta_\mathrm{astro}}
P_\mathrm{astro-astro}\left(\frac{\ell}{\chi}, z\right), \label{eq:pk_astro_astro} \\
&&
C_{\mathrm{astro-ALP}}(\ell|\nu, \nu^{\prime})
=
\sum_{Q} \int \mathrm{d}\chi\, 
\Biggl(\frac{W_Q(z|\nu)\, W_\mathrm{DM}(z|\nu^\prime)}{\chi^2} \nonumber \\
&&
\qquad \qquad \qquad
+\frac{W_Q(z|\nu^{\prime})\, W_\mathrm{DM}(z|\nu)}{\chi^2} \Biggr)
{\cal A}_\mathrm{asto} \, (1+z)^{\eta_\mathrm{astro}} \nonumber \\
&&
\qquad \qquad \qquad
\times \left(\frac{\Omega_\mathrm{DM}}{\Omega_\mathrm{m0}}\right)
P_\mathrm{astro-m}\left(\frac{\ell}{\chi}, z\right), \label{eq:pk_astro_ALP} \\
&&
C_{\mathrm{ALP-ALP}}(\ell|\nu, \nu^{\prime})
=
\int \mathrm{d}\chi\, \frac{W_\mathrm{DM}(z|\nu)\, W_\mathrm{DM}(z|\nu^\prime)}{\chi^2}  \nonumber \\
&&
\qquad \qquad \qquad
\times
\left(\frac{\Omega_\mathrm{DM}}{\Omega_\mathrm{m0}}\right)^2\, P_\mathrm{m}\left(\frac{\ell}{\chi}, z\right), \label{eq:pk_ALP_ALP}
\\
&&
C_{\mathrm{N}}(\ell|\nu, \nu^{\prime})
=
\delta^{K}_{\nu\nu^{\prime}} \frac{\sigma^2_n}{n_\mathrm{pix}}, 
\eeqa
where 
$Q = \{\mathrm{H}_{\alpha}, 
[\mathrm{O}_\mathrm{I\hspace{-.1em}I}],
[\mathrm{O}_\mathrm{I\hspace{-.1em}I\hspace{-.1em}I}],
\mathrm{H}_{\beta},
\mathrm{Ly}\alpha\}$,
$P_\mathrm{astro-astro}$ is the three-dimensional auto power spectrum of the field $\delta_\mathrm{astro}$ (see Eq.~[\ref{eq:delta_astro}]), 
$\sigma_n$ is the observational noise in line intensity maps in units of $\mathrm{erg}/\mathrm{s}/\mathrm{cm}^2/\mathrm{Hz}/\mathrm{str}$, and $n_\mathrm{pix}$ is the number density of pixels in intensity maps. For a given $5\sigma$-limiting AB magnitude $m_\mathrm{lim}$ per each pixel, $\sigma_n$ is given by
\beqa
\sigma_{n} = \frac{10^{-(m_\mathrm{lim}+48.60)/2.5}}{5\theta^2_\mathrm{pix}}\, [\mathrm{erg}/\mathrm{s}/\mathrm{cm}^2/\mathrm{Hz}/\mathrm{str}],
\eeqa
where $\theta_{\rm pix}$ is the angular size of each pixel. Note that $n_\mathrm{pix} = \theta^{-2}_\mathrm{pix}$.
Within the halo model, we can compute $P_\mathrm{astro-astro}$ as
\beqa
&&P_{\mathrm{astro-astro}}(k,z) = P^\mathrm{1h}_{\mathrm{astro-astro}}(k,z) \nonumber \\
&&
\qquad \qquad \qquad \qquad \qquad \qquad 
+ P^\mathrm{2h}_{\mathrm{astro-astro}}(k,z), \label{eq:P_astro_astro} \\
&&P^\mathrm{1h}_{\mathrm{astro-astro}}(k,z) =
{\scriptstyle {\cal N}^{-2}_\mathrm{astro}(z)}
\int\mathrm{d}M \frac{\mathrm{d}n}{\mathrm{d}M} \,\dot{M}^2_{*}(M,z), \\
&&P^\mathrm{2h}_{\mathrm{astro-astro}}(k,z) =
\left({\scriptstyle {\cal N}^{-1}_\mathrm{astro}(z)}
\int\mathrm{d}M \frac{\mathrm{d}n}{\mathrm{d}M} b_h(M,z) 
\dot{M}_{*}(M,z)
\right)^2
\nonumber \\
&&
\qquad \qquad \qquad \qquad \qquad \qquad
\times
P_L(k,z).
\eeqa
In practice, the integrand of Eq.~(\ref{eq:pk_LIM_LIM}) includes the product of delta functions, leading a divergence. Hence, we approximate the integral over $\chi$ by a discrete sum such as 
\beqa
&&
\int \mathrm{d}\chi \delta^{(1)}_\mathrm{D}(\chi-\chi_{Q}) \delta^{(1)}_\mathrm{D}(\chi-\chi_{Q^\prime}) \cdots \nonumber \\
&&\rightarrow
\sum^{\infty}_{i=0} \Delta \chi\, \Theta(\chi_i-\chi_{Q} | \Delta \chi)
\Theta(\chi_i-\chi_{Q^\prime} | \Delta \chi) \cdots,
\eeqa
where $\chi_i = (i+1/2)\, \Delta \chi$, and the function of $\Theta(\chi-\chi_Q | \Delta \chi)$ returns $1/\Delta \chi$ for $|\chi-\chi_Q| \le \Delta \chi/2$ and zero otherwise.
The width of comoving distance $\Delta \chi$ is set by
\beqa
\Delta \chi &=& 
\Biggl[
\left(\frac{\mathrm{d}\chi}{\mathrm{d}z}\right)\Bigg|_{z_Q}
\frac{1+z_Q}{\nu}\Delta \nu
\nonumber \\
&&
\qquad
\times
\left(\frac{\mathrm{d}\chi}{\mathrm{d}z}\right)\Bigg|_{z_{Q^\prime}}
\frac{1+z_{Q^\prime}}{\nu^\prime} \Delta \nu^{\prime}
\Biggr]^{1/2},
\eeqa
where $\Delta \nu = \nu/R$, $\Delta \nu^\prime = \nu^\prime/R$,
$z_Q = \nu_Q/\nu -1$, and $z_{Q^\prime} = \nu_{Q^\prime}/\nu^{\prime} -1$.
Note that $\nu_{Q}$ is the rest-frame frequency for the line $Q$ and $R$ is the frequency resolution in line intensity mapping measurements.

The auto power spectrum of the observed convergence field, $C^{(\mathrm{obs})}_{\kappa-\kappa}(\ell)$, is then given by
\beqa
C^{(\mathrm{obs})}_{\kappa-\kappa}(\ell) 
&=& 
\int \mathrm{d}\chi\, \frac{W^2_{\kappa}(z)}{\chi^2}\, P_\mathrm{m}\left(\frac{\ell}{\chi}, z\right) \nonumber \\
&&
+
\int \mathrm{d}\chi\, \frac{2W_{\kappa}(z)W_{\rm IA}(z)}{\chi^2}\, P_\mathrm{m}\left(\frac{\ell}{\chi}, z\right) \nonumber \\
&&
+
\int \mathrm{d}\chi\, \frac{W^2_{\rm IA}(z)}{\chi^2}\, P_\mathrm{m}\left(\frac{\ell}{\chi}, z\right) \nonumber \\
&&
+ \frac{\sigma^{2}_e}{n_\mathrm{g}},
\eeqa
where $\sigma_e$ is the intrinsic scatter of galaxy ellipticies per components, 
and $n_\mathrm{g}$ is the mean number density of source galaxies.

\section{Expected upper limits of dark matter lifetime}
\label{apdx:lifetime}

\begin{figure}[!t]
\includegraphics[clip, width=0.95\columnwidth]{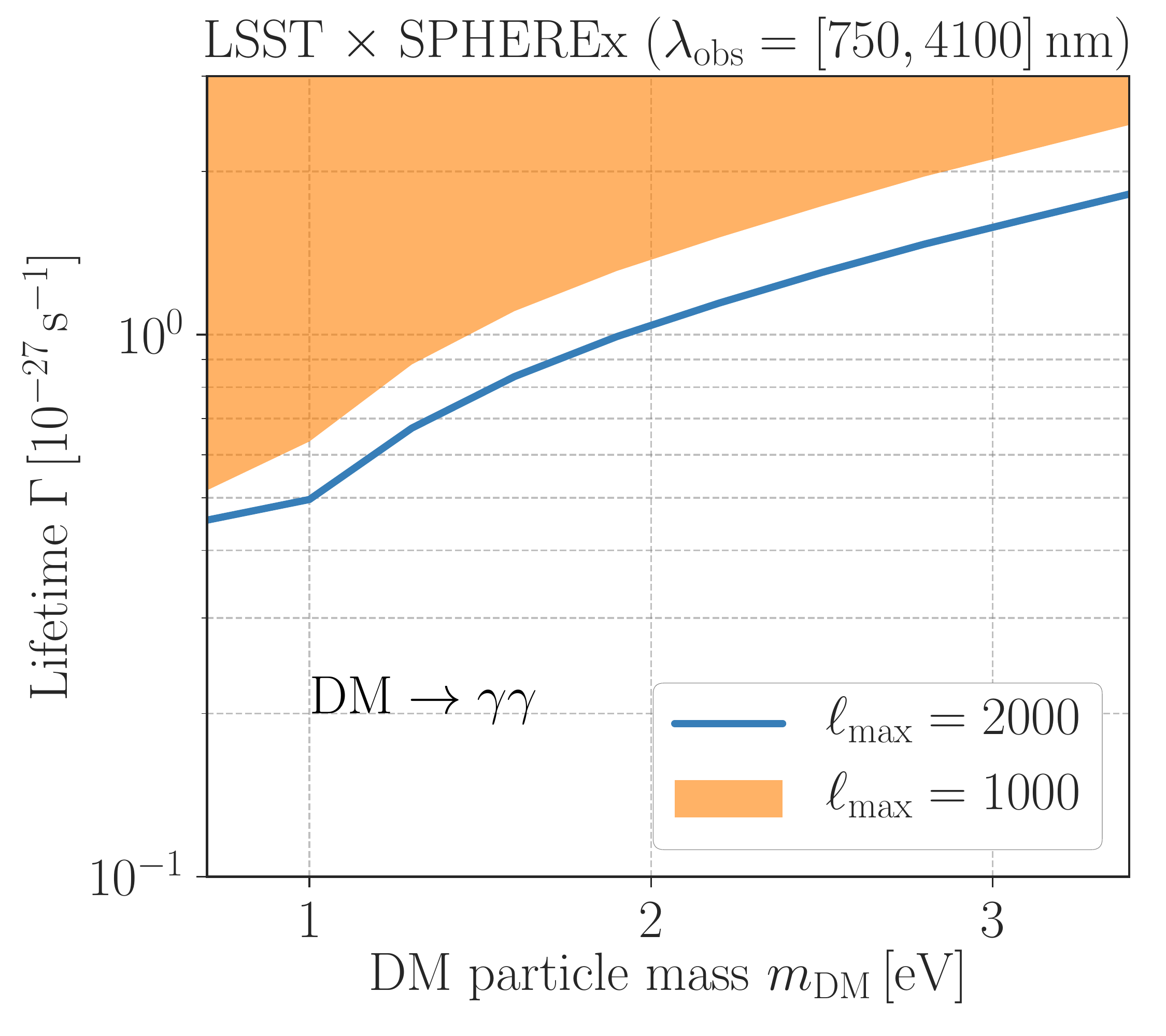}
\caption{\label{fig:fig8}
\rev{$2\sigma$-level upper limits of the lifetime of decaying dark matter (DM) $\Gamma$
as a function of DM mass $m_\mathrm{DM}$. 
The orange shaded region can be excluded by the cross correlation analysis 
with SPHEREx line intensity maps and LSST lensing when one uses the power spectra with $\ell\le \ell_\mathrm{max}=1000$.
The blue line shows the limits when one uses the information from smaller angular scales with $\ell_\mathrm{max}=2000$.}
}
\end{figure}

\rev{Fig.~\ref{fig:fig8} provides an expected $2\sigma$ upper limit of the dark matter lifetime $\Gamma$ by the cross power spectrum between line intensity and weak lensing maps in LSST and SPHEREx.}

\section{Full two-dimensional error circles in our Fisher analysis}
\label{apdx:full_fisher}

\begin{figure*}[!ht]
\begin{center}
\includegraphics[clip, width=2.1\columnwidth, bb=0 0 901 901]{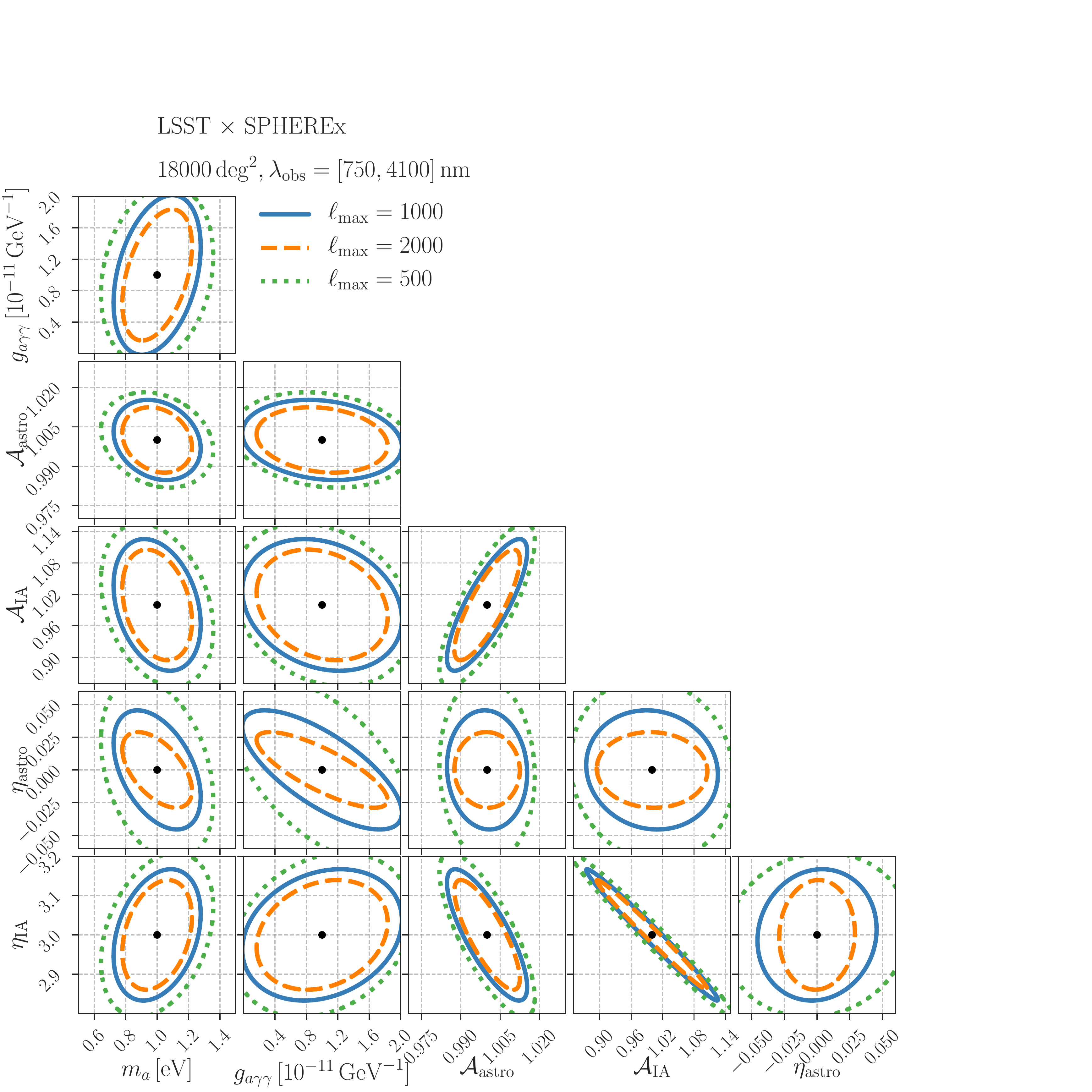}
\caption{\label{fig:fig9} 
Similar to Fig.~\ref{fig:fig7}, but we include uncertainties of redshift evolution in astrophysical line intensity and intrinsic alignments of galaxy shapes.
}
\end{center}
\end{figure*}

Fig.~\ref{fig:fig9} provides an expected 68\%-level confidence level of our six parameters by the cross power spectrum between line intensity and weak lensing maps in LSST and SPHEREx.

\bibliography{apssamp}

\begin{thebibliography}{58}%
\makeatletter
\providecommand \@ifxundefined [1]{%
 \@ifx{#1\undefined}
}%
\providecommand \@ifnum [1]{%
 \ifnum #1\expandafter \@firstoftwo
 \else \expandafter \@secondoftwo
 \fi
}%
\providecommand \@ifx [1]{%
 \ifx #1\expandafter \@firstoftwo
 \else \expandafter \@secondoftwo
 \fi
}%
\providecommand \natexlab [1]{#1}%
\providecommand \enquote  [1]{``#1''}%
\providecommand \bibnamefont  [1]{#1}%
\providecommand \bibfnamefont [1]{#1}%
\providecommand \citenamefont [1]{#1}%
\providecommand \href@noop [0]{\@secondoftwo}%
\providecommand \href [0]{\begingroup \@sanitize@url \@href}%
\providecommand \@href[1]{\@@startlink{#1}\@@href}%
\providecommand \@@href[1]{\endgroup#1\@@endlink}%
\providecommand \@sanitize@url [0]{\catcode `\\12\catcode `\$12\catcode
  `\&12\catcode `\#12\catcode `\^12\catcode `\_12\catcode `\%12\relax}%
\providecommand \@@startlink[1]{}%
\providecommand \@@endlink[0]{}%
\providecommand \url  [0]{\begingroup\@sanitize@url \@url }%
\providecommand \@url [1]{\endgroup\@href {#1}{\urlprefix }}%
\providecommand \urlprefix  [0]{URL }%
\providecommand \Eprint [0]{\href }%
\providecommand \doibase [0]{https://doi.org/}%
\providecommand \selectlanguage [0]{\@gobble}%
\providecommand \bibinfo  [0]{\@secondoftwo}%
\providecommand \bibfield  [0]{\@secondoftwo}%
\providecommand \translation [1]{[#1]}%
\providecommand \BibitemOpen [0]{}%
\providecommand \bibitemStop [0]{}%
\providecommand \bibitemNoStop [0]{.\EOS\space}%
\providecommand \EOS [0]{\spacefactor3000\relax}%
\providecommand \BibitemShut  [1]{\csname bibitem#1\endcsname}%
\let\auto@bib@innerbib\@empty
\bibitem [{\citenamefont {{Planck Collaboration}}\ \emph
  {et~al.}(2020)\citenamefont {{Planck Collaboration}}, \citenamefont
  {{Aghanim}}, \citenamefont {{Akrami}}, \citenamefont {{Ashdown}},
  \citenamefont {{Aumont}}, \citenamefont {{Baccigalupi}}, \citenamefont
  {{Ballardini}}, \citenamefont {{Banday}}, \citenamefont {{Barreiro}},
  \citenamefont {{Bartolo}}, \citenamefont {{Basak}}, \citenamefont {{Battye}},
  \citenamefont {{Benabed}}, \citenamefont {{Bernard}}, \citenamefont
  {{Bersanelli}}, \citenamefont {{Bielewicz}}, \citenamefont {{Bock}},
  \citenamefont {{Bond}}, \citenamefont {{Borrill}}, \citenamefont {{Bouchet}},
  \citenamefont {{Boulanger}}, \citenamefont {{Bucher}}, \citenamefont
  {{Burigana}}, \citenamefont {{Butler}}, \citenamefont {{Calabrese}},
  \citenamefont {{Cardoso}}, \citenamefont {{Carron}}, \citenamefont
  {{Challinor}}, \citenamefont {{Chiang}}, \citenamefont {{Chluba}},
  \citenamefont {{Colombo}}, \citenamefont {{Combet}}, \citenamefont
  {{Contreras}}, \citenamefont {{Crill}}, \citenamefont {{Cuttaia}},
  \citenamefont {{de Bernardis}}, \citenamefont {{de Zotti}}, \citenamefont
  {{Delabrouille}}, \citenamefont {{Delouis}}, \citenamefont {{Di Valentino}},
  \citenamefont {{Diego}}, \citenamefont {{Dor{\'e}}}, \citenamefont
  {{Douspis}}, \citenamefont {{Ducout}}, \citenamefont {{Dupac}}, \citenamefont
  {{Dusini}}, \citenamefont {{Efstathiou}}, \citenamefont {{Elsner}},
  \citenamefont {{En{\ss}lin}}, \citenamefont {{Eriksen}}, \citenamefont
  {{Fantaye}}, \citenamefont {{Farhang}}, \citenamefont {{Fergusson}},
  \citenamefont {{Fernandez-Cobos}}, \citenamefont {{Finelli}}, \citenamefont
  {{Forastieri}}, \citenamefont {{Frailis}}, \citenamefont {{Fraisse}},
  \citenamefont {{Franceschi}}, \citenamefont {{Frolov}}, \citenamefont
  {{Galeotta}}, \citenamefont {{Galli}}, \citenamefont {{Ganga}}, \citenamefont
  {{G{\'e}nova-Santos}}, \citenamefont {{Gerbino}}, \citenamefont {{Ghosh}},
  \citenamefont {{Gonz{\'a}lez-Nuevo}}, \citenamefont {{G{\'o}rski}},
  \citenamefont {{Gratton}}, \citenamefont {{Gruppuso}}, \citenamefont
  {{Gudmundsson}}, \citenamefont {{Hamann}}, \citenamefont {{Handley}},
  \citenamefont {{Hansen}}, \citenamefont {{Herranz}}, \citenamefont
  {{Hildebrandt}}, \citenamefont {{Hivon}}, \citenamefont {{Huang}},
  \citenamefont {{Jaffe}}, \citenamefont {{Jones}}, \citenamefont {{Karakci}},
  \citenamefont {{Keih{\"a}nen}}, \citenamefont {{Keskitalo}}, \citenamefont
  {{Kiiveri}}, \citenamefont {{Kim}}, \citenamefont {{Kisner}}, \citenamefont
  {{Knox}}, \citenamefont {{Krachmalnicoff}}, \citenamefont {{Kunz}},
  \citenamefont {{Kurki-Suonio}}, \citenamefont {{Lagache}}, \citenamefont
  {{Lamarre}}, \citenamefont {{Lasenby}}, \citenamefont {{Lattanzi}},
  \citenamefont {{Lawrence}}, \citenamefont {{Le Jeune}}, \citenamefont
  {{Lemos}}, \citenamefont {{Lesgourgues}}, \citenamefont {{Levrier}},
  \citenamefont {{Lewis}}, \citenamefont {{Liguori}}, \citenamefont {{Lilje}},
  \citenamefont {{Lilley}}, \citenamefont {{Lindholm}}, \citenamefont
  {{L{\'o}pez-Caniego}}, \citenamefont {{Lubin}}, \citenamefont {{Ma}},
  \citenamefont {{Mac{\'\i}as-P{\'e}rez}}, \citenamefont {{Maggio}},
  \citenamefont {{Maino}}, \citenamefont {{Mandolesi}}, \citenamefont
  {{Mangilli}}, \citenamefont {{Marcos-Caballero}}, \citenamefont {{Maris}},
  \citenamefont {{Martin}}, \citenamefont {{Martinelli}}, \citenamefont
  {{Mart{\'\i}nez-Gonz{\'a}lez}}, \citenamefont {{Matarrese}}, \citenamefont
  {{Mauri}}, \citenamefont {{McEwen}}, \citenamefont {{Meinhold}},
  \citenamefont {{Melchiorri}}, \citenamefont {{Mennella}}, \citenamefont
  {{Migliaccio}}, \citenamefont {{Millea}}, \citenamefont {{Mitra}},
  \citenamefont {{Miville-Desch{\^e}nes}}, \citenamefont {{Molinari}},
  \citenamefont {{Montier}}, \citenamefont {{Morgante}}, \citenamefont
  {{Moss}}, \citenamefont {{Natoli}}, \citenamefont {{N{\o}rgaard-Nielsen}},
  \citenamefont {{Pagano}}, \citenamefont {{Paoletti}}, \citenamefont
  {{Partridge}}, \citenamefont {{Patanchon}}, \citenamefont {{Peiris}},
  \citenamefont {{Perrotta}}, \citenamefont {{Pettorino}}, \citenamefont
  {{Piacentini}}, \citenamefont {{Polastri}}, \citenamefont {{Polenta}},
  \citenamefont {{Puget}}, \citenamefont {{Rachen}}, \citenamefont
  {{Reinecke}}, \citenamefont {{Remazeilles}}, \citenamefont {{Renzi}},
  \citenamefont {{Rocha}}, \citenamefont {{Rosset}}, \citenamefont {{Roudier}},
  \citenamefont {{Rubi{\~n}o-Mart{\'\i}n}}, \citenamefont {{Ruiz-Granados}},
  \citenamefont {{Salvati}}, \citenamefont {{Sandri}}, \citenamefont
  {{Savelainen}}, \citenamefont {{Scott}}, \citenamefont {{Shellard}},
  \citenamefont {{Sirignano}}, \citenamefont {{Sirri}}, \citenamefont
  {{Spencer}}, \citenamefont {{Sunyaev}}, \citenamefont {{Suur-Uski}},
  \citenamefont {{Tauber}}, \citenamefont {{Tavagnacco}}, \citenamefont
  {{Tenti}}, \citenamefont {{Toffolatti}}, \citenamefont {{Tomasi}},
  \citenamefont {{Trombetti}}, \citenamefont {{Valenziano}}, \citenamefont
  {{Valiviita}}, \citenamefont {{Van Tent}}, \citenamefont {{Vibert}},
  \citenamefont {{Vielva}}, \citenamefont {{Villa}}, \citenamefont
  {{Vittorio}}, \citenamefont {{Wandelt}}, \citenamefont {{Wehus}},
  \citenamefont {{White}}, \citenamefont {{White}}, \citenamefont {{Zacchei}},\
  and\ \citenamefont {{Zonca}}}]{2020A&A...641A...6P}%
  \BibitemOpen
  \bibfield  {author} {\bibinfo {author} {\bibnamefont {{Planck
  Collaboration}}}, \bibinfo {author} {\bibfnamefont {N.}~\bibnamefont
  {{Aghanim}}}, \bibinfo {author} {\bibfnamefont {Y.}~\bibnamefont {{Akrami}}},
  \bibinfo {author} {\bibfnamefont {M.}~\bibnamefont {{Ashdown}}}, \bibinfo
  {author} {\bibfnamefont {J.}~\bibnamefont {{Aumont}}}, \bibinfo {author}
  {\bibfnamefont {C.}~\bibnamefont {{Baccigalupi}}}, \bibinfo {author}
  {\bibfnamefont {M.}~\bibnamefont {{Ballardini}}}, \bibinfo {author}
  {\bibfnamefont {A.~J.}\ \bibnamefont {{Banday}}}, \bibinfo {author}
  {\bibfnamefont {R.~B.}\ \bibnamefont {{Barreiro}}}, \bibinfo {author}
  {\bibfnamefont {N.}~\bibnamefont {{Bartolo}}}, \bibinfo {author}
  {\bibfnamefont {S.}~\bibnamefont {{Basak}}}, \bibinfo {author} {\bibfnamefont
  {R.}~\bibnamefont {{Battye}}}, \bibinfo {author} {\bibfnamefont
  {K.}~\bibnamefont {{Benabed}}}, \bibinfo {author} {\bibfnamefont {J.~P.}\
  \bibnamefont {{Bernard}}}, \bibinfo {author} {\bibfnamefont {M.}~\bibnamefont
  {{Bersanelli}}}, \bibinfo {author} {\bibfnamefont {P.}~\bibnamefont
  {{Bielewicz}}}, \bibinfo {author} {\bibfnamefont {J.~J.}\ \bibnamefont
  {{Bock}}}, \bibinfo {author} {\bibfnamefont {J.~R.}\ \bibnamefont {{Bond}}},
  \bibinfo {author} {\bibfnamefont {J.}~\bibnamefont {{Borrill}}}, \bibinfo
  {author} {\bibfnamefont {F.~R.}\ \bibnamefont {{Bouchet}}}, \bibinfo {author}
  {\bibfnamefont {F.}~\bibnamefont {{Boulanger}}}, \bibinfo {author}
  {\bibfnamefont {M.}~\bibnamefont {{Bucher}}}, \bibinfo {author}
  {\bibfnamefont {C.}~\bibnamefont {{Burigana}}}, \bibinfo {author}
  {\bibfnamefont {R.~C.}\ \bibnamefont {{Butler}}}, \bibinfo {author}
  {\bibfnamefont {E.}~\bibnamefont {{Calabrese}}}, \bibinfo {author}
  {\bibfnamefont {J.~F.}\ \bibnamefont {{Cardoso}}}, \bibinfo {author}
  {\bibfnamefont {J.}~\bibnamefont {{Carron}}}, \bibinfo {author}
  {\bibfnamefont {A.}~\bibnamefont {{Challinor}}}, \bibinfo {author}
  {\bibfnamefont {H.~C.}\ \bibnamefont {{Chiang}}}, \bibinfo {author}
  {\bibfnamefont {J.}~\bibnamefont {{Chluba}}}, \bibinfo {author}
  {\bibfnamefont {L.~P.~L.}\ \bibnamefont {{Colombo}}}, \bibinfo {author}
  {\bibfnamefont {C.}~\bibnamefont {{Combet}}}, \bibinfo {author}
  {\bibfnamefont {D.}~\bibnamefont {{Contreras}}}, \bibinfo {author}
  {\bibfnamefont {B.~P.}\ \bibnamefont {{Crill}}}, \bibinfo {author}
  {\bibfnamefont {F.}~\bibnamefont {{Cuttaia}}}, \bibinfo {author}
  {\bibfnamefont {P.}~\bibnamefont {{de Bernardis}}}, \bibinfo {author}
  {\bibfnamefont {G.}~\bibnamefont {{de Zotti}}}, \bibinfo {author}
  {\bibfnamefont {J.}~\bibnamefont {{Delabrouille}}}, \bibinfo {author}
  {\bibfnamefont {J.~M.}\ \bibnamefont {{Delouis}}}, \bibinfo {author}
  {\bibfnamefont {E.}~\bibnamefont {{Di Valentino}}}, \bibinfo {author}
  {\bibfnamefont {J.~M.}\ \bibnamefont {{Diego}}}, \bibinfo {author}
  {\bibfnamefont {O.}~\bibnamefont {{Dor{\'e}}}}, \bibinfo {author}
  {\bibfnamefont {M.}~\bibnamefont {{Douspis}}}, \bibinfo {author}
  {\bibfnamefont {A.}~\bibnamefont {{Ducout}}}, \bibinfo {author}
  {\bibfnamefont {X.}~\bibnamefont {{Dupac}}}, \bibinfo {author} {\bibfnamefont
  {S.}~\bibnamefont {{Dusini}}}, \bibinfo {author} {\bibfnamefont
  {G.}~\bibnamefont {{Efstathiou}}}, \bibinfo {author} {\bibfnamefont
  {F.}~\bibnamefont {{Elsner}}}, \bibinfo {author} {\bibfnamefont {T.~A.}\
  \bibnamefont {{En{\ss}lin}}}, \bibinfo {author} {\bibfnamefont {H.~K.}\
  \bibnamefont {{Eriksen}}}, \bibinfo {author} {\bibfnamefont {Y.}~\bibnamefont
  {{Fantaye}}}, \bibinfo {author} {\bibfnamefont {M.}~\bibnamefont
  {{Farhang}}}, \bibinfo {author} {\bibfnamefont {J.}~\bibnamefont
  {{Fergusson}}}, \bibinfo {author} {\bibfnamefont {R.}~\bibnamefont
  {{Fernandez-Cobos}}}, \bibinfo {author} {\bibfnamefont {F.}~\bibnamefont
  {{Finelli}}}, \bibinfo {author} {\bibfnamefont {F.}~\bibnamefont
  {{Forastieri}}}, \bibinfo {author} {\bibfnamefont {M.}~\bibnamefont
  {{Frailis}}}, \bibinfo {author} {\bibfnamefont {A.~A.}\ \bibnamefont
  {{Fraisse}}}, \bibinfo {author} {\bibfnamefont {E.}~\bibnamefont
  {{Franceschi}}}, \bibinfo {author} {\bibfnamefont {A.}~\bibnamefont
  {{Frolov}}}, \bibinfo {author} {\bibfnamefont {S.}~\bibnamefont
  {{Galeotta}}}, \bibinfo {author} {\bibfnamefont {S.}~\bibnamefont {{Galli}}},
  \bibinfo {author} {\bibfnamefont {K.}~\bibnamefont {{Ganga}}}, \bibinfo
  {author} {\bibfnamefont {R.~T.}\ \bibnamefont {{G{\'e}nova-Santos}}},
  \bibinfo {author} {\bibfnamefont {M.}~\bibnamefont {{Gerbino}}}, \bibinfo
  {author} {\bibfnamefont {T.}~\bibnamefont {{Ghosh}}}, \bibinfo {author}
  {\bibfnamefont {J.}~\bibnamefont {{Gonz{\'a}lez-Nuevo}}}, \bibinfo {author}
  {\bibfnamefont {K.~M.}\ \bibnamefont {{G{\'o}rski}}}, \bibinfo {author}
  {\bibfnamefont {S.}~\bibnamefont {{Gratton}}}, \bibinfo {author}
  {\bibfnamefont {A.}~\bibnamefont {{Gruppuso}}}, \bibinfo {author}
  {\bibfnamefont {J.~E.}\ \bibnamefont {{Gudmundsson}}}, \bibinfo {author}
  {\bibfnamefont {J.}~\bibnamefont {{Hamann}}}, \bibinfo {author}
  {\bibfnamefont {W.}~\bibnamefont {{Handley}}}, \bibinfo {author}
  {\bibfnamefont {F.~K.}\ \bibnamefont {{Hansen}}}, \bibinfo {author}
  {\bibfnamefont {D.}~\bibnamefont {{Herranz}}}, \bibinfo {author}
  {\bibfnamefont {S.~R.}\ \bibnamefont {{Hildebrandt}}}, \bibinfo {author}
  {\bibfnamefont {E.}~\bibnamefont {{Hivon}}}, \bibinfo {author} {\bibfnamefont
  {Z.}~\bibnamefont {{Huang}}}, \bibinfo {author} {\bibfnamefont {A.~H.}\
  \bibnamefont {{Jaffe}}}, \bibinfo {author} {\bibfnamefont {W.~C.}\
  \bibnamefont {{Jones}}}, \bibinfo {author} {\bibfnamefont {A.}~\bibnamefont
  {{Karakci}}}, \bibinfo {author} {\bibfnamefont {E.}~\bibnamefont
  {{Keih{\"a}nen}}}, \bibinfo {author} {\bibfnamefont {R.}~\bibnamefont
  {{Keskitalo}}}, \bibinfo {author} {\bibfnamefont {K.}~\bibnamefont
  {{Kiiveri}}}, \bibinfo {author} {\bibfnamefont {J.}~\bibnamefont {{Kim}}},
  \bibinfo {author} {\bibfnamefont {T.~S.}\ \bibnamefont {{Kisner}}}, \bibinfo
  {author} {\bibfnamefont {L.}~\bibnamefont {{Knox}}}, \bibinfo {author}
  {\bibfnamefont {N.}~\bibnamefont {{Krachmalnicoff}}}, \bibinfo {author}
  {\bibfnamefont {M.}~\bibnamefont {{Kunz}}}, \bibinfo {author} {\bibfnamefont
  {H.}~\bibnamefont {{Kurki-Suonio}}}, \bibinfo {author} {\bibfnamefont
  {G.}~\bibnamefont {{Lagache}}}, \bibinfo {author} {\bibfnamefont {J.~M.}\
  \bibnamefont {{Lamarre}}}, \bibinfo {author} {\bibfnamefont {A.}~\bibnamefont
  {{Lasenby}}}, \bibinfo {author} {\bibfnamefont {M.}~\bibnamefont
  {{Lattanzi}}}, \bibinfo {author} {\bibfnamefont {C.~R.}\ \bibnamefont
  {{Lawrence}}}, \bibinfo {author} {\bibfnamefont {M.}~\bibnamefont {{Le
  Jeune}}}, \bibinfo {author} {\bibfnamefont {P.}~\bibnamefont {{Lemos}}},
  \bibinfo {author} {\bibfnamefont {J.}~\bibnamefont {{Lesgourgues}}}, \bibinfo
  {author} {\bibfnamefont {F.}~\bibnamefont {{Levrier}}}, \bibinfo {author}
  {\bibfnamefont {A.}~\bibnamefont {{Lewis}}}, \bibinfo {author} {\bibfnamefont
  {M.}~\bibnamefont {{Liguori}}}, \bibinfo {author} {\bibfnamefont {P.~B.}\
  \bibnamefont {{Lilje}}}, \bibinfo {author} {\bibfnamefont {M.}~\bibnamefont
  {{Lilley}}}, \bibinfo {author} {\bibfnamefont {V.}~\bibnamefont
  {{Lindholm}}}, \bibinfo {author} {\bibfnamefont {M.}~\bibnamefont
  {{L{\'o}pez-Caniego}}}, \bibinfo {author} {\bibfnamefont {P.~M.}\
  \bibnamefont {{Lubin}}}, \bibinfo {author} {\bibfnamefont {Y.~Z.}\
  \bibnamefont {{Ma}}}, \bibinfo {author} {\bibfnamefont {J.~F.}\ \bibnamefont
  {{Mac{\'\i}as-P{\'e}rez}}}, \bibinfo {author} {\bibfnamefont
  {G.}~\bibnamefont {{Maggio}}}, \bibinfo {author} {\bibfnamefont
  {D.}~\bibnamefont {{Maino}}}, \bibinfo {author} {\bibfnamefont
  {N.}~\bibnamefont {{Mandolesi}}}, \bibinfo {author} {\bibfnamefont
  {A.}~\bibnamefont {{Mangilli}}}, \bibinfo {author} {\bibfnamefont
  {A.}~\bibnamefont {{Marcos-Caballero}}}, \bibinfo {author} {\bibfnamefont
  {M.}~\bibnamefont {{Maris}}}, \bibinfo {author} {\bibfnamefont {P.~G.}\
  \bibnamefont {{Martin}}}, \bibinfo {author} {\bibfnamefont {M.}~\bibnamefont
  {{Martinelli}}}, \bibinfo {author} {\bibfnamefont {E.}~\bibnamefont
  {{Mart{\'\i}nez-Gonz{\'a}lez}}}, \bibinfo {author} {\bibfnamefont
  {S.}~\bibnamefont {{Matarrese}}}, \bibinfo {author} {\bibfnamefont
  {N.}~\bibnamefont {{Mauri}}}, \bibinfo {author} {\bibfnamefont {J.~D.}\
  \bibnamefont {{McEwen}}}, \bibinfo {author} {\bibfnamefont {P.~R.}\
  \bibnamefont {{Meinhold}}}, \bibinfo {author} {\bibfnamefont
  {A.}~\bibnamefont {{Melchiorri}}}, \bibinfo {author} {\bibfnamefont
  {A.}~\bibnamefont {{Mennella}}}, \bibinfo {author} {\bibfnamefont
  {M.}~\bibnamefont {{Migliaccio}}}, \bibinfo {author} {\bibfnamefont
  {M.}~\bibnamefont {{Millea}}}, \bibinfo {author} {\bibfnamefont
  {S.}~\bibnamefont {{Mitra}}}, \bibinfo {author} {\bibfnamefont {M.~A.}\
  \bibnamefont {{Miville-Desch{\^e}nes}}}, \bibinfo {author} {\bibfnamefont
  {D.}~\bibnamefont {{Molinari}}}, \bibinfo {author} {\bibfnamefont
  {L.}~\bibnamefont {{Montier}}}, \bibinfo {author} {\bibfnamefont
  {G.}~\bibnamefont {{Morgante}}}, \bibinfo {author} {\bibfnamefont
  {A.}~\bibnamefont {{Moss}}}, \bibinfo {author} {\bibfnamefont
  {P.}~\bibnamefont {{Natoli}}}, \bibinfo {author} {\bibfnamefont {H.~U.}\
  \bibnamefont {{N{\o}rgaard-Nielsen}}}, \bibinfo {author} {\bibfnamefont
  {L.}~\bibnamefont {{Pagano}}}, \bibinfo {author} {\bibfnamefont
  {D.}~\bibnamefont {{Paoletti}}}, \bibinfo {author} {\bibfnamefont
  {B.}~\bibnamefont {{Partridge}}}, \bibinfo {author} {\bibfnamefont
  {G.}~\bibnamefont {{Patanchon}}}, \bibinfo {author} {\bibfnamefont {H.~V.}\
  \bibnamefont {{Peiris}}}, \bibinfo {author} {\bibfnamefont {F.}~\bibnamefont
  {{Perrotta}}}, \bibinfo {author} {\bibfnamefont {V.}~\bibnamefont
  {{Pettorino}}}, \bibinfo {author} {\bibfnamefont {F.}~\bibnamefont
  {{Piacentini}}}, \bibinfo {author} {\bibfnamefont {L.}~\bibnamefont
  {{Polastri}}}, \bibinfo {author} {\bibfnamefont {G.}~\bibnamefont
  {{Polenta}}}, \bibinfo {author} {\bibfnamefont {J.~L.}\ \bibnamefont
  {{Puget}}}, \bibinfo {author} {\bibfnamefont {J.~P.}\ \bibnamefont
  {{Rachen}}}, \bibinfo {author} {\bibfnamefont {M.}~\bibnamefont
  {{Reinecke}}}, \bibinfo {author} {\bibfnamefont {M.}~\bibnamefont
  {{Remazeilles}}}, \bibinfo {author} {\bibfnamefont {A.}~\bibnamefont
  {{Renzi}}}, \bibinfo {author} {\bibfnamefont {G.}~\bibnamefont {{Rocha}}},
  \bibinfo {author} {\bibfnamefont {C.}~\bibnamefont {{Rosset}}}, \bibinfo
  {author} {\bibfnamefont {G.}~\bibnamefont {{Roudier}}}, \bibinfo {author}
  {\bibfnamefont {J.~A.}\ \bibnamefont {{Rubi{\~n}o-Mart{\'\i}n}}}, \bibinfo
  {author} {\bibfnamefont {B.}~\bibnamefont {{Ruiz-Granados}}}, \bibinfo
  {author} {\bibfnamefont {L.}~\bibnamefont {{Salvati}}}, \bibinfo {author}
  {\bibfnamefont {M.}~\bibnamefont {{Sandri}}}, \bibinfo {author}
  {\bibfnamefont {M.}~\bibnamefont {{Savelainen}}}, \bibinfo {author}
  {\bibfnamefont {D.}~\bibnamefont {{Scott}}}, \bibinfo {author} {\bibfnamefont
  {E.~P.~S.}\ \bibnamefont {{Shellard}}}, \bibinfo {author} {\bibfnamefont
  {C.}~\bibnamefont {{Sirignano}}}, \bibinfo {author} {\bibfnamefont
  {G.}~\bibnamefont {{Sirri}}}, \bibinfo {author} {\bibfnamefont {L.~D.}\
  \bibnamefont {{Spencer}}}, \bibinfo {author} {\bibfnamefont {R.}~\bibnamefont
  {{Sunyaev}}}, \bibinfo {author} {\bibfnamefont {A.~S.}\ \bibnamefont
  {{Suur-Uski}}}, \bibinfo {author} {\bibfnamefont {J.~A.}\ \bibnamefont
  {{Tauber}}}, \bibinfo {author} {\bibfnamefont {D.}~\bibnamefont
  {{Tavagnacco}}}, \bibinfo {author} {\bibfnamefont {M.}~\bibnamefont
  {{Tenti}}}, \bibinfo {author} {\bibfnamefont {L.}~\bibnamefont
  {{Toffolatti}}}, \bibinfo {author} {\bibfnamefont {M.}~\bibnamefont
  {{Tomasi}}}, \bibinfo {author} {\bibfnamefont {T.}~\bibnamefont
  {{Trombetti}}}, \bibinfo {author} {\bibfnamefont {L.}~\bibnamefont
  {{Valenziano}}}, \bibinfo {author} {\bibfnamefont {J.}~\bibnamefont
  {{Valiviita}}}, \bibinfo {author} {\bibfnamefont {B.}~\bibnamefont {{Van
  Tent}}}, \bibinfo {author} {\bibfnamefont {L.}~\bibnamefont {{Vibert}}},
  \bibinfo {author} {\bibfnamefont {P.}~\bibnamefont {{Vielva}}}, \bibinfo
  {author} {\bibfnamefont {F.}~\bibnamefont {{Villa}}}, \bibinfo {author}
  {\bibfnamefont {N.}~\bibnamefont {{Vittorio}}}, \bibinfo {author}
  {\bibfnamefont {B.~D.}\ \bibnamefont {{Wandelt}}}, \bibinfo {author}
  {\bibfnamefont {I.~K.}\ \bibnamefont {{Wehus}}}, \bibinfo {author}
  {\bibfnamefont {M.}~\bibnamefont {{White}}}, \bibinfo {author} {\bibfnamefont
  {S.~D.~M.}\ \bibnamefont {{White}}}, \bibinfo {author} {\bibfnamefont
  {A.}~\bibnamefont {{Zacchei}}},\ and\ \bibinfo {author} {\bibfnamefont
  {A.}~\bibnamefont {{Zonca}}},\ }\href
  {https://doi.org/10.1051/0004-6361/201833910} {\bibfield  {journal} {\bibinfo
   {journal} {\aap}\ }\textbf {\bibinfo {volume} {641}},\ \bibinfo {eid} {A6}
  (\bibinfo {year} {2020})},\ \Eprint {https://arxiv.org/abs/1807.06209}
  {arXiv:1807.06209 [astro-ph.CO]} \BibitemShut {NoStop}%
\bibitem [{\citenamefont {{Preskill}}\ \emph {et~al.}(1983)\citenamefont
  {{Preskill}}, \citenamefont {{Wise}},\ and\ \citenamefont
  {{Wilczek}}}]{1983PhLB..120..127P}%
  \BibitemOpen
  \bibfield  {author} {\bibinfo {author} {\bibfnamefont {J.}~\bibnamefont
  {{Preskill}}}, \bibinfo {author} {\bibfnamefont {M.~B.}\ \bibnamefont
  {{Wise}}},\ and\ \bibinfo {author} {\bibfnamefont {F.}~\bibnamefont
  {{Wilczek}}},\ }\href {https://doi.org/10.1016/0370-2693(83)90637-8}
  {\bibfield  {journal} {\bibinfo  {journal} {Physics Letters B}\ }\textbf
  {\bibinfo {volume} {120}},\ \bibinfo {pages} {127} (\bibinfo {year}
  {1983})}\BibitemShut {NoStop}%
\bibitem [{\citenamefont {{Abbott}}\ and\ \citenamefont
  {{Sikivie}}(1983)}]{1983PhLB..120..133A}%
  \BibitemOpen
  \bibfield  {author} {\bibinfo {author} {\bibfnamefont {L.~F.}\ \bibnamefont
  {{Abbott}}}\ and\ \bibinfo {author} {\bibfnamefont {P.}~\bibnamefont
  {{Sikivie}}},\ }\href {https://doi.org/10.1016/0370-2693(83)90638-X}
  {\bibfield  {journal} {\bibinfo  {journal} {Physics Letters B}\ }\textbf
  {\bibinfo {volume} {120}},\ \bibinfo {pages} {133} (\bibinfo {year}
  {1983})}\BibitemShut {NoStop}%
\bibitem [{\citenamefont {{Dine}}\ and\ \citenamefont
  {{Fischler}}(1983)}]{1983PhLB..120..137D}%
  \BibitemOpen
  \bibfield  {author} {\bibinfo {author} {\bibfnamefont {M.}~\bibnamefont
  {{Dine}}}\ and\ \bibinfo {author} {\bibfnamefont {W.}~\bibnamefont
  {{Fischler}}},\ }\href {https://doi.org/10.1016/0370-2693(83)90639-1}
  {\bibfield  {journal} {\bibinfo  {journal} {Physics Letters B}\ }\textbf
  {\bibinfo {volume} {120}},\ \bibinfo {pages} {137} (\bibinfo {year}
  {1983})}\BibitemShut {NoStop}%
\bibitem [{\citenamefont {{Peccei}}\ and\ \citenamefont
  {{Quinn}}(1977{\natexlab{a}})}]{1977PhRvL..38.1440P}%
  \BibitemOpen
  \bibfield  {author} {\bibinfo {author} {\bibfnamefont {R.~D.}\ \bibnamefont
  {{Peccei}}}\ and\ \bibinfo {author} {\bibfnamefont {H.~R.}\ \bibnamefont
  {{Quinn}}},\ }\href {https://doi.org/10.1103/PhysRevLett.38.1440} {\bibfield
  {journal} {\bibinfo  {journal} {\prl}\ }\textbf {\bibinfo {volume} {38}},\
  \bibinfo {pages} {1440} (\bibinfo {year} {1977}{\natexlab{a}})}\BibitemShut
  {NoStop}%
\bibitem [{\citenamefont {{Peccei}}\ and\ \citenamefont
  {{Quinn}}(1977{\natexlab{b}})}]{1977PhRvD..16.1791P}%
  \BibitemOpen
  \bibfield  {author} {\bibinfo {author} {\bibfnamefont {R.~D.}\ \bibnamefont
  {{Peccei}}}\ and\ \bibinfo {author} {\bibfnamefont {H.~R.}\ \bibnamefont
  {{Quinn}}},\ }\href {https://doi.org/10.1103/PhysRevD.16.1791} {\bibfield
  {journal} {\bibinfo  {journal} {\prd}\ }\textbf {\bibinfo {volume} {16}},\
  \bibinfo {pages} {1791} (\bibinfo {year} {1977}{\natexlab{b}})}\BibitemShut
  {NoStop}%
\bibitem [{\citenamefont {{Wilczek}}(1978)}]{1978PhRvL..40..279W}%
  \BibitemOpen
  \bibfield  {author} {\bibinfo {author} {\bibfnamefont {F.}~\bibnamefont
  {{Wilczek}}},\ }\href {https://doi.org/10.1103/PhysRevLett.40.279} {\bibfield
   {journal} {\bibinfo  {journal} {\prl}\ }\textbf {\bibinfo {volume} {40}},\
  \bibinfo {pages} {279} (\bibinfo {year} {1978})}\BibitemShut {NoStop}%
\bibitem [{\citenamefont {{Weinberg}}(1978)}]{1978PhRvL..40..223W}%
  \BibitemOpen
  \bibfield  {author} {\bibinfo {author} {\bibfnamefont {S.}~\bibnamefont
  {{Weinberg}}},\ }\href {https://doi.org/10.1103/PhysRevLett.40.223}
  {\bibfield  {journal} {\bibinfo  {journal} {\prl}\ }\textbf {\bibinfo
  {volume} {40}},\ \bibinfo {pages} {223} (\bibinfo {year} {1978})}\BibitemShut
  {NoStop}%
\bibitem [{\citenamefont {{Svrcek}}\ and\ \citenamefont
  {{Witten}}(2006)}]{2006JHEP...06..051S}%
  \BibitemOpen
  \bibfield  {author} {\bibinfo {author} {\bibfnamefont {P.}~\bibnamefont
  {{Svrcek}}}\ and\ \bibinfo {author} {\bibfnamefont {E.}~\bibnamefont
  {{Witten}}},\ }\href {https://doi.org/10.1088/1126-6708/2006/06/051}
  {\bibfield  {journal} {\bibinfo  {journal} {Journal of High Energy Physics}\
  }\textbf {\bibinfo {volume} {2006}},\ \bibinfo {eid} {051} (\bibinfo {year}
  {2006})},\ \Eprint {https://arxiv.org/abs/hep-th/0605206}
  {arXiv:hep-th/0605206 [hep-th]} \BibitemShut {NoStop}%
\bibitem [{\citenamefont {{Arvanitaki}}\ \emph {et~al.}(2010)\citenamefont
  {{Arvanitaki}}, \citenamefont {{Dimopoulos}}, \citenamefont {{Dubovsky}},
  \citenamefont {{Kaloper}},\ and\ \citenamefont
  {{March-Russell}}}]{2010PhRvD..81l3530A}%
  \BibitemOpen
  \bibfield  {author} {\bibinfo {author} {\bibfnamefont {A.}~\bibnamefont
  {{Arvanitaki}}}, \bibinfo {author} {\bibfnamefont {S.}~\bibnamefont
  {{Dimopoulos}}}, \bibinfo {author} {\bibfnamefont {S.}~\bibnamefont
  {{Dubovsky}}}, \bibinfo {author} {\bibfnamefont {N.}~\bibnamefont
  {{Kaloper}}},\ and\ \bibinfo {author} {\bibfnamefont {J.}~\bibnamefont
  {{March-Russell}}},\ }\href {https://doi.org/10.1103/PhysRevD.81.123530}
  {\bibfield  {journal} {\bibinfo  {journal} {\prd}\ }\textbf {\bibinfo
  {volume} {81}},\ \bibinfo {eid} {123530} (\bibinfo {year} {2010})},\ \Eprint
  {https://arxiv.org/abs/0905.4720} {arXiv:0905.4720 [hep-th]} \BibitemShut
  {NoStop}%
\bibitem [{\citenamefont {{Arias}}\ \emph {et~al.}(2012)\citenamefont
  {{Arias}}, \citenamefont {{Cadamuro}}, \citenamefont {{Goodsell}},
  \citenamefont {{Jaeckel}}, \citenamefont {{Redondo}},\ and\ \citenamefont
  {{Ringwald}}}]{2012JCAP...06..013A}%
  \BibitemOpen
  \bibfield  {author} {\bibinfo {author} {\bibfnamefont {P.}~\bibnamefont
  {{Arias}}}, \bibinfo {author} {\bibfnamefont {D.}~\bibnamefont {{Cadamuro}}},
  \bibinfo {author} {\bibfnamefont {M.}~\bibnamefont {{Goodsell}}}, \bibinfo
  {author} {\bibfnamefont {J.}~\bibnamefont {{Jaeckel}}}, \bibinfo {author}
  {\bibfnamefont {J.}~\bibnamefont {{Redondo}}},\ and\ \bibinfo {author}
  {\bibfnamefont {A.}~\bibnamefont {{Ringwald}}},\ }\href
  {https://doi.org/10.1088/1475-7516/2012/06/013} {\bibfield  {journal}
  {\bibinfo  {journal} {\jcap}\ }\textbf {\bibinfo {volume} {2012}},\ \bibinfo
  {eid} {013} (\bibinfo {year} {2012})},\ \Eprint
  {https://arxiv.org/abs/1201.5902} {arXiv:1201.5902 [hep-ph]} \BibitemShut
  {NoStop}%
\bibitem [{\citenamefont {{Cicoli}}\ \emph {et~al.}(2012)\citenamefont
  {{Cicoli}}, \citenamefont {{Goodsell}},\ and\ \citenamefont
  {{Ringwald}}}]{2012JHEP...10..146C}%
  \BibitemOpen
  \bibfield  {author} {\bibinfo {author} {\bibfnamefont {M.}~\bibnamefont
  {{Cicoli}}}, \bibinfo {author} {\bibfnamefont {M.~D.}\ \bibnamefont
  {{Goodsell}}},\ and\ \bibinfo {author} {\bibfnamefont {A.}~\bibnamefont
  {{Ringwald}}},\ }\href {https://doi.org/10.1007/JHEP10(2012)146} {\bibfield
  {journal} {\bibinfo  {journal} {Journal of High Energy Physics}\ }\textbf
  {\bibinfo {volume} {2012}},\ \bibinfo {eid} {146} (\bibinfo {year} {2012})},\
  \Eprint {https://arxiv.org/abs/1206.0819} {arXiv:1206.0819 [hep-th]}
  \BibitemShut {NoStop}%
\bibitem [{\citenamefont {{Blinov}}\ \emph {et~al.}(2019)\citenamefont
  {{Blinov}}, \citenamefont {{Dolan}}, \citenamefont {{Draper}},\ and\
  \citenamefont {{Kozaczuk}}}]{2019PhRvD.100a5049B}%
  \BibitemOpen
  \bibfield  {author} {\bibinfo {author} {\bibfnamefont {N.}~\bibnamefont
  {{Blinov}}}, \bibinfo {author} {\bibfnamefont {M.~J.}\ \bibnamefont
  {{Dolan}}}, \bibinfo {author} {\bibfnamefont {P.}~\bibnamefont {{Draper}}},\
  and\ \bibinfo {author} {\bibfnamefont {J.}~\bibnamefont {{Kozaczuk}}},\
  }\href {https://doi.org/10.1103/PhysRevD.100.015049} {\bibfield  {journal}
  {\bibinfo  {journal} {\prd}\ }\textbf {\bibinfo {volume} {100}},\ \bibinfo
  {eid} {015049} (\bibinfo {year} {2019})},\ \Eprint
  {https://arxiv.org/abs/1905.06952} {arXiv:1905.06952 [hep-ph]} \BibitemShut
  {NoStop}%
\bibitem [{\citenamefont {Bartelmann}\ and\ \citenamefont
  {Schneider}(2001)}]{Bartelmann2001}%
  \BibitemOpen
  \bibfield  {author} {\bibinfo {author} {\bibfnamefont {M.}~\bibnamefont
  {Bartelmann}}\ and\ \bibinfo {author} {\bibfnamefont {P.}~\bibnamefont
  {Schneider}},\ }\href {https://doi.org/10.1016/S0370-1573(00)00082-X}
  {\bibfield  {journal} {\bibinfo  {journal} {Physics Reports}\ }\textbf
  {\bibinfo {volume} {340}},\ \bibinfo {pages} {291} (\bibinfo {year}
  {2001})}\BibitemShut {NoStop}%
\bibitem [{\citenamefont {{Creque-Sarbinowski}}\ and\ \citenamefont
  {{Kamionkowski}}(2018)}]{2018PhRvD..98f3524C}%
  \BibitemOpen
  \bibfield  {author} {\bibinfo {author} {\bibfnamefont {C.}~\bibnamefont
  {{Creque-Sarbinowski}}}\ and\ \bibinfo {author} {\bibfnamefont
  {M.}~\bibnamefont {{Kamionkowski}}},\ }\href
  {https://doi.org/10.1103/PhysRevD.98.063524} {\bibfield  {journal} {\bibinfo
  {journal} {\prd}\ }\textbf {\bibinfo {volume} {98}},\ \bibinfo {eid} {063524}
  (\bibinfo {year} {2018})},\ \Eprint {https://arxiv.org/abs/1806.11119}
  {arXiv:1806.11119 [astro-ph.CO]} \BibitemShut {NoStop}%
\bibitem [{\citenamefont {{Bernal}}\ \emph {et~al.}(2020)\citenamefont
  {{Bernal}}, \citenamefont {{Caputo}},\ and\ \citenamefont
  {{Kamionkowski}}}]{2020arXiv201200771B}%
  \BibitemOpen
  \bibfield  {author} {\bibinfo {author} {\bibfnamefont {J.~L.}\ \bibnamefont
  {{Bernal}}}, \bibinfo {author} {\bibfnamefont {A.}~\bibnamefont {{Caputo}}},\
  and\ \bibinfo {author} {\bibfnamefont {M.}~\bibnamefont {{Kamionkowski}}},\
  }\href@noop {} {\bibfield  {journal} {\bibinfo  {journal} {arXiv e-prints}\
  ,\ \bibinfo {eid} {arXiv:2012.00771}} (\bibinfo {year} {2020})},\ \Eprint
  {https://arxiv.org/abs/2012.00771} {arXiv:2012.00771 [astro-ph.CO]}
  \BibitemShut {NoStop}%
\bibitem [{\citenamefont {{Kennicutt}}(1998)}]{1998ARA&A..36..189K}%
  \BibitemOpen
  \bibfield  {author} {\bibinfo {author} {\bibfnamefont {J.}~\bibnamefont
  {{Kennicutt}}, \bibfnamefont {Robert~C.}},\ }\href
  {https://doi.org/10.1146/annurev.astro.36.1.189} {\bibfield  {journal}
  {\bibinfo  {journal} {\araa}\ }\textbf {\bibinfo {volume} {36}},\ \bibinfo
  {pages} {189} (\bibinfo {year} {1998})},\ \Eprint
  {https://arxiv.org/abs/astro-ph/9807187} {arXiv:astro-ph/9807187 [astro-ph]}
  \BibitemShut {NoStop}%
\bibitem [{\citenamefont {{Sobral}}\ \emph {et~al.}(2013)\citenamefont
  {{Sobral}}, \citenamefont {{Smail}}, \citenamefont {{Best}}, \citenamefont
  {{Geach}}, \citenamefont {{Matsuda}}, \citenamefont {{Stott}}, \citenamefont
  {{Cirasuolo}},\ and\ \citenamefont {{Kurk}}}]{2013MNRAS.428.1128S}%
  \BibitemOpen
  \bibfield  {author} {\bibinfo {author} {\bibfnamefont {D.}~\bibnamefont
  {{Sobral}}}, \bibinfo {author} {\bibfnamefont {I.}~\bibnamefont {{Smail}}},
  \bibinfo {author} {\bibfnamefont {P.~N.}\ \bibnamefont {{Best}}}, \bibinfo
  {author} {\bibfnamefont {J.~E.}\ \bibnamefont {{Geach}}}, \bibinfo {author}
  {\bibfnamefont {Y.}~\bibnamefont {{Matsuda}}}, \bibinfo {author}
  {\bibfnamefont {J.~P.}\ \bibnamefont {{Stott}}}, \bibinfo {author}
  {\bibfnamefont {M.}~\bibnamefont {{Cirasuolo}}},\ and\ \bibinfo {author}
  {\bibfnamefont {J.}~\bibnamefont {{Kurk}}},\ }\href
  {https://doi.org/10.1093/mnras/sts096} {\bibfield  {journal} {\bibinfo
  {journal} {\mnras}\ }\textbf {\bibinfo {volume} {428}},\ \bibinfo {pages}
  {1128} (\bibinfo {year} {2013})},\ \Eprint {https://arxiv.org/abs/1202.3436}
  {arXiv:1202.3436 [astro-ph.CO]} \BibitemShut {NoStop}%
\bibitem [{\citenamefont {{Silva}}\ \emph {et~al.}(2017)\citenamefont
  {{Silva}}, \citenamefont {{Zaroubi}}, \citenamefont {{Kooistra}},\ and\
  \citenamefont {{Cooray}}}]{2017arXiv171109902S}%
  \BibitemOpen
  \bibfield  {author} {\bibinfo {author} {\bibfnamefont {M.~B.}\ \bibnamefont
  {{Silva}}}, \bibinfo {author} {\bibfnamefont {S.}~\bibnamefont {{Zaroubi}}},
  \bibinfo {author} {\bibfnamefont {R.}~\bibnamefont {{Kooistra}}},\ and\
  \bibinfo {author} {\bibfnamefont {A.}~\bibnamefont {{Cooray}}},\ }\href@noop
  {} {\bibfield  {journal} {\bibinfo  {journal} {arXiv e-prints}\ ,\ \bibinfo
  {eid} {arXiv:1711.09902}} (\bibinfo {year} {2017})},\ \Eprint
  {https://arxiv.org/abs/1711.09902} {arXiv:1711.09902 [astro-ph.GA]}
  \BibitemShut {NoStop}%
\bibitem [{\citenamefont {{Anders}}\ and\ \citenamefont {{Fritze-v.
  Alvensleben}}(2003)}]{2003A&A...401.1063A}%
  \BibitemOpen
  \bibfield  {author} {\bibinfo {author} {\bibfnamefont {P.}~\bibnamefont
  {{Anders}}}\ and\ \bibinfo {author} {\bibfnamefont {U.}~\bibnamefont
  {{Fritze-v. Alvensleben}}},\ }\href
  {https://doi.org/10.1051/0004-6361:20030151} {\bibfield  {journal} {\bibinfo
  {journal} {\aap}\ }\textbf {\bibinfo {volume} {401}},\ \bibinfo {pages}
  {1063} (\bibinfo {year} {2003})},\ \Eprint
  {https://arxiv.org/abs/astro-ph/0302146} {arXiv:astro-ph/0302146 [astro-ph]}
  \BibitemShut {NoStop}%
\bibitem [{\citenamefont {{M{\'a}rmol-Queralt{\'o}}}\ \emph
  {et~al.}(2016)\citenamefont {{M{\'a}rmol-Queralt{\'o}}}, \citenamefont
  {{McLure}}, \citenamefont {{Cullen}}, \citenamefont {{Dunlop}}, \citenamefont
  {{Fontana}},\ and\ \citenamefont {{McLeod}}}]{2016MNRAS.460.3587M}%
  \BibitemOpen
  \bibfield  {author} {\bibinfo {author} {\bibfnamefont {E.}~\bibnamefont
  {{M{\'a}rmol-Queralt{\'o}}}}, \bibinfo {author} {\bibfnamefont {R.~J.}\
  \bibnamefont {{McLure}}}, \bibinfo {author} {\bibfnamefont {F.}~\bibnamefont
  {{Cullen}}}, \bibinfo {author} {\bibfnamefont {J.~S.}\ \bibnamefont
  {{Dunlop}}}, \bibinfo {author} {\bibfnamefont {A.}~\bibnamefont
  {{Fontana}}},\ and\ \bibinfo {author} {\bibfnamefont {D.~J.}\ \bibnamefont
  {{McLeod}}},\ }\href {https://doi.org/10.1093/mnras/stw1212} {\bibfield
  {journal} {\bibinfo  {journal} {\mnras}\ }\textbf {\bibinfo {volume} {460}},\
  \bibinfo {pages} {3587} (\bibinfo {year} {2016})},\ \Eprint
  {https://arxiv.org/abs/1511.01911} {arXiv:1511.01911 [astro-ph.GA]}
  \BibitemShut {NoStop}%
\bibitem [{\citenamefont {{Ly}}\ \emph {et~al.}(2007)\citenamefont {{Ly}},
  \citenamefont {{Malkan}}, \citenamefont {{Kashikawa}}, \citenamefont
  {{Shimasaku}}, \citenamefont {{Doi}}, \citenamefont {{Nagao}}, \citenamefont
  {{Iye}}, \citenamefont {{Kodama}}, \citenamefont {{Morokuma}},\ and\
  \citenamefont {{Motohara}}}]{2007ApJ...657..738L}%
  \BibitemOpen
  \bibfield  {author} {\bibinfo {author} {\bibfnamefont {C.}~\bibnamefont
  {{Ly}}}, \bibinfo {author} {\bibfnamefont {M.~A.}\ \bibnamefont {{Malkan}}},
  \bibinfo {author} {\bibfnamefont {N.}~\bibnamefont {{Kashikawa}}}, \bibinfo
  {author} {\bibfnamefont {K.}~\bibnamefont {{Shimasaku}}}, \bibinfo {author}
  {\bibfnamefont {M.}~\bibnamefont {{Doi}}}, \bibinfo {author} {\bibfnamefont
  {T.}~\bibnamefont {{Nagao}}}, \bibinfo {author} {\bibfnamefont
  {M.}~\bibnamefont {{Iye}}}, \bibinfo {author} {\bibfnamefont
  {T.}~\bibnamefont {{Kodama}}}, \bibinfo {author} {\bibfnamefont
  {T.}~\bibnamefont {{Morokuma}}},\ and\ \bibinfo {author} {\bibfnamefont
  {K.}~\bibnamefont {{Motohara}}},\ }\href {https://doi.org/10.1086/510828}
  {\bibfield  {journal} {\bibinfo  {journal} {\apj}\ }\textbf {\bibinfo
  {volume} {657}},\ \bibinfo {pages} {738} (\bibinfo {year} {2007})},\ \Eprint
  {https://arxiv.org/abs/astro-ph/0610846} {arXiv:astro-ph/0610846 [astro-ph]}
  \BibitemShut {NoStop}%
\bibitem [{\citenamefont {{Khostovan}}\ \emph {et~al.}(2015)\citenamefont
  {{Khostovan}}, \citenamefont {{Sobral}}, \citenamefont {{Mobasher}},
  \citenamefont {{Best}}, \citenamefont {{Smail}}, \citenamefont {{Stott}},
  \citenamefont {{Hemmati}},\ and\ \citenamefont
  {{Nayyeri}}}]{2015MNRAS.452.3948K}%
  \BibitemOpen
  \bibfield  {author} {\bibinfo {author} {\bibfnamefont {A.~A.}\ \bibnamefont
  {{Khostovan}}}, \bibinfo {author} {\bibfnamefont {D.}~\bibnamefont
  {{Sobral}}}, \bibinfo {author} {\bibfnamefont {B.}~\bibnamefont
  {{Mobasher}}}, \bibinfo {author} {\bibfnamefont {P.~N.}\ \bibnamefont
  {{Best}}}, \bibinfo {author} {\bibfnamefont {I.}~\bibnamefont {{Smail}}},
  \bibinfo {author} {\bibfnamefont {J.~P.}\ \bibnamefont {{Stott}}}, \bibinfo
  {author} {\bibfnamefont {S.}~\bibnamefont {{Hemmati}}},\ and\ \bibinfo
  {author} {\bibfnamefont {H.}~\bibnamefont {{Nayyeri}}},\ }\href
  {https://doi.org/10.1093/mnras/stv1474} {\bibfield  {journal} {\bibinfo
  {journal} {\mnras}\ }\textbf {\bibinfo {volume} {452}},\ \bibinfo {pages}
  {3948} (\bibinfo {year} {2015})},\ \Eprint {https://arxiv.org/abs/1503.00004}
  {arXiv:1503.00004 [astro-ph.GA]} \BibitemShut {NoStop}%
\bibitem [{\citenamefont {{Calzetti}}\ \emph {et~al.}(2000)\citenamefont
  {{Calzetti}}, \citenamefont {{Armus}}, \citenamefont {{Bohlin}},
  \citenamefont {{Kinney}}, \citenamefont {{Koornneef}},\ and\ \citenamefont
  {{Storchi-Bergmann}}}]{2000ApJ...533..682C}%
  \BibitemOpen
  \bibfield  {author} {\bibinfo {author} {\bibfnamefont {D.}~\bibnamefont
  {{Calzetti}}}, \bibinfo {author} {\bibfnamefont {L.}~\bibnamefont {{Armus}}},
  \bibinfo {author} {\bibfnamefont {R.~C.}\ \bibnamefont {{Bohlin}}}, \bibinfo
  {author} {\bibfnamefont {A.~L.}\ \bibnamefont {{Kinney}}}, \bibinfo {author}
  {\bibfnamefont {J.}~\bibnamefont {{Koornneef}}},\ and\ \bibinfo {author}
  {\bibfnamefont {T.}~\bibnamefont {{Storchi-Bergmann}}},\ }\href
  {https://doi.org/10.1086/308692} {\bibfield  {journal} {\bibinfo  {journal}
  {\apj}\ }\textbf {\bibinfo {volume} {533}},\ \bibinfo {pages} {682} (\bibinfo
  {year} {2000})},\ \Eprint {https://arxiv.org/abs/astro-ph/9911459}
  {arXiv:astro-ph/9911459 [astro-ph]} \BibitemShut {NoStop}%
\bibitem [{\citenamefont {{Scoville}}\ \emph {et~al.}(2015)\citenamefont
  {{Scoville}}, \citenamefont {{Faisst}}, \citenamefont {{Capak}},
  \citenamefont {{Kakazu}}, \citenamefont {{Li}},\ and\ \citenamefont
  {{Steinhardt}}}]{2015ApJ...800..108S}%
  \BibitemOpen
  \bibfield  {author} {\bibinfo {author} {\bibfnamefont {N.}~\bibnamefont
  {{Scoville}}}, \bibinfo {author} {\bibfnamefont {A.}~\bibnamefont
  {{Faisst}}}, \bibinfo {author} {\bibfnamefont {P.}~\bibnamefont {{Capak}}},
  \bibinfo {author} {\bibfnamefont {Y.}~\bibnamefont {{Kakazu}}}, \bibinfo
  {author} {\bibfnamefont {G.}~\bibnamefont {{Li}}},\ and\ \bibinfo {author}
  {\bibfnamefont {C.}~\bibnamefont {{Steinhardt}}},\ }\href
  {https://doi.org/10.1088/0004-637X/800/2/108} {\bibfield  {journal} {\bibinfo
   {journal} {\apj}\ }\textbf {\bibinfo {volume} {800}},\ \bibinfo {eid} {108}
  (\bibinfo {year} {2015})},\ \Eprint {https://arxiv.org/abs/1412.8219}
  {arXiv:1412.8219 [astro-ph.GA]} \BibitemShut {NoStop}%
\bibitem [{\citenamefont {{Hayashi}}\ \emph {et~al.}(2013)\citenamefont
  {{Hayashi}}, \citenamefont {{Sobral}}, \citenamefont {{Best}}, \citenamefont
  {{Smail}},\ and\ \citenamefont {{Kodama}}}]{2013MNRAS.430.1042H}%
  \BibitemOpen
  \bibfield  {author} {\bibinfo {author} {\bibfnamefont {M.}~\bibnamefont
  {{Hayashi}}}, \bibinfo {author} {\bibfnamefont {D.}~\bibnamefont {{Sobral}}},
  \bibinfo {author} {\bibfnamefont {P.~N.}\ \bibnamefont {{Best}}}, \bibinfo
  {author} {\bibfnamefont {I.}~\bibnamefont {{Smail}}},\ and\ \bibinfo {author}
  {\bibfnamefont {T.}~\bibnamefont {{Kodama}}},\ }\href
  {https://doi.org/10.1093/mnras/sts676} {\bibfield  {journal} {\bibinfo
  {journal} {\mnras}\ }\textbf {\bibinfo {volume} {430}},\ \bibinfo {pages}
  {1042} (\bibinfo {year} {2013})},\ \Eprint {https://arxiv.org/abs/1212.4905}
  {arXiv:1212.4905 [astro-ph.CO]} \BibitemShut {NoStop}%
\bibitem [{\citenamefont {{Chang}}\ \emph {et~al.}(2013)\citenamefont
  {{Chang}}, \citenamefont {{Jarvis}}, \citenamefont {{Jain}}, \citenamefont
  {{Kahn}}, \citenamefont {{Kirkby}}, \citenamefont {{Connolly}}, \citenamefont
  {{Krughoff}}, \citenamefont {{Peng}},\ and\ \citenamefont
  {{Peterson}}}]{2013MNRAS.434.2121C}%
  \BibitemOpen
  \bibfield  {author} {\bibinfo {author} {\bibfnamefont {C.}~\bibnamefont
  {{Chang}}}, \bibinfo {author} {\bibfnamefont {M.}~\bibnamefont {{Jarvis}}},
  \bibinfo {author} {\bibfnamefont {B.}~\bibnamefont {{Jain}}}, \bibinfo
  {author} {\bibfnamefont {S.~M.}\ \bibnamefont {{Kahn}}}, \bibinfo {author}
  {\bibfnamefont {D.}~\bibnamefont {{Kirkby}}}, \bibinfo {author}
  {\bibfnamefont {A.}~\bibnamefont {{Connolly}}}, \bibinfo {author}
  {\bibfnamefont {S.}~\bibnamefont {{Krughoff}}}, \bibinfo {author}
  {\bibfnamefont {E.~H.}\ \bibnamefont {{Peng}}},\ and\ \bibinfo {author}
  {\bibfnamefont {J.~R.}\ \bibnamefont {{Peterson}}},\ }\href
  {https://doi.org/10.1093/mnras/stt1156} {\bibfield  {journal} {\bibinfo
  {journal} {\mnras}\ }\textbf {\bibinfo {volume} {434}},\ \bibinfo {pages}
  {2121} (\bibinfo {year} {2013})},\ \Eprint {https://arxiv.org/abs/1305.0793}
  {arXiv:1305.0793 [astro-ph.CO]} \BibitemShut {NoStop}%
\bibitem [{\citenamefont {{West}}\ \emph {et~al.}(1991)\citenamefont {{West}},
  \citenamefont {{Villumsen}},\ and\ \citenamefont
  {{Dekel}}}]{1991ApJ...369..287W}%
  \BibitemOpen
  \bibfield  {author} {\bibinfo {author} {\bibfnamefont {M.~J.}\ \bibnamefont
  {{West}}}, \bibinfo {author} {\bibfnamefont {J.~V.}\ \bibnamefont
  {{Villumsen}}},\ and\ \bibinfo {author} {\bibfnamefont {A.}~\bibnamefont
  {{Dekel}}},\ }\href {https://doi.org/10.1086/169760} {\bibfield  {journal}
  {\bibinfo  {journal} {\apj}\ }\textbf {\bibinfo {volume} {369}},\ \bibinfo
  {pages} {287} (\bibinfo {year} {1991})}\BibitemShut {NoStop}%
\bibitem [{\citenamefont {{Tormen}}(1997)}]{1997MNRAS.290..411T}%
  \BibitemOpen
  \bibfield  {author} {\bibinfo {author} {\bibfnamefont {G.}~\bibnamefont
  {{Tormen}}},\ }\href {https://doi.org/10.1093/mnras/290.3.411} {\bibfield
  {journal} {\bibinfo  {journal} {\mnras}\ }\textbf {\bibinfo {volume} {290}},\
  \bibinfo {pages} {411} (\bibinfo {year} {1997})},\ \Eprint
  {https://arxiv.org/abs/astro-ph/9611078} {arXiv:astro-ph/9611078 [astro-ph]}
  \BibitemShut {NoStop}%
\bibitem [{\citenamefont {{Troxel}}\ and\ \citenamefont
  {{Ishak}}(2015)}]{2015PhR...558....1T}%
  \BibitemOpen
  \bibfield  {author} {\bibinfo {author} {\bibfnamefont {M.~A.}\ \bibnamefont
  {{Troxel}}}\ and\ \bibinfo {author} {\bibfnamefont {M.}~\bibnamefont
  {{Ishak}}},\ }\href {https://doi.org/10.1016/j.physrep.2014.11.001}
  {\bibfield  {journal} {\bibinfo  {journal} {\physrep}\ }\textbf {\bibinfo
  {volume} {558}},\ \bibinfo {pages} {1} (\bibinfo {year} {2015})},\ \Eprint
  {https://arxiv.org/abs/1407.6990} {arXiv:1407.6990 [astro-ph.CO]}
  \BibitemShut {NoStop}%
\bibitem [{\citenamefont {{Catelan}}\ \emph {et~al.}(2001)\citenamefont
  {{Catelan}}, \citenamefont {{Kamionkowski}},\ and\ \citenamefont
  {{Blandford}}}]{2001MNRAS.320L...7C}%
  \BibitemOpen
  \bibfield  {author} {\bibinfo {author} {\bibfnamefont {P.}~\bibnamefont
  {{Catelan}}}, \bibinfo {author} {\bibfnamefont {M.}~\bibnamefont
  {{Kamionkowski}}},\ and\ \bibinfo {author} {\bibfnamefont {R.~D.}\
  \bibnamefont {{Blandford}}},\ }\href
  {https://doi.org/10.1046/j.1365-8711.2001.04105.x} {\bibfield  {journal}
  {\bibinfo  {journal} {\mnras}\ }\textbf {\bibinfo {volume} {320}},\ \bibinfo
  {pages} {L7} (\bibinfo {year} {2001})},\ \Eprint
  {https://arxiv.org/abs/astro-ph/0005470} {arXiv:astro-ph/0005470 [astro-ph]}
  \BibitemShut {NoStop}%
\bibitem [{\citenamefont {{Hirata}}\ and\ \citenamefont
  {{Seljak}}(2004)}]{2004PhRvD..70f3526H}%
  \BibitemOpen
  \bibfield  {author} {\bibinfo {author} {\bibfnamefont {C.~M.}\ \bibnamefont
  {{Hirata}}}\ and\ \bibinfo {author} {\bibfnamefont {U.}~\bibnamefont
  {{Seljak}}},\ }\href {https://doi.org/10.1103/PhysRevD.70.063526} {\bibfield
  {journal} {\bibinfo  {journal} {\prd}\ }\textbf {\bibinfo {volume} {70}},\
  \bibinfo {eid} {063526} (\bibinfo {year} {2004})},\ \Eprint
  {https://arxiv.org/abs/astro-ph/0406275} {arXiv:astro-ph/0406275 [astro-ph]}
  \BibitemShut {NoStop}%
\bibitem [{\citenamefont {{Fluri}}\ \emph {et~al.}(2019)\citenamefont
  {{Fluri}}, \citenamefont {{Kacprzak}}, \citenamefont {{Lucchi}},
  \citenamefont {{Refregier}}, \citenamefont {{Amara}}, \citenamefont
  {{Hofmann}},\ and\ \citenamefont {{Schneider}}}]{2019PhRvD.100f3514F}%
  \BibitemOpen
  \bibfield  {author} {\bibinfo {author} {\bibfnamefont {J.}~\bibnamefont
  {{Fluri}}}, \bibinfo {author} {\bibfnamefont {T.}~\bibnamefont {{Kacprzak}}},
  \bibinfo {author} {\bibfnamefont {A.}~\bibnamefont {{Lucchi}}}, \bibinfo
  {author} {\bibfnamefont {A.}~\bibnamefont {{Refregier}}}, \bibinfo {author}
  {\bibfnamefont {A.}~\bibnamefont {{Amara}}}, \bibinfo {author} {\bibfnamefont
  {T.}~\bibnamefont {{Hofmann}}},\ and\ \bibinfo {author} {\bibfnamefont
  {A.}~\bibnamefont {{Schneider}}},\ }\href
  {https://doi.org/10.1103/PhysRevD.100.063514} {\bibfield  {journal} {\bibinfo
   {journal} {\prd}\ }\textbf {\bibinfo {volume} {100}},\ \bibinfo {eid}
  {063514} (\bibinfo {year} {2019})},\ \Eprint
  {https://arxiv.org/abs/1906.03156} {arXiv:1906.03156 [astro-ph.CO]}
  \BibitemShut {NoStop}%
\bibitem [{\citenamefont {{Brown}}\ \emph {et~al.}(2002)\citenamefont
  {{Brown}}, \citenamefont {{Taylor}}, \citenamefont {{Hambly}},\ and\
  \citenamefont {{Dye}}}]{2002MNRAS.333..501B}%
  \BibitemOpen
  \bibfield  {author} {\bibinfo {author} {\bibfnamefont {M.~L.}\ \bibnamefont
  {{Brown}}}, \bibinfo {author} {\bibfnamefont {A.~N.}\ \bibnamefont
  {{Taylor}}}, \bibinfo {author} {\bibfnamefont {N.~C.}\ \bibnamefont
  {{Hambly}}},\ and\ \bibinfo {author} {\bibfnamefont {S.}~\bibnamefont
  {{Dye}}},\ }\href {https://doi.org/10.1046/j.1365-8711.2002.05354.x}
  {\bibfield  {journal} {\bibinfo  {journal} {\mnras}\ }\textbf {\bibinfo
  {volume} {333}},\ \bibinfo {pages} {501} (\bibinfo {year} {2002})},\ \Eprint
  {https://arxiv.org/abs/astro-ph/0009499} {arXiv:astro-ph/0009499 [astro-ph]}
  \BibitemShut {NoStop}%
\bibitem [{\citenamefont {{Hikage}}\ \emph {et~al.}(2019)\citenamefont
  {{Hikage}}, \citenamefont {{Oguri}}, \citenamefont {{Hamana}}, \citenamefont
  {{More}}, \citenamefont {{Mandelbaum}}, \citenamefont {{Takada}},
  \citenamefont {{K{\"o}hlinger}}, \citenamefont {{Miyatake}}, \citenamefont
  {{Nishizawa}}, \citenamefont {{Aihara}}, \citenamefont {{Armstrong}},
  \citenamefont {{Bosch}}, \citenamefont {{Coupon}}, \citenamefont {{Ducout}},
  \citenamefont {{Ho}}, \citenamefont {{Hsieh}}, \citenamefont {{Komiyama}},
  \citenamefont {{Lanusse}}, \citenamefont {{Leauthaud}}, \citenamefont
  {{Lupton}}, \citenamefont {{Medezinski}}, \citenamefont {{Mineo}},
  \citenamefont {{Miyama}}, \citenamefont {{Miyazaki}}, \citenamefont
  {{Murata}}, \citenamefont {{Murayama}}, \citenamefont {{Shirasaki}},
  \citenamefont {{Sif{\'o}n}}, \citenamefont {{Simet}}, \citenamefont
  {{Speagle}}, \citenamefont {{Spergel}}, \citenamefont {{Strauss}},
  \citenamefont {{Sugiyama}}, \citenamefont {{Tanaka}}, \citenamefont
  {{Utsumi}}, \citenamefont {{Wang}},\ and\ \citenamefont
  {{Yamada}}}]{2019PASJ...71...43H}%
  \BibitemOpen
  \bibfield  {author} {\bibinfo {author} {\bibfnamefont {C.}~\bibnamefont
  {{Hikage}}}, \bibinfo {author} {\bibfnamefont {M.}~\bibnamefont {{Oguri}}},
  \bibinfo {author} {\bibfnamefont {T.}~\bibnamefont {{Hamana}}}, \bibinfo
  {author} {\bibfnamefont {S.}~\bibnamefont {{More}}}, \bibinfo {author}
  {\bibfnamefont {R.}~\bibnamefont {{Mandelbaum}}}, \bibinfo {author}
  {\bibfnamefont {M.}~\bibnamefont {{Takada}}}, \bibinfo {author}
  {\bibfnamefont {F.}~\bibnamefont {{K{\"o}hlinger}}}, \bibinfo {author}
  {\bibfnamefont {H.}~\bibnamefont {{Miyatake}}}, \bibinfo {author}
  {\bibfnamefont {A.~J.}\ \bibnamefont {{Nishizawa}}}, \bibinfo {author}
  {\bibfnamefont {H.}~\bibnamefont {{Aihara}}}, \bibinfo {author}
  {\bibfnamefont {R.}~\bibnamefont {{Armstrong}}}, \bibinfo {author}
  {\bibfnamefont {J.}~\bibnamefont {{Bosch}}}, \bibinfo {author} {\bibfnamefont
  {J.}~\bibnamefont {{Coupon}}}, \bibinfo {author} {\bibfnamefont
  {A.}~\bibnamefont {{Ducout}}}, \bibinfo {author} {\bibfnamefont
  {P.}~\bibnamefont {{Ho}}}, \bibinfo {author} {\bibfnamefont {B.-C.}\
  \bibnamefont {{Hsieh}}}, \bibinfo {author} {\bibfnamefont {Y.}~\bibnamefont
  {{Komiyama}}}, \bibinfo {author} {\bibfnamefont {F.}~\bibnamefont
  {{Lanusse}}}, \bibinfo {author} {\bibfnamefont {A.}~\bibnamefont
  {{Leauthaud}}}, \bibinfo {author} {\bibfnamefont {R.~H.}\ \bibnamefont
  {{Lupton}}}, \bibinfo {author} {\bibfnamefont {E.}~\bibnamefont
  {{Medezinski}}}, \bibinfo {author} {\bibfnamefont {S.}~\bibnamefont
  {{Mineo}}}, \bibinfo {author} {\bibfnamefont {S.}~\bibnamefont {{Miyama}}},
  \bibinfo {author} {\bibfnamefont {S.}~\bibnamefont {{Miyazaki}}}, \bibinfo
  {author} {\bibfnamefont {R.}~\bibnamefont {{Murata}}}, \bibinfo {author}
  {\bibfnamefont {H.}~\bibnamefont {{Murayama}}}, \bibinfo {author}
  {\bibfnamefont {M.}~\bibnamefont {{Shirasaki}}}, \bibinfo {author}
  {\bibfnamefont {C.}~\bibnamefont {{Sif{\'o}n}}}, \bibinfo {author}
  {\bibfnamefont {M.}~\bibnamefont {{Simet}}}, \bibinfo {author} {\bibfnamefont
  {J.}~\bibnamefont {{Speagle}}}, \bibinfo {author} {\bibfnamefont {D.~N.}\
  \bibnamefont {{Spergel}}}, \bibinfo {author} {\bibfnamefont {M.~A.}\
  \bibnamefont {{Strauss}}}, \bibinfo {author} {\bibfnamefont {N.}~\bibnamefont
  {{Sugiyama}}}, \bibinfo {author} {\bibfnamefont {M.}~\bibnamefont
  {{Tanaka}}}, \bibinfo {author} {\bibfnamefont {Y.}~\bibnamefont {{Utsumi}}},
  \bibinfo {author} {\bibfnamefont {S.-Y.}\ \bibnamefont {{Wang}}},\ and\
  \bibinfo {author} {\bibfnamefont {Y.}~\bibnamefont {{Yamada}}},\ }\href
  {https://doi.org/10.1093/pasj/psz010} {\bibfield  {journal} {\bibinfo
  {journal} {\pasj}\ }\textbf {\bibinfo {volume} {71}},\ \bibinfo {eid} {43}
  (\bibinfo {year} {2019})},\ \Eprint {https://arxiv.org/abs/1809.09148}
  {arXiv:1809.09148 [astro-ph.CO]} \BibitemShut {NoStop}%
\bibitem [{\citenamefont {{Troxel}}\ \emph {et~al.}(2018)\citenamefont
  {{Troxel}}, \citenamefont {{MacCrann}}, \citenamefont {{Zuntz}},
  \citenamefont {{Eifler}}, \citenamefont {{Krause}}, \citenamefont
  {{Dodelson}}, \citenamefont {{Gruen}}, \citenamefont {{Blazek}},
  \citenamefont {{Friedrich}}, \citenamefont {{Samuroff}}, \citenamefont
  {{Prat}}, \citenamefont {{Secco}}, \citenamefont {{Davis}}, \citenamefont
  {{Fert{\'e}}}, \citenamefont {{DeRose}}, \citenamefont {{Alarcon}},
  \citenamefont {{Amara}}, \citenamefont {{Baxter}}, \citenamefont {{Becker}},
  \citenamefont {{Bernstein}}, \citenamefont {{Bridle}}, \citenamefont
  {{Cawthon}}, \citenamefont {{Chang}}, \citenamefont {{Choi}}, \citenamefont
  {{De Vicente}}, \citenamefont {{Drlica-Wagner}}, \citenamefont
  {{Elvin-Poole}}, \citenamefont {{Frieman}}, \citenamefont {{Gatti}},
  \citenamefont {{Hartley}}, \citenamefont {{Honscheid}}, \citenamefont
  {{Hoyle}}, \citenamefont {{Huff}}, \citenamefont {{Huterer}}, \citenamefont
  {{Jain}}, \citenamefont {{Jarvis}}, \citenamefont {{Kacprzak}}, \citenamefont
  {{Kirk}}, \citenamefont {{Kokron}}, \citenamefont {{Krawiec}}, \citenamefont
  {{Lahav}}, \citenamefont {{Liddle}}, \citenamefont {{Peacock}}, \citenamefont
  {{Rau}}, \citenamefont {{Refregier}}, \citenamefont {{Rollins}},
  \citenamefont {{Rozo}}, \citenamefont {{Rykoff}}, \citenamefont
  {{S{\'a}nchez}}, \citenamefont {{Sevilla-Noarbe}}, \citenamefont {{Sheldon}},
  \citenamefont {{Stebbins}}, \citenamefont {{Varga}}, \citenamefont
  {{Vielzeuf}}, \citenamefont {{Wang}}, \citenamefont {{Wechsler}},
  \citenamefont {{Yanny}}, \citenamefont {{Abbott}}, \citenamefont {{Abdalla}},
  \citenamefont {{Allam}}, \citenamefont {{Annis}}, \citenamefont {{Bechtol}},
  \citenamefont {{Benoit-L{\'e}vy}}, \citenamefont {{Bertin}}, \citenamefont
  {{Brooks}}, \citenamefont {{Buckley-Geer}}, \citenamefont {{Burke}},
  \citenamefont {{Carnero Rosell}}, \citenamefont {{Carrasco Kind}},
  \citenamefont {{Carretero}}, \citenamefont {{Castander}}, \citenamefont
  {{Crocce}}, \citenamefont {{Cunha}}, \citenamefont {{D'Andrea}},
  \citenamefont {{da Costa}}, \citenamefont {{DePoy}}, \citenamefont {{Desai}},
  \citenamefont {{Diehl}}, \citenamefont {{Dietrich}}, \citenamefont {{Doel}},
  \citenamefont {{Fernandez}}, \citenamefont {{Flaugher}}, \citenamefont
  {{Fosalba}}, \citenamefont {{Garc{\'\i}a-Bellido}}, \citenamefont
  {{Gaztanaga}}, \citenamefont {{Gerdes}}, \citenamefont {{Giannantonio}},
  \citenamefont {{Goldstein}}, \citenamefont {{Gruendl}}, \citenamefont
  {{Gschwend}}, \citenamefont {{Gutierrez}}, \citenamefont {{James}},
  \citenamefont {{Jeltema}}, \citenamefont {{Johnson}}, \citenamefont
  {{Johnson}}, \citenamefont {{Kent}}, \citenamefont {{Kuehn}}, \citenamefont
  {{Kuhlmann}}, \citenamefont {{Kuropatkin}}, \citenamefont {{Li}},
  \citenamefont {{Lima}}, \citenamefont {{Lin}}, \citenamefont {{Maia}},
  \citenamefont {{March}}, \citenamefont {{Marshall}}, \citenamefont
  {{Martini}}, \citenamefont {{Melchior}}, \citenamefont {{Menanteau}},
  \citenamefont {{Miquel}}, \citenamefont {{Mohr}}, \citenamefont {{Neilsen}},
  \citenamefont {{Nichol}}, \citenamefont {{Nord}}, \citenamefont
  {{Petravick}}, \citenamefont {{Plazas}}, \citenamefont {{Romer}},
  \citenamefont {{Roodman}}, \citenamefont {{Sako}}, \citenamefont {{Sanchez}},
  \citenamefont {{Scarpine}}, \citenamefont {{Schindler}}, \citenamefont
  {{Schubnell}}, \citenamefont {{Smith}}, \citenamefont {{Smith}},
  \citenamefont {{Soares-Santos}}, \citenamefont {{Sobreira}}, \citenamefont
  {{Suchyta}}, \citenamefont {{Swanson}}, \citenamefont {{Tarle}},
  \citenamefont {{Thomas}}, \citenamefont {{Tucker}}, \citenamefont {{Vikram}},
  \citenamefont {{Walker}}, \citenamefont {{Weller}}, \citenamefont {{Zhang}},\
  and\ \citenamefont {{DES Collaboration}}}]{2018PhRvD..98d3528T}%
  \BibitemOpen
  \bibfield  {author} {\bibinfo {author} {\bibfnamefont {M.~A.}\ \bibnamefont
  {{Troxel}}}, \bibinfo {author} {\bibfnamefont {N.}~\bibnamefont
  {{MacCrann}}}, \bibinfo {author} {\bibfnamefont {J.}~\bibnamefont {{Zuntz}}},
  \bibinfo {author} {\bibfnamefont {T.~F.}\ \bibnamefont {{Eifler}}}, \bibinfo
  {author} {\bibfnamefont {E.}~\bibnamefont {{Krause}}}, \bibinfo {author}
  {\bibfnamefont {S.}~\bibnamefont {{Dodelson}}}, \bibinfo {author}
  {\bibfnamefont {D.}~\bibnamefont {{Gruen}}}, \bibinfo {author} {\bibfnamefont
  {J.}~\bibnamefont {{Blazek}}}, \bibinfo {author} {\bibfnamefont
  {O.}~\bibnamefont {{Friedrich}}}, \bibinfo {author} {\bibfnamefont
  {S.}~\bibnamefont {{Samuroff}}}, \bibinfo {author} {\bibfnamefont
  {J.}~\bibnamefont {{Prat}}}, \bibinfo {author} {\bibfnamefont {L.~F.}\
  \bibnamefont {{Secco}}}, \bibinfo {author} {\bibfnamefont {C.}~\bibnamefont
  {{Davis}}}, \bibinfo {author} {\bibfnamefont {A.}~\bibnamefont
  {{Fert{\'e}}}}, \bibinfo {author} {\bibfnamefont {J.}~\bibnamefont
  {{DeRose}}}, \bibinfo {author} {\bibfnamefont {A.}~\bibnamefont {{Alarcon}}},
  \bibinfo {author} {\bibfnamefont {A.}~\bibnamefont {{Amara}}}, \bibinfo
  {author} {\bibfnamefont {E.}~\bibnamefont {{Baxter}}}, \bibinfo {author}
  {\bibfnamefont {M.~R.}\ \bibnamefont {{Becker}}}, \bibinfo {author}
  {\bibfnamefont {G.~M.}\ \bibnamefont {{Bernstein}}}, \bibinfo {author}
  {\bibfnamefont {S.~L.}\ \bibnamefont {{Bridle}}}, \bibinfo {author}
  {\bibfnamefont {R.}~\bibnamefont {{Cawthon}}}, \bibinfo {author}
  {\bibfnamefont {C.}~\bibnamefont {{Chang}}}, \bibinfo {author} {\bibfnamefont
  {A.}~\bibnamefont {{Choi}}}, \bibinfo {author} {\bibfnamefont
  {J.}~\bibnamefont {{De Vicente}}}, \bibinfo {author} {\bibfnamefont
  {A.}~\bibnamefont {{Drlica-Wagner}}}, \bibinfo {author} {\bibfnamefont
  {J.}~\bibnamefont {{Elvin-Poole}}}, \bibinfo {author} {\bibfnamefont
  {J.}~\bibnamefont {{Frieman}}}, \bibinfo {author} {\bibfnamefont
  {M.}~\bibnamefont {{Gatti}}}, \bibinfo {author} {\bibfnamefont {W.~G.}\
  \bibnamefont {{Hartley}}}, \bibinfo {author} {\bibfnamefont {K.}~\bibnamefont
  {{Honscheid}}}, \bibinfo {author} {\bibfnamefont {B.}~\bibnamefont
  {{Hoyle}}}, \bibinfo {author} {\bibfnamefont {E.~M.}\ \bibnamefont {{Huff}}},
  \bibinfo {author} {\bibfnamefont {D.}~\bibnamefont {{Huterer}}}, \bibinfo
  {author} {\bibfnamefont {B.}~\bibnamefont {{Jain}}}, \bibinfo {author}
  {\bibfnamefont {M.}~\bibnamefont {{Jarvis}}}, \bibinfo {author}
  {\bibfnamefont {T.}~\bibnamefont {{Kacprzak}}}, \bibinfo {author}
  {\bibfnamefont {D.}~\bibnamefont {{Kirk}}}, \bibinfo {author} {\bibfnamefont
  {N.}~\bibnamefont {{Kokron}}}, \bibinfo {author} {\bibfnamefont
  {C.}~\bibnamefont {{Krawiec}}}, \bibinfo {author} {\bibfnamefont
  {O.}~\bibnamefont {{Lahav}}}, \bibinfo {author} {\bibfnamefont {A.~R.}\
  \bibnamefont {{Liddle}}}, \bibinfo {author} {\bibfnamefont {J.}~\bibnamefont
  {{Peacock}}}, \bibinfo {author} {\bibfnamefont {M.~M.}\ \bibnamefont
  {{Rau}}}, \bibinfo {author} {\bibfnamefont {A.}~\bibnamefont {{Refregier}}},
  \bibinfo {author} {\bibfnamefont {R.~P.}\ \bibnamefont {{Rollins}}}, \bibinfo
  {author} {\bibfnamefont {E.}~\bibnamefont {{Rozo}}}, \bibinfo {author}
  {\bibfnamefont {E.~S.}\ \bibnamefont {{Rykoff}}}, \bibinfo {author}
  {\bibfnamefont {C.}~\bibnamefont {{S{\'a}nchez}}}, \bibinfo {author}
  {\bibfnamefont {I.}~\bibnamefont {{Sevilla-Noarbe}}}, \bibinfo {author}
  {\bibfnamefont {E.}~\bibnamefont {{Sheldon}}}, \bibinfo {author}
  {\bibfnamefont {A.}~\bibnamefont {{Stebbins}}}, \bibinfo {author}
  {\bibfnamefont {T.~N.}\ \bibnamefont {{Varga}}}, \bibinfo {author}
  {\bibfnamefont {P.}~\bibnamefont {{Vielzeuf}}}, \bibinfo {author}
  {\bibfnamefont {M.}~\bibnamefont {{Wang}}}, \bibinfo {author} {\bibfnamefont
  {R.~H.}\ \bibnamefont {{Wechsler}}}, \bibinfo {author} {\bibfnamefont
  {B.}~\bibnamefont {{Yanny}}}, \bibinfo {author} {\bibfnamefont {T.~M.~C.}\
  \bibnamefont {{Abbott}}}, \bibinfo {author} {\bibfnamefont {F.~B.}\
  \bibnamefont {{Abdalla}}}, \bibinfo {author} {\bibfnamefont {S.}~\bibnamefont
  {{Allam}}}, \bibinfo {author} {\bibfnamefont {J.}~\bibnamefont {{Annis}}},
  \bibinfo {author} {\bibfnamefont {K.}~\bibnamefont {{Bechtol}}}, \bibinfo
  {author} {\bibfnamefont {A.}~\bibnamefont {{Benoit-L{\'e}vy}}}, \bibinfo
  {author} {\bibfnamefont {E.}~\bibnamefont {{Bertin}}}, \bibinfo {author}
  {\bibfnamefont {D.}~\bibnamefont {{Brooks}}}, \bibinfo {author}
  {\bibfnamefont {E.}~\bibnamefont {{Buckley-Geer}}}, \bibinfo {author}
  {\bibfnamefont {D.~L.}\ \bibnamefont {{Burke}}}, \bibinfo {author}
  {\bibfnamefont {A.}~\bibnamefont {{Carnero Rosell}}}, \bibinfo {author}
  {\bibfnamefont {M.}~\bibnamefont {{Carrasco Kind}}}, \bibinfo {author}
  {\bibfnamefont {J.}~\bibnamefont {{Carretero}}}, \bibinfo {author}
  {\bibfnamefont {F.~J.}\ \bibnamefont {{Castander}}}, \bibinfo {author}
  {\bibfnamefont {M.}~\bibnamefont {{Crocce}}}, \bibinfo {author}
  {\bibfnamefont {C.~E.}\ \bibnamefont {{Cunha}}}, \bibinfo {author}
  {\bibfnamefont {C.~B.}\ \bibnamefont {{D'Andrea}}}, \bibinfo {author}
  {\bibfnamefont {L.~N.}\ \bibnamefont {{da Costa}}}, \bibinfo {author}
  {\bibfnamefont {D.~L.}\ \bibnamefont {{DePoy}}}, \bibinfo {author}
  {\bibfnamefont {S.}~\bibnamefont {{Desai}}}, \bibinfo {author} {\bibfnamefont
  {H.~T.}\ \bibnamefont {{Diehl}}}, \bibinfo {author} {\bibfnamefont {J.~P.}\
  \bibnamefont {{Dietrich}}}, \bibinfo {author} {\bibfnamefont
  {P.}~\bibnamefont {{Doel}}}, \bibinfo {author} {\bibfnamefont
  {E.}~\bibnamefont {{Fernandez}}}, \bibinfo {author} {\bibfnamefont
  {B.}~\bibnamefont {{Flaugher}}}, \bibinfo {author} {\bibfnamefont
  {P.}~\bibnamefont {{Fosalba}}}, \bibinfo {author} {\bibfnamefont
  {J.}~\bibnamefont {{Garc{\'\i}a-Bellido}}}, \bibinfo {author} {\bibfnamefont
  {E.}~\bibnamefont {{Gaztanaga}}}, \bibinfo {author} {\bibfnamefont {D.~W.}\
  \bibnamefont {{Gerdes}}}, \bibinfo {author} {\bibfnamefont {T.}~\bibnamefont
  {{Giannantonio}}}, \bibinfo {author} {\bibfnamefont {D.~A.}\ \bibnamefont
  {{Goldstein}}}, \bibinfo {author} {\bibfnamefont {R.~A.}\ \bibnamefont
  {{Gruendl}}}, \bibinfo {author} {\bibfnamefont {J.}~\bibnamefont
  {{Gschwend}}}, \bibinfo {author} {\bibfnamefont {G.}~\bibnamefont
  {{Gutierrez}}}, \bibinfo {author} {\bibfnamefont {D.~J.}\ \bibnamefont
  {{James}}}, \bibinfo {author} {\bibfnamefont {T.}~\bibnamefont {{Jeltema}}},
  \bibinfo {author} {\bibfnamefont {M.~W.~G.}\ \bibnamefont {{Johnson}}},
  \bibinfo {author} {\bibfnamefont {M.~D.}\ \bibnamefont {{Johnson}}}, \bibinfo
  {author} {\bibfnamefont {S.}~\bibnamefont {{Kent}}}, \bibinfo {author}
  {\bibfnamefont {K.}~\bibnamefont {{Kuehn}}}, \bibinfo {author} {\bibfnamefont
  {S.}~\bibnamefont {{Kuhlmann}}}, \bibinfo {author} {\bibfnamefont
  {N.}~\bibnamefont {{Kuropatkin}}}, \bibinfo {author} {\bibfnamefont {T.~S.}\
  \bibnamefont {{Li}}}, \bibinfo {author} {\bibfnamefont {M.}~\bibnamefont
  {{Lima}}}, \bibinfo {author} {\bibfnamefont {H.}~\bibnamefont {{Lin}}},
  \bibinfo {author} {\bibfnamefont {M.~A.~G.}\ \bibnamefont {{Maia}}}, \bibinfo
  {author} {\bibfnamefont {M.}~\bibnamefont {{March}}}, \bibinfo {author}
  {\bibfnamefont {J.~L.}\ \bibnamefont {{Marshall}}}, \bibinfo {author}
  {\bibfnamefont {P.}~\bibnamefont {{Martini}}}, \bibinfo {author}
  {\bibfnamefont {P.}~\bibnamefont {{Melchior}}}, \bibinfo {author}
  {\bibfnamefont {F.}~\bibnamefont {{Menanteau}}}, \bibinfo {author}
  {\bibfnamefont {R.}~\bibnamefont {{Miquel}}}, \bibinfo {author}
  {\bibfnamefont {J.~J.}\ \bibnamefont {{Mohr}}}, \bibinfo {author}
  {\bibfnamefont {E.}~\bibnamefont {{Neilsen}}}, \bibinfo {author}
  {\bibfnamefont {R.~C.}\ \bibnamefont {{Nichol}}}, \bibinfo {author}
  {\bibfnamefont {B.}~\bibnamefont {{Nord}}}, \bibinfo {author} {\bibfnamefont
  {D.}~\bibnamefont {{Petravick}}}, \bibinfo {author} {\bibfnamefont {A.~A.}\
  \bibnamefont {{Plazas}}}, \bibinfo {author} {\bibfnamefont {A.~K.}\
  \bibnamefont {{Romer}}}, \bibinfo {author} {\bibfnamefont {A.}~\bibnamefont
  {{Roodman}}}, \bibinfo {author} {\bibfnamefont {M.}~\bibnamefont {{Sako}}},
  \bibinfo {author} {\bibfnamefont {E.}~\bibnamefont {{Sanchez}}}, \bibinfo
  {author} {\bibfnamefont {V.}~\bibnamefont {{Scarpine}}}, \bibinfo {author}
  {\bibfnamefont {R.}~\bibnamefont {{Schindler}}}, \bibinfo {author}
  {\bibfnamefont {M.}~\bibnamefont {{Schubnell}}}, \bibinfo {author}
  {\bibfnamefont {M.}~\bibnamefont {{Smith}}}, \bibinfo {author} {\bibfnamefont
  {R.~C.}\ \bibnamefont {{Smith}}}, \bibinfo {author} {\bibfnamefont
  {M.}~\bibnamefont {{Soares-Santos}}}, \bibinfo {author} {\bibfnamefont
  {F.}~\bibnamefont {{Sobreira}}}, \bibinfo {author} {\bibfnamefont
  {E.}~\bibnamefont {{Suchyta}}}, \bibinfo {author} {\bibfnamefont {M.~E.~C.}\
  \bibnamefont {{Swanson}}}, \bibinfo {author} {\bibfnamefont {G.}~\bibnamefont
  {{Tarle}}}, \bibinfo {author} {\bibfnamefont {D.}~\bibnamefont {{Thomas}}},
  \bibinfo {author} {\bibfnamefont {D.~L.}\ \bibnamefont {{Tucker}}}, \bibinfo
  {author} {\bibfnamefont {V.}~\bibnamefont {{Vikram}}}, \bibinfo {author}
  {\bibfnamefont {A.~R.}\ \bibnamefont {{Walker}}}, \bibinfo {author}
  {\bibfnamefont {J.}~\bibnamefont {{Weller}}}, \bibinfo {author}
  {\bibfnamefont {Y.}~\bibnamefont {{Zhang}}},\ and\ \bibinfo {author}
  {\bibnamefont {{DES Collaboration}}},\ }\href
  {https://doi.org/10.1103/PhysRevD.98.043528} {\bibfield  {journal} {\bibinfo
  {journal} {\prd}\ }\textbf {\bibinfo {volume} {98}},\ \bibinfo {eid} {043528}
  (\bibinfo {year} {2018})},\ \Eprint {https://arxiv.org/abs/1708.01538}
  {arXiv:1708.01538 [astro-ph.CO]} \BibitemShut {NoStop}%
\bibitem [{\citenamefont {{Hamana}}\ \emph {et~al.}(2020)\citenamefont
  {{Hamana}}, \citenamefont {{Shirasaki}}, \citenamefont {{Miyazaki}},
  \citenamefont {{Hikage}}, \citenamefont {{Oguri}}, \citenamefont {{More}},
  \citenamefont {{Armstrong}}, \citenamefont {{Leauthaud}}, \citenamefont
  {{Mandelbaum}}, \citenamefont {{Miyatake}}, \citenamefont {{Nishizawa}},
  \citenamefont {{Simet}}, \citenamefont {{Takada}}, \citenamefont {{Aihara}},
  \citenamefont {{Bosch}}, \citenamefont {{Komiyama}}, \citenamefont
  {{Lupton}}, \citenamefont {{Murayama}}, \citenamefont {{Strauss}},\ and\
  \citenamefont {{Tanaka}}}]{2020PASJ...72...16H}%
  \BibitemOpen
  \bibfield  {author} {\bibinfo {author} {\bibfnamefont {T.}~\bibnamefont
  {{Hamana}}}, \bibinfo {author} {\bibfnamefont {M.}~\bibnamefont
  {{Shirasaki}}}, \bibinfo {author} {\bibfnamefont {S.}~\bibnamefont
  {{Miyazaki}}}, \bibinfo {author} {\bibfnamefont {C.}~\bibnamefont
  {{Hikage}}}, \bibinfo {author} {\bibfnamefont {M.}~\bibnamefont {{Oguri}}},
  \bibinfo {author} {\bibfnamefont {S.}~\bibnamefont {{More}}}, \bibinfo
  {author} {\bibfnamefont {R.}~\bibnamefont {{Armstrong}}}, \bibinfo {author}
  {\bibfnamefont {A.}~\bibnamefont {{Leauthaud}}}, \bibinfo {author}
  {\bibfnamefont {R.}~\bibnamefont {{Mandelbaum}}}, \bibinfo {author}
  {\bibfnamefont {H.}~\bibnamefont {{Miyatake}}}, \bibinfo {author}
  {\bibfnamefont {A.~J.}\ \bibnamefont {{Nishizawa}}}, \bibinfo {author}
  {\bibfnamefont {M.}~\bibnamefont {{Simet}}}, \bibinfo {author} {\bibfnamefont
  {M.}~\bibnamefont {{Takada}}}, \bibinfo {author} {\bibfnamefont
  {H.}~\bibnamefont {{Aihara}}}, \bibinfo {author} {\bibfnamefont
  {J.}~\bibnamefont {{Bosch}}}, \bibinfo {author} {\bibfnamefont
  {Y.}~\bibnamefont {{Komiyama}}}, \bibinfo {author} {\bibfnamefont
  {R.}~\bibnamefont {{Lupton}}}, \bibinfo {author} {\bibfnamefont
  {H.}~\bibnamefont {{Murayama}}}, \bibinfo {author} {\bibfnamefont {M.~A.}\
  \bibnamefont {{Strauss}}},\ and\ \bibinfo {author} {\bibfnamefont
  {M.}~\bibnamefont {{Tanaka}}},\ }\href {https://doi.org/10.1093/pasj/psz138}
  {\bibfield  {journal} {\bibinfo  {journal} {\pasj}\ }\textbf {\bibinfo
  {volume} {72}},\ \bibinfo {eid} {16} (\bibinfo {year} {2020})},\ \Eprint
  {https://arxiv.org/abs/1906.06041} {arXiv:1906.06041 [astro-ph.CO]}
  \BibitemShut {NoStop}%
\bibitem [{\citenamefont {{Dor{\'e}}}\ \emph {et~al.}(2014)\citenamefont
  {{Dor{\'e}}}, \citenamefont {{Bock}}, \citenamefont {{Ashby}}, \citenamefont
  {{Capak}}, \citenamefont {{Cooray}}, \citenamefont {{de Putter}},
  \citenamefont {{Eifler}}, \citenamefont {{Flagey}}, \citenamefont {{Gong}},
  \citenamefont {{Habib}}, \citenamefont {{Heitmann}}, \citenamefont
  {{Hirata}}, \citenamefont {{Jeong}}, \citenamefont {{Katti}}, \citenamefont
  {{Korngut}}, \citenamefont {{Krause}}, \citenamefont {{Lee}}, \citenamefont
  {{Masters}}, \citenamefont {{Mauskopf}}, \citenamefont {{Melnick}},
  \citenamefont {{Mennesson}}, \citenamefont {{Nguyen}}, \citenamefont
  {{{\"O}berg}}, \citenamefont {{Pullen}}, \citenamefont {{Raccanelli}},
  \citenamefont {{Smith}}, \citenamefont {{Song}}, \citenamefont {{Tolls}},
  \citenamefont {{Unwin}}, \citenamefont {{Venumadhav}}, \citenamefont
  {{Viero}}, \citenamefont {{Werner}},\ and\ \citenamefont
  {{Zemcov}}}]{2014arXiv1412.4872D}%
  \BibitemOpen
  \bibfield  {author} {\bibinfo {author} {\bibfnamefont {O.}~\bibnamefont
  {{Dor{\'e}}}}, \bibinfo {author} {\bibfnamefont {J.}~\bibnamefont {{Bock}}},
  \bibinfo {author} {\bibfnamefont {M.}~\bibnamefont {{Ashby}}}, \bibinfo
  {author} {\bibfnamefont {P.}~\bibnamefont {{Capak}}}, \bibinfo {author}
  {\bibfnamefont {A.}~\bibnamefont {{Cooray}}}, \bibinfo {author}
  {\bibfnamefont {R.}~\bibnamefont {{de Putter}}}, \bibinfo {author}
  {\bibfnamefont {T.}~\bibnamefont {{Eifler}}}, \bibinfo {author}
  {\bibfnamefont {N.}~\bibnamefont {{Flagey}}}, \bibinfo {author}
  {\bibfnamefont {Y.}~\bibnamefont {{Gong}}}, \bibinfo {author} {\bibfnamefont
  {S.}~\bibnamefont {{Habib}}}, \bibinfo {author} {\bibfnamefont
  {K.}~\bibnamefont {{Heitmann}}}, \bibinfo {author} {\bibfnamefont
  {C.}~\bibnamefont {{Hirata}}}, \bibinfo {author} {\bibfnamefont {W.-S.}\
  \bibnamefont {{Jeong}}}, \bibinfo {author} {\bibfnamefont {R.}~\bibnamefont
  {{Katti}}}, \bibinfo {author} {\bibfnamefont {P.}~\bibnamefont {{Korngut}}},
  \bibinfo {author} {\bibfnamefont {E.}~\bibnamefont {{Krause}}}, \bibinfo
  {author} {\bibfnamefont {D.-H.}\ \bibnamefont {{Lee}}}, \bibinfo {author}
  {\bibfnamefont {D.}~\bibnamefont {{Masters}}}, \bibinfo {author}
  {\bibfnamefont {P.}~\bibnamefont {{Mauskopf}}}, \bibinfo {author}
  {\bibfnamefont {G.}~\bibnamefont {{Melnick}}}, \bibinfo {author}
  {\bibfnamefont {B.}~\bibnamefont {{Mennesson}}}, \bibinfo {author}
  {\bibfnamefont {H.}~\bibnamefont {{Nguyen}}}, \bibinfo {author}
  {\bibfnamefont {K.}~\bibnamefont {{{\"O}berg}}}, \bibinfo {author}
  {\bibfnamefont {A.}~\bibnamefont {{Pullen}}}, \bibinfo {author}
  {\bibfnamefont {A.}~\bibnamefont {{Raccanelli}}}, \bibinfo {author}
  {\bibfnamefont {R.}~\bibnamefont {{Smith}}}, \bibinfo {author} {\bibfnamefont
  {Y.-S.}\ \bibnamefont {{Song}}}, \bibinfo {author} {\bibfnamefont
  {V.}~\bibnamefont {{Tolls}}}, \bibinfo {author} {\bibfnamefont
  {S.}~\bibnamefont {{Unwin}}}, \bibinfo {author} {\bibfnamefont
  {T.}~\bibnamefont {{Venumadhav}}}, \bibinfo {author} {\bibfnamefont
  {M.}~\bibnamefont {{Viero}}}, \bibinfo {author} {\bibfnamefont
  {M.}~\bibnamefont {{Werner}}},\ and\ \bibinfo {author} {\bibfnamefont
  {M.}~\bibnamefont {{Zemcov}}},\ }\href@noop {} {\bibfield  {journal}
  {\bibinfo  {journal} {arXiv e-prints}\ ,\ \bibinfo {eid} {arXiv:1412.4872}}
  (\bibinfo {year} {2014})},\ \Eprint {https://arxiv.org/abs/1412.4872}
  {arXiv:1412.4872 [astro-ph.CO]} \BibitemShut {NoStop}%
\bibitem [{\citenamefont {Limber}(1954)}]{Limber:1954zz}%
  \BibitemOpen
  \bibfield  {author} {\bibinfo {author} {\bibfnamefont {D.~N.}\ \bibnamefont
  {Limber}},\ }\href@noop {} {\bibfield  {journal} {\bibinfo  {journal} {ApJ}\
  }\textbf {\bibinfo {volume} {119}},\ \bibinfo {pages} {655} (\bibinfo {year}
  {1954})}\BibitemShut {NoStop}%
\bibitem [{\citenamefont {{Takahashi}}\ \emph {et~al.}(2012)\citenamefont
  {{Takahashi}}, \citenamefont {{Sato}}, \citenamefont {{Nishimichi}},
  \citenamefont {{Taruya}},\ and\ \citenamefont
  {{Oguri}}}]{2012ApJ...761..152T}%
  \BibitemOpen
  \bibfield  {author} {\bibinfo {author} {\bibfnamefont {R.}~\bibnamefont
  {{Takahashi}}}, \bibinfo {author} {\bibfnamefont {M.}~\bibnamefont {{Sato}}},
  \bibinfo {author} {\bibfnamefont {T.}~\bibnamefont {{Nishimichi}}}, \bibinfo
  {author} {\bibfnamefont {A.}~\bibnamefont {{Taruya}}},\ and\ \bibinfo
  {author} {\bibfnamefont {M.}~\bibnamefont {{Oguri}}},\ }\href
  {https://doi.org/10.1088/0004-637X/761/2/152} {\bibfield  {journal} {\bibinfo
   {journal} {\apj}\ }\textbf {\bibinfo {volume} {761}},\ \bibinfo {eid} {152}
  (\bibinfo {year} {2012})},\ \Eprint {https://arxiv.org/abs/1208.2701}
  {arXiv:1208.2701 [astro-ph.CO]} \BibitemShut {NoStop}%
\bibitem [{\citenamefont {{Chisari}}\ \emph {et~al.}(2019)\citenamefont
  {{Chisari}}, \citenamefont {{Mead}}, \citenamefont {{Joudaki}}, \citenamefont
  {{Ferreira}}, \citenamefont {{Schneider}}, \citenamefont {{Mohr}},
  \citenamefont {{Tr{\"o}ster}}, \citenamefont {{Alonso}}, \citenamefont
  {{McCarthy}}, \citenamefont {{Martin-Alvarez}}, \citenamefont {{Devriendt}},
  \citenamefont {{Slyz}},\ and\ \citenamefont {{van
  Daalen}}}]{2019OJAp....2E...4C}%
  \BibitemOpen
  \bibfield  {author} {\bibinfo {author} {\bibfnamefont {N.~E.}\ \bibnamefont
  {{Chisari}}}, \bibinfo {author} {\bibfnamefont {A.~J.}\ \bibnamefont
  {{Mead}}}, \bibinfo {author} {\bibfnamefont {S.}~\bibnamefont {{Joudaki}}},
  \bibinfo {author} {\bibfnamefont {P.~G.}\ \bibnamefont {{Ferreira}}},
  \bibinfo {author} {\bibfnamefont {A.}~\bibnamefont {{Schneider}}}, \bibinfo
  {author} {\bibfnamefont {J.}~\bibnamefont {{Mohr}}}, \bibinfo {author}
  {\bibfnamefont {T.}~\bibnamefont {{Tr{\"o}ster}}}, \bibinfo {author}
  {\bibfnamefont {D.}~\bibnamefont {{Alonso}}}, \bibinfo {author}
  {\bibfnamefont {I.~G.}\ \bibnamefont {{McCarthy}}}, \bibinfo {author}
  {\bibfnamefont {S.}~\bibnamefont {{Martin-Alvarez}}}, \bibinfo {author}
  {\bibfnamefont {J.}~\bibnamefont {{Devriendt}}}, \bibinfo {author}
  {\bibfnamefont {A.}~\bibnamefont {{Slyz}}},\ and\ \bibinfo {author}
  {\bibfnamefont {M.~P.}\ \bibnamefont {{van Daalen}}},\ }\href
  {https://doi.org/10.21105/astro.1905.06082} {\bibfield  {journal} {\bibinfo
  {journal} {The Open Journal of Astrophysics}\ }\textbf {\bibinfo {volume}
  {2}},\ \bibinfo {eid} {4} (\bibinfo {year} {2019})},\ \Eprint
  {https://arxiv.org/abs/1905.06082} {arXiv:1905.06082 [astro-ph.CO]}
  \BibitemShut {NoStop}%
\bibitem [{\citenamefont {{Cooray}}\ and\ \citenamefont
  {{Sheth}}(2002)}]{2002PhR...372....1C}%
  \BibitemOpen
  \bibfield  {author} {\bibinfo {author} {\bibfnamefont {A.}~\bibnamefont
  {{Cooray}}}\ and\ \bibinfo {author} {\bibfnamefont {R.}~\bibnamefont
  {{Sheth}}},\ }\href {https://doi.org/10.1016/S0370-1573(02)00276-4}
  {\bibfield  {journal} {\bibinfo  {journal} {\physrep}\ }\textbf {\bibinfo
  {volume} {372}},\ \bibinfo {pages} {1} (\bibinfo {year} {2002})},\ \Eprint
  {https://arxiv.org/abs/astro-ph/0206508} {arXiv:astro-ph/0206508 [astro-ph]}
  \BibitemShut {NoStop}%
\bibitem [{\citenamefont {{Navarro}}\ \emph {et~al.}(1996)\citenamefont
  {{Navarro}}, \citenamefont {{Frenk}},\ and\ \citenamefont
  {{White}}}]{1996ApJ...462..563N}%
  \BibitemOpen
  \bibfield  {author} {\bibinfo {author} {\bibfnamefont {J.~F.}\ \bibnamefont
  {{Navarro}}}, \bibinfo {author} {\bibfnamefont {C.~S.}\ \bibnamefont
  {{Frenk}}},\ and\ \bibinfo {author} {\bibfnamefont {S.~D.~M.}\ \bibnamefont
  {{White}}},\ }\href {https://doi.org/10.1086/177173} {\bibfield  {journal}
  {\bibinfo  {journal} {\apj}\ }\textbf {\bibinfo {volume} {462}},\ \bibinfo
  {pages} {563} (\bibinfo {year} {1996})},\ \Eprint
  {https://arxiv.org/abs/astro-ph/9508025} {arXiv:astro-ph/9508025 [astro-ph]}
  \BibitemShut {NoStop}%
\bibitem [{\citenamefont {{Diemer}}\ and\ \citenamefont
  {{Kravtsov}}(2015)}]{2015ApJ...799..108D}%
  \BibitemOpen
  \bibfield  {author} {\bibinfo {author} {\bibfnamefont {B.}~\bibnamefont
  {{Diemer}}}\ and\ \bibinfo {author} {\bibfnamefont {A.~V.}\ \bibnamefont
  {{Kravtsov}}},\ }\href {https://doi.org/10.1088/0004-637X/799/1/108}
  {\bibfield  {journal} {\bibinfo  {journal} {\apj}\ }\textbf {\bibinfo
  {volume} {799}},\ \bibinfo {eid} {108} (\bibinfo {year} {2015})},\ \Eprint
  {https://arxiv.org/abs/1407.4730} {arXiv:1407.4730 [astro-ph.CO]}
  \BibitemShut {NoStop}%
\bibitem [{\citenamefont {{Tinker}}\ \emph {et~al.}(2008)\citenamefont
  {{Tinker}}, \citenamefont {{Kravtsov}}, \citenamefont {{Klypin}},
  \citenamefont {{Abazajian}}, \citenamefont {{Warren}}, \citenamefont
  {{Yepes}}, \citenamefont {{Gottl{\"o}ber}},\ and\ \citenamefont
  {{Holz}}}]{2008ApJ...688..709T}%
  \BibitemOpen
  \bibfield  {author} {\bibinfo {author} {\bibfnamefont {J.}~\bibnamefont
  {{Tinker}}}, \bibinfo {author} {\bibfnamefont {A.~V.}\ \bibnamefont
  {{Kravtsov}}}, \bibinfo {author} {\bibfnamefont {A.}~\bibnamefont
  {{Klypin}}}, \bibinfo {author} {\bibfnamefont {K.}~\bibnamefont
  {{Abazajian}}}, \bibinfo {author} {\bibfnamefont {M.}~\bibnamefont
  {{Warren}}}, \bibinfo {author} {\bibfnamefont {G.}~\bibnamefont {{Yepes}}},
  \bibinfo {author} {\bibfnamefont {S.}~\bibnamefont {{Gottl{\"o}ber}}},\ and\
  \bibinfo {author} {\bibfnamefont {D.~E.}\ \bibnamefont {{Holz}}},\ }\href
  {https://doi.org/10.1086/591439} {\bibfield  {journal} {\bibinfo  {journal}
  {\apj}\ }\textbf {\bibinfo {volume} {688}},\ \bibinfo {pages} {709} (\bibinfo
  {year} {2008})},\ \Eprint {https://arxiv.org/abs/0803.2706} {arXiv:0803.2706
  [astro-ph]} \BibitemShut {NoStop}%
\bibitem [{\citenamefont {{Tinker}}\ \emph {et~al.}(2010)\citenamefont
  {{Tinker}}, \citenamefont {{Robertson}}, \citenamefont {{Kravtsov}},
  \citenamefont {{Klypin}}, \citenamefont {{Warren}}, \citenamefont {{Yepes}},\
  and\ \citenamefont {{Gottl{\"o}ber}}}]{2010ApJ...724..878T}%
  \BibitemOpen
  \bibfield  {author} {\bibinfo {author} {\bibfnamefont {J.~L.}\ \bibnamefont
  {{Tinker}}}, \bibinfo {author} {\bibfnamefont {B.~E.}\ \bibnamefont
  {{Robertson}}}, \bibinfo {author} {\bibfnamefont {A.~V.}\ \bibnamefont
  {{Kravtsov}}}, \bibinfo {author} {\bibfnamefont {A.}~\bibnamefont
  {{Klypin}}}, \bibinfo {author} {\bibfnamefont {M.~S.}\ \bibnamefont
  {{Warren}}}, \bibinfo {author} {\bibfnamefont {G.}~\bibnamefont {{Yepes}}},\
  and\ \bibinfo {author} {\bibfnamefont {S.}~\bibnamefont {{Gottl{\"o}ber}}},\
  }\href {https://doi.org/10.1088/0004-637X/724/2/878} {\bibfield  {journal}
  {\bibinfo  {journal} {\apj}\ }\textbf {\bibinfo {volume} {724}},\ \bibinfo
  {pages} {878} (\bibinfo {year} {2010})},\ \Eprint
  {https://arxiv.org/abs/1001.3162} {arXiv:1001.3162 [astro-ph.CO]}
  \BibitemShut {NoStop}%
\bibitem [{\citenamefont {{Guo}}\ \emph {et~al.}(2013)\citenamefont {{Guo}},
  \citenamefont {{White}}, \citenamefont {{Angulo}}, \citenamefont
  {{Henriques}}, \citenamefont {{Lemson}}, \citenamefont {{Boylan-Kolchin}},
  \citenamefont {{Thomas}},\ and\ \citenamefont
  {{Short}}}]{2013MNRAS.428.1351G}%
  \BibitemOpen
  \bibfield  {author} {\bibinfo {author} {\bibfnamefont {Q.}~\bibnamefont
  {{Guo}}}, \bibinfo {author} {\bibfnamefont {S.}~\bibnamefont {{White}}},
  \bibinfo {author} {\bibfnamefont {R.~E.}\ \bibnamefont {{Angulo}}}, \bibinfo
  {author} {\bibfnamefont {B.}~\bibnamefont {{Henriques}}}, \bibinfo {author}
  {\bibfnamefont {G.}~\bibnamefont {{Lemson}}}, \bibinfo {author}
  {\bibfnamefont {M.}~\bibnamefont {{Boylan-Kolchin}}}, \bibinfo {author}
  {\bibfnamefont {P.}~\bibnamefont {{Thomas}}},\ and\ \bibinfo {author}
  {\bibfnamefont {C.}~\bibnamefont {{Short}}},\ }\href
  {https://doi.org/10.1093/mnras/sts115} {\bibfield  {journal} {\bibinfo
  {journal} {\mnras}\ }\textbf {\bibinfo {volume} {428}},\ \bibinfo {pages}
  {1351} (\bibinfo {year} {2013})},\ \Eprint {https://arxiv.org/abs/1206.0052}
  {arXiv:1206.0052 [astro-ph.CO]} \BibitemShut {NoStop}%
\bibitem [{\citenamefont {{Ayala}}\ \emph {et~al.}(2014)\citenamefont
  {{Ayala}}, \citenamefont {{Dom{\'\i}nguez}}, \citenamefont {{Giannotti}},
  \citenamefont {{Mirizzi}},\ and\ \citenamefont
  {{Straniero}}}]{2014PhRvL.113s1302A}%
  \BibitemOpen
  \bibfield  {author} {\bibinfo {author} {\bibfnamefont {A.}~\bibnamefont
  {{Ayala}}}, \bibinfo {author} {\bibfnamefont {I.}~\bibnamefont
  {{Dom{\'\i}nguez}}}, \bibinfo {author} {\bibfnamefont {M.}~\bibnamefont
  {{Giannotti}}}, \bibinfo {author} {\bibfnamefont {A.}~\bibnamefont
  {{Mirizzi}}},\ and\ \bibinfo {author} {\bibfnamefont {O.}~\bibnamefont
  {{Straniero}}},\ }\href {https://doi.org/10.1103/PhysRevLett.113.191302}
  {\bibfield  {journal} {\bibinfo  {journal} {\prl}\ }\textbf {\bibinfo
  {volume} {113}},\ \bibinfo {eid} {191302} (\bibinfo {year} {2014})},\ \Eprint
  {https://arxiv.org/abs/1406.6053} {arXiv:1406.6053 [astro-ph.SR]}
  \BibitemShut {NoStop}%
\bibitem [{\citenamefont {{Arik}}\ \emph {et~al.}(2014)\citenamefont {{Arik}},
  \citenamefont {{Aune}}, \citenamefont {{Barth}}, \citenamefont {{Belov}},
  \citenamefont {{Borghi}}, \citenamefont {{Br{\"a}uninger}}, \citenamefont
  {{Cantatore}}, \citenamefont {{Carmona}}, \citenamefont {{Cetin}},
  \citenamefont {{Collar}}, \citenamefont {{Da Riva}}, \citenamefont {{Dafni}},
  \citenamefont {{Davenport}}, \citenamefont {{Eleftheriadis}}, \citenamefont
  {{Elias}}, \citenamefont {{Fanourakis}}, \citenamefont {{Ferrer-Ribas}},
  \citenamefont {{Friedrich}}, \citenamefont {{Gal{\'a}n}}, \citenamefont
  {{Garc{\'\i}a}}, \citenamefont {{Gardikiotis}}, \citenamefont {{Garza}},
  \citenamefont {{Gazis}}, \citenamefont {{Geralis}}, \citenamefont
  {{Georgiopoulou}}, \citenamefont {{Giomataris}}, \citenamefont {{Gninenko}},
  \citenamefont {{G{\'o}mez}}, \citenamefont {{G{\'o}mez Marzoa}},
  \citenamefont {{Gruber}}, \citenamefont {{Guth{\"o}rl}}, \citenamefont
  {{Hartmann}}, \citenamefont {{Hauf}}, \citenamefont {{Haug}}, \citenamefont
  {{Hasinoff}}, \citenamefont {{Hoffmann}}, \citenamefont {{Iguaz}},
  \citenamefont {{Irastorza}}, \citenamefont {{Jacoby}}, \citenamefont
  {{Jakov{\v{c}}i{\'c}}}, \citenamefont {{Karuza}}, \citenamefont
  {{K{\"o}nigsmann}}, \citenamefont {{Kotthaus}}, \citenamefont
  {{Kr{\v{c}}mar}}, \citenamefont {{Kuster}}, \citenamefont {{Laki{\'c}}},
  \citenamefont {{Lang}}, \citenamefont {{Laurent}}, \citenamefont {{Liolios}},
  \citenamefont {{Ljubi{\v{c}}i{\'c}}}, \citenamefont {{Luz{\'o}n}},
  \citenamefont {{Neff}}, \citenamefont {{Niinikoski}}, \citenamefont
  {{Nordt}}, \citenamefont {{Papaevangelou}}, \citenamefont {{Pivovaroff}},
  \citenamefont {{Raffelt}}, \citenamefont {{Riege}}, \citenamefont
  {{Rodr{\'\i}guez}}, \citenamefont {{Rosu}}, \citenamefont {{Ruz}},
  \citenamefont {{Savvidis}}, \citenamefont {{Shilon}}, \citenamefont
  {{Silva}}, \citenamefont {{Solanki}}, \citenamefont {{Stewart}},
  \citenamefont {{Tom{\'a}s}}, \citenamefont {{Tsagri}}, \citenamefont {{van
  Bibber}}, \citenamefont {{Vafeiadis}}, \citenamefont {{Villar}},
  \citenamefont {{Vogel}}, \citenamefont {{Yildiz}}, \citenamefont
  {{Zioutas}},\ and\ \citenamefont {{CAST
  Collaboration}}}]{2014PhRvL.112i1302A}%
  \BibitemOpen
  \bibfield  {author} {\bibinfo {author} {\bibfnamefont {M.}~\bibnamefont
  {{Arik}}}, \bibinfo {author} {\bibfnamefont {S.}~\bibnamefont {{Aune}}},
  \bibinfo {author} {\bibfnamefont {K.}~\bibnamefont {{Barth}}}, \bibinfo
  {author} {\bibfnamefont {A.}~\bibnamefont {{Belov}}}, \bibinfo {author}
  {\bibfnamefont {S.}~\bibnamefont {{Borghi}}}, \bibinfo {author}
  {\bibfnamefont {H.}~\bibnamefont {{Br{\"a}uninger}}}, \bibinfo {author}
  {\bibfnamefont {G.}~\bibnamefont {{Cantatore}}}, \bibinfo {author}
  {\bibfnamefont {J.~M.}\ \bibnamefont {{Carmona}}}, \bibinfo {author}
  {\bibfnamefont {S.~A.}\ \bibnamefont {{Cetin}}}, \bibinfo {author}
  {\bibfnamefont {J.~I.}\ \bibnamefont {{Collar}}}, \bibinfo {author}
  {\bibfnamefont {E.}~\bibnamefont {{Da Riva}}}, \bibinfo {author}
  {\bibfnamefont {T.}~\bibnamefont {{Dafni}}}, \bibinfo {author} {\bibfnamefont
  {M.}~\bibnamefont {{Davenport}}}, \bibinfo {author} {\bibfnamefont
  {C.}~\bibnamefont {{Eleftheriadis}}}, \bibinfo {author} {\bibfnamefont
  {N.}~\bibnamefont {{Elias}}}, \bibinfo {author} {\bibfnamefont
  {G.}~\bibnamefont {{Fanourakis}}}, \bibinfo {author} {\bibfnamefont
  {E.}~\bibnamefont {{Ferrer-Ribas}}}, \bibinfo {author} {\bibfnamefont
  {P.}~\bibnamefont {{Friedrich}}}, \bibinfo {author} {\bibfnamefont
  {J.}~\bibnamefont {{Gal{\'a}n}}}, \bibinfo {author} {\bibfnamefont {J.~A.}\
  \bibnamefont {{Garc{\'\i}a}}}, \bibinfo {author} {\bibfnamefont
  {A.}~\bibnamefont {{Gardikiotis}}}, \bibinfo {author} {\bibfnamefont {J.~G.}\
  \bibnamefont {{Garza}}}, \bibinfo {author} {\bibfnamefont {E.~N.}\
  \bibnamefont {{Gazis}}}, \bibinfo {author} {\bibfnamefont {T.}~\bibnamefont
  {{Geralis}}}, \bibinfo {author} {\bibfnamefont {E.}~\bibnamefont
  {{Georgiopoulou}}}, \bibinfo {author} {\bibfnamefont {I.}~\bibnamefont
  {{Giomataris}}}, \bibinfo {author} {\bibfnamefont {S.}~\bibnamefont
  {{Gninenko}}}, \bibinfo {author} {\bibfnamefont {H.}~\bibnamefont
  {{G{\'o}mez}}}, \bibinfo {author} {\bibfnamefont {M.}~\bibnamefont
  {{G{\'o}mez Marzoa}}}, \bibinfo {author} {\bibfnamefont {E.}~\bibnamefont
  {{Gruber}}}, \bibinfo {author} {\bibfnamefont {T.}~\bibnamefont
  {{Guth{\"o}rl}}}, \bibinfo {author} {\bibfnamefont {R.}~\bibnamefont
  {{Hartmann}}}, \bibinfo {author} {\bibfnamefont {S.}~\bibnamefont {{Hauf}}},
  \bibinfo {author} {\bibfnamefont {F.}~\bibnamefont {{Haug}}}, \bibinfo
  {author} {\bibfnamefont {M.~D.}\ \bibnamefont {{Hasinoff}}}, \bibinfo
  {author} {\bibfnamefont {D.~H.~H.}\ \bibnamefont {{Hoffmann}}}, \bibinfo
  {author} {\bibfnamefont {F.~J.}\ \bibnamefont {{Iguaz}}}, \bibinfo {author}
  {\bibfnamefont {I.~G.}\ \bibnamefont {{Irastorza}}}, \bibinfo {author}
  {\bibfnamefont {J.}~\bibnamefont {{Jacoby}}}, \bibinfo {author}
  {\bibfnamefont {K.}~\bibnamefont {{Jakov{\v{c}}i{\'c}}}}, \bibinfo {author}
  {\bibfnamefont {M.}~\bibnamefont {{Karuza}}}, \bibinfo {author}
  {\bibfnamefont {K.}~\bibnamefont {{K{\"o}nigsmann}}}, \bibinfo {author}
  {\bibfnamefont {R.}~\bibnamefont {{Kotthaus}}}, \bibinfo {author}
  {\bibfnamefont {M.}~\bibnamefont {{Kr{\v{c}}mar}}}, \bibinfo {author}
  {\bibfnamefont {M.}~\bibnamefont {{Kuster}}}, \bibinfo {author}
  {\bibfnamefont {B.}~\bibnamefont {{Laki{\'c}}}}, \bibinfo {author}
  {\bibfnamefont {P.~M.}\ \bibnamefont {{Lang}}}, \bibinfo {author}
  {\bibfnamefont {J.~M.}\ \bibnamefont {{Laurent}}}, \bibinfo {author}
  {\bibfnamefont {A.}~\bibnamefont {{Liolios}}}, \bibinfo {author}
  {\bibfnamefont {A.}~\bibnamefont {{Ljubi{\v{c}}i{\'c}}}}, \bibinfo {author}
  {\bibfnamefont {G.}~\bibnamefont {{Luz{\'o}n}}}, \bibinfo {author}
  {\bibfnamefont {S.}~\bibnamefont {{Neff}}}, \bibinfo {author} {\bibfnamefont
  {T.}~\bibnamefont {{Niinikoski}}}, \bibinfo {author} {\bibfnamefont
  {A.}~\bibnamefont {{Nordt}}}, \bibinfo {author} {\bibfnamefont
  {T.}~\bibnamefont {{Papaevangelou}}}, \bibinfo {author} {\bibfnamefont
  {M.~J.}\ \bibnamefont {{Pivovaroff}}}, \bibinfo {author} {\bibfnamefont
  {G.}~\bibnamefont {{Raffelt}}}, \bibinfo {author} {\bibfnamefont
  {H.}~\bibnamefont {{Riege}}}, \bibinfo {author} {\bibfnamefont
  {A.}~\bibnamefont {{Rodr{\'\i}guez}}}, \bibinfo {author} {\bibfnamefont
  {M.}~\bibnamefont {{Rosu}}}, \bibinfo {author} {\bibfnamefont
  {J.}~\bibnamefont {{Ruz}}}, \bibinfo {author} {\bibfnamefont
  {I.}~\bibnamefont {{Savvidis}}}, \bibinfo {author} {\bibfnamefont
  {I.}~\bibnamefont {{Shilon}}}, \bibinfo {author} {\bibfnamefont {P.~S.}\
  \bibnamefont {{Silva}}}, \bibinfo {author} {\bibfnamefont {S.~K.}\
  \bibnamefont {{Solanki}}}, \bibinfo {author} {\bibfnamefont {L.}~\bibnamefont
  {{Stewart}}}, \bibinfo {author} {\bibfnamefont {A.}~\bibnamefont
  {{Tom{\'a}s}}}, \bibinfo {author} {\bibfnamefont {M.}~\bibnamefont
  {{Tsagri}}}, \bibinfo {author} {\bibfnamefont {K.}~\bibnamefont {{van
  Bibber}}}, \bibinfo {author} {\bibfnamefont {T.}~\bibnamefont {{Vafeiadis}}},
  \bibinfo {author} {\bibfnamefont {J.}~\bibnamefont {{Villar}}}, \bibinfo
  {author} {\bibfnamefont {J.~K.}\ \bibnamefont {{Vogel}}}, \bibinfo {author}
  {\bibfnamefont {S.~C.}\ \bibnamefont {{Yildiz}}}, \bibinfo {author}
  {\bibfnamefont {K.}~\bibnamefont {{Zioutas}}},\ and\ \bibinfo {author}
  {\bibnamefont {{CAST Collaboration}}},\ }\href
  {https://doi.org/10.1103/PhysRevLett.112.091302} {\bibfield  {journal}
  {\bibinfo  {journal} {\prl}\ }\textbf {\bibinfo {volume} {112}},\ \bibinfo
  {eid} {091302} (\bibinfo {year} {2014})},\ \Eprint
  {https://arxiv.org/abs/1307.1985} {arXiv:1307.1985 [hep-ex]} \BibitemShut
  {NoStop}%
\bibitem [{\citenamefont {{Regis}}\ \emph {et~al.}(2020)\citenamefont
  {{Regis}}, \citenamefont {{Taoso}}, \citenamefont {{Vaz}}, \citenamefont
  {{Brinchmann}}, \citenamefont {{Zoutendijk}}, \citenamefont {{Bouch{\'e}}},\
  and\ \citenamefont {{Steinmetz}}}]{2020arXiv200901310R}%
  \BibitemOpen
  \bibfield  {author} {\bibinfo {author} {\bibfnamefont {M.}~\bibnamefont
  {{Regis}}}, \bibinfo {author} {\bibfnamefont {M.}~\bibnamefont {{Taoso}}},
  \bibinfo {author} {\bibfnamefont {D.}~\bibnamefont {{Vaz}}}, \bibinfo
  {author} {\bibfnamefont {J.}~\bibnamefont {{Brinchmann}}}, \bibinfo {author}
  {\bibfnamefont {S.~L.}\ \bibnamefont {{Zoutendijk}}}, \bibinfo {author}
  {\bibfnamefont {N.}~\bibnamefont {{Bouch{\'e}}}},\ and\ \bibinfo {author}
  {\bibfnamefont {M.}~\bibnamefont {{Steinmetz}}},\ }\href@noop {} {\bibfield
  {journal} {\bibinfo  {journal} {arXiv e-prints}\ ,\ \bibinfo {eid}
  {arXiv:2009.01310}} (\bibinfo {year} {2020})},\ \Eprint
  {https://arxiv.org/abs/2009.01310} {arXiv:2009.01310 [astro-ph.CO]}
  \BibitemShut {NoStop}%
\bibitem [{\citenamefont {{Hu}}(1999)}]{1999ApJ...522L..21H}%
  \BibitemOpen
  \bibfield  {author} {\bibinfo {author} {\bibfnamefont {W.}~\bibnamefont
  {{Hu}}},\ }\href {https://doi.org/10.1086/312210} {\bibfield  {journal}
  {\bibinfo  {journal} {\apjl}\ }\textbf {\bibinfo {volume} {522}},\ \bibinfo
  {pages} {L21} (\bibinfo {year} {1999})},\ \Eprint
  {https://arxiv.org/abs/astro-ph/9904153} {arXiv:astro-ph/9904153 [astro-ph]}
  \BibitemShut {NoStop}%
\bibitem [{\citenamefont {{Caputo}}\ \emph {et~al.}(2020)\citenamefont
  {{Caputo}}, \citenamefont {{Vittino}}, \citenamefont {{Fornengo}},
  \citenamefont {{Regis}},\ and\ \citenamefont
  {{Taoso}}}]{2020arXiv201209179C}%
  \BibitemOpen
  \bibfield  {author} {\bibinfo {author} {\bibfnamefont {A.}~\bibnamefont
  {{Caputo}}}, \bibinfo {author} {\bibfnamefont {A.}~\bibnamefont {{Vittino}}},
  \bibinfo {author} {\bibfnamefont {N.}~\bibnamefont {{Fornengo}}}, \bibinfo
  {author} {\bibfnamefont {M.}~\bibnamefont {{Regis}}},\ and\ \bibinfo {author}
  {\bibfnamefont {M.}~\bibnamefont {{Taoso}}},\ }\href@noop {} {\bibfield
  {journal} {\bibinfo  {journal} {arXiv e-prints}\ ,\ \bibinfo {eid}
  {arXiv:2012.09179}} (\bibinfo {year} {2020})},\ \Eprint
  {https://arxiv.org/abs/2012.09179} {arXiv:2012.09179 [astro-ph.CO]}
  \BibitemShut {NoStop}%
\bibitem [{\citenamefont {{Aghanim}}\ \emph {et~al.}(2008)\citenamefont
  {{Aghanim}}, \citenamefont {{Majumdar}},\ and\ \citenamefont
  {{Silk}}}]{2008RPPh...71f6902A}%
  \BibitemOpen
  \bibfield  {author} {\bibinfo {author} {\bibfnamefont {N.}~\bibnamefont
  {{Aghanim}}}, \bibinfo {author} {\bibfnamefont {S.}~\bibnamefont
  {{Majumdar}}},\ and\ \bibinfo {author} {\bibfnamefont {J.}~\bibnamefont
  {{Silk}}},\ }\href {https://doi.org/10.1088/0034-4885/71/6/066902} {\bibfield
   {journal} {\bibinfo  {journal} {Reports on Progress in Physics}\ }\textbf
  {\bibinfo {volume} {71}},\ \bibinfo {eid} {066902} (\bibinfo {year}
  {2008})},\ \Eprint {https://arxiv.org/abs/0711.0518} {arXiv:0711.0518
  [astro-ph]} \BibitemShut {NoStop}%
\bibitem [{\citenamefont {{Desjacques}}\ \emph {et~al.}(2018)\citenamefont
  {{Desjacques}}, \citenamefont {{Jeong}},\ and\ \citenamefont
  {{Schmidt}}}]{2018PhR...733....1D}%
  \BibitemOpen
  \bibfield  {author} {\bibinfo {author} {\bibfnamefont {V.}~\bibnamefont
  {{Desjacques}}}, \bibinfo {author} {\bibfnamefont {D.}~\bibnamefont
  {{Jeong}}},\ and\ \bibinfo {author} {\bibfnamefont {F.}~\bibnamefont
  {{Schmidt}}},\ }\href {https://doi.org/10.1016/j.physrep.2017.12.002}
  {\bibfield  {journal} {\bibinfo  {journal} {\physrep}\ }\textbf {\bibinfo
  {volume} {733}},\ \bibinfo {pages} {1} (\bibinfo {year} {2018})},\ \Eprint
  {https://arxiv.org/abs/1611.09787} {arXiv:1611.09787 [astro-ph.CO]}
  \BibitemShut {NoStop}%
\bibitem [{\citenamefont {{Peebles}}(1980)}]{1980lssu.book.....P}%
  \BibitemOpen
  \bibfield  {author} {\bibinfo {author} {\bibfnamefont {P.~J.~E.}\
  \bibnamefont {{Peebles}}},\ }\href@noop {} {\emph {\bibinfo {title} {{The
  large-scale structure of the universe}}}}\ (\bibinfo {year}
  {1980})\BibitemShut {NoStop}%
\bibitem [{\citenamefont {{Hamilton}}(1998)}]{1998ASSL..231..185H}%
  \BibitemOpen
  \bibfield  {author} {\bibinfo {author} {\bibfnamefont {A.~J.~S.}\
  \bibnamefont {{Hamilton}}},\ }\bibinfo {title} {{Linear Redshift Distortions:
  a Review}},\ in\ \href {https://doi.org/10.1007/978-94-011-4960-0_17} {\emph
  {\bibinfo {booktitle} {The Evolving Universe}}},\ Vol.\ \bibinfo {volume}
  {231},\ \bibinfo {editor} {edited by\ \bibinfo {editor} {\bibfnamefont
  {D.}~\bibnamefont {{Hamilton}}}}\ (\bibinfo {year} {1998})\ p.\ \bibinfo
  {pages} {185}\BibitemShut {NoStop}%
\bibitem [{\citenamefont {{Scoccimarro}}(2004)}]{2004PhRvD..70h3007S}%
  \BibitemOpen
  \bibfield  {author} {\bibinfo {author} {\bibfnamefont {R.}~\bibnamefont
  {{Scoccimarro}}},\ }\href {https://doi.org/10.1103/PhysRevD.70.083007}
  {\bibfield  {journal} {\bibinfo  {journal} {\prd}\ }\textbf {\bibinfo
  {volume} {70}},\ \bibinfo {eid} {083007} (\bibinfo {year} {2004})},\ \Eprint
  {https://arxiv.org/abs/astro-ph/0407214} {arXiv:astro-ph/0407214 [astro-ph]}
  \BibitemShut {NoStop}%
\bibitem [{\citenamefont {{Lewis}}\ and\ \citenamefont
  {{Challinor}}(2006)}]{2006PhR...429....1L}%
  \BibitemOpen
  \bibfield  {author} {\bibinfo {author} {\bibfnamefont {A.}~\bibnamefont
  {{Lewis}}}\ and\ \bibinfo {author} {\bibfnamefont {A.}~\bibnamefont
  {{Challinor}}},\ }\href {https://doi.org/10.1016/j.physrep.2006.03.002}
  {\bibfield  {journal} {\bibinfo  {journal} {\physrep}\ }\textbf {\bibinfo
  {volume} {429}},\ \bibinfo {pages} {1} (\bibinfo {year} {2006})},\ \Eprint
  {https://arxiv.org/abs/astro-ph/0601594} {arXiv:astro-ph/0601594 [astro-ph]}
  \BibitemShut {NoStop}%
\end{thebibliography}%

\end{document}